\documentclass[12pt]{article}
\usepackage[margin=1in]{geometry}
\usepackage[utf8]{inputenc}
\usepackage{setspace}
\usepackage{xcolor}
\usepackage{amsmath}
\usepackage{amsfonts}
\usepackage{amssymb}
\usepackage{graphicx}
\usepackage{pdflscape}
\usepackage{multirow}
\usepackage{natbib}
\usepackage{tabularx}
\usepackage{authblk}
\usepackage{hyperref}
\usepackage{circledsteps}

\title{Bayesian Superpopulation Models for Population-Based Diabetes Surveillance Using Multi-Site Electronic Health Record Data}

\author[1]{Rebecca Anthopolos\thanks{rebecca.anthopolos@nyulangone.org}}
\author[2]{Bo Cai}
\author[1]{Lorna E. Thorpe}
\author[3]{Hui Xie}
\author[1]{Xiaochun Li}
\author[1]{Sarah Conderino}
\author[1,4]{David C. Lee}
\author[5]{Hui Shao}
\author[3]{Ibrahim Zaganjor}
\author[6,7]{Marc Rosenman}
\author[8]{H. Lester Kirchner}
\author[9]{Muchiri E. Wandai}
\author[10,11]{Levon Utidjian}
\author[12]{Tessa Crume}
\author[13]{Kristi Reynolds}
\author[14]{Jasmin Divers}
\author[1,14]{DiCAYA Study Group}

\affil[1]{Department of Population Health, NYU Grossman School of Medicine, New York, USA}
\affil[2]{Department of Epidemiology and Biostatistics, University of South Carolina, South Carolina, USA}
\affil[3]{Division of Diabetes Translation, Centers for Disease Control and Prevention, Georgia, USA}
\affil[4]{Department of Emergency Medicine, NYU Grossman School of Medicine, New York, USA}
\affil[5]{Hubert Department of Global Health, Emory Rollins School of Public Health, Georgia, USA}
\affil[6]{Department of Pediatrics, Ann \& Robert H. Lurie Children's Hospital of Chicago, Illinois, USA}
\affil[7]{Department of Pediatrics, Northwestern University Feinberg School of Medicine, Illinois, USA}
\affil[8]{Department of Population Health Sciences, Geisinger Health System, Pennsylvania, USA}
\affil[9]{Center for Biomedical Informatics, Regenstrief Institute, Indiana, USA}
\affil[10]{Applied Clinical Research Center, Children's Hospital of Philadelphia, Pennsylvania, USA}
\affil[11]{Department of Pediatrics, Children's Hospital of Philadelphia, Pennsylvania, USA}
\affil[12]{Lifecourse Epidemiology of Adiposity and Diabetes, University of Colorado Anschutz Medical Campus, Colorado, USA}
\affil[13]{Department of Research \& Evaluation, Kaiser Permanente Southern California, California, USA}
\affil[14]{Department of Foundations of Medicine, NYU Grossman Long Island School of Medicine, New York, USA}

\date{}

\begin{document}

\maketitle

\begin{center}
\begin{minipage}{0.95\textwidth}
\small
\textbf{Funding:} This work was supported by the Centers for Disease Control and Prevention’s National Center for Chronic Disease Prevention and Health Promotion and the National Institute for Diabetes and Digestive and Kidney Diseases (awards U18DP006521 to Children’s Hospital of Philadelphia; U18DP006512 to University of Florida; U18DP006509 to Geisinger; U18DP006500 to Indiana University Indianapolis; U18DP006513 to University of South Carolina; U18DP006506 to Kaiser Foundation Hospitals; U18DP006693 and U18DP006694 to Lurie Children’s Hospital; U18DP006517 to University of Colorado Component-A; U18DP006518 to University of Colorado Component-B; and U18DP006633 to New York University Long Island School of Medicine).
\vspace{0.75em}

\textbf{Disclaimer:} The findings and conclusions in this report are those of the authors and do not necessarily represent the official position of the CDC.

\vspace{0.75em}

\end{minipage}
\end{center}

\begin{abstract}
    Electronic health records (EHRs) are increasingly used for disease surveillance. However, data collected in EHRs may not be a representative sample of the target population for inference. Methods for EHR-based surveillance remain in their nascence. In addition to simple weighting approaches, researchers have used multilevel regression and poststratification (MRP), a popular technique for finite population quantity estimation that remains understudied in EHRs. We are motivated by a US-based collaborative to modernize diabetes surveillance using EHR data from different health system centers across the country. To estimate type 1 diabetes (T1D) prevalence in children and adolescents, we propose and evaluate three Bayesian superpopulation models that are distinguished by how they account for two important features of EHR data: unobserved heterogeneity and data sparsity. The three models include an analysis of variance-inspired approach; a model with geographic effects varying by demographic subgroup; and MRP, which we consider as the reference model. We apply the models to estimate prevalence overall and in demographic- and geographic-based subgroups. Using plasmode simulations, we evaluate model estimation performance while preserving the complexity of the EHR data. Our results show that the superpopulation-based prevalence estimates were higher than those observed in the EHR sample alone, with the magnitude of the differences varying by health system center. The bias and coverage probabilities of MRP markedly worsened when the EHR sample size was smaller or the selection mechanism was more complex. Overall, the superpopulation model based on analysis of variance demonstrated the best estimation performance.
    \end{abstract}
    
\section{Introduction}

In public health surveillance, the objective is to estimate some finite population quantity of interest (e.g., disease prevalence) in a target population for inference. Unlike the surveillance for many infectious diseases that are reportable by law, the gold standard for chronic disease surveillance has been to use a probability sample collected from a complex survey design whereby all individuals are contained in the sampling frame and have a known non-zero probability of selection. Sampling weights enable projecting the sample to the target population to derive estimates of population quantities. For example, using sampling weights in the US National Health and Nutrition Examination Surveys (NHANES), \citet{Cheng2019} recently provided nationwide estimates of diabetes prevalence by race and ethnicity. Despite the ease of estimation, disease surveillance with probability samples suffers from long time lags and faces growing challenges related to rising costs and worsening problems of nonresponse \citep{Kohut2012,Brick2013,Elliott2017}.

Routinely collected data in electronic health records (EHRs) have received increasing attention as a potential data source for public health surveillance \citep{Paul2015,Bower2017,Klompas2017}. EHRs may offer the opportunity for timely, low-cost surveillance, including of rare diseases and in understudied demographic groups. We are motivated by the Assessing the Burden of \underline{\textbf{D}}iabetes by Type in \underline{\textbf{C}}hildren, \underline{\textbf{A}}dolescents and \underline{\textbf{Y}}oung \underline{\textbf{A}}dults (DiCAYA) Network. DiCAYA is a U.S. multi-site collaboration funded by the Centers for Disease Control and Prevention  and the National Institute of Diabetes and Digestive and Kidney Diseases that aims to modernize diabetes surveillance through young adulthood by capitalizing on large-volume EHR data \citep{Hirsch2024}. A primary goal of DiCAYA is to generate population-based prevalence estimates of diabetes by type, overall and by demographic- and geographic-based subgroups. Using EHRs for population-based surveillance, however, introduces a formidable statistical challenge. Unlike a probability sample, data collected in EHRs constitute a convenience sample -- a type of nonprobability sample -- in which no sampling weights are available. Akin to the missing data mechanism in the missing data lexicon, the selection mechanism (defined as the process by which individuals from the target population select into the EHR sample) is unknown \citep{Elliott2017}. Prior research indicates that compared to the population, patients in EHRs are more likely to be older and skew towards overrepresenting female, higher socioeconomic status, and sicker individuals \citep{Romo2016}. Thus, valid population-based disease surveillance with EHRs rests on statistical methodologies that can correct for the fact that patients in EHRs may be nonrepresentative of the target population for inference.

In nonprobability samples, two broad frameworks for finite population quantity estimation have been explicated \citep{Elliott2017,Valliant2020}. First, in superpopulation modeling, a prediction model is built, and population estimates are typically derived from observed and predicted responses. Second, in the quasirandomization approach, the probability that an individual from the target population is included in the sample is estimated and used with survey design-based formulas to obtain point estimates and variances. Specifically for EHR-based surveillance, few statistical methods have been employed. Initial studies used simple weighting methods, such as raking or poststratification, to obtain population-based estimates of obesity and other chronic diseases \citep{Flood2015,Daley2020,Klompas2017,Thorpe2016,McVeigh2016}. However, as these methods are able to account for only a small number of covariates, the implicit assumption of a selection mechanism that is selection at random (akin to missing at random) may not be tenable in EHRs. For example, the validity of estimates with raking based on race, sex, and age assumes that given only these demographic characteristics, the probability of selecting into the EHR sample is conditionally independent of missing disease status from individuals in the population who were not included in the EHR. More recent EHR-based surveillance \citep{Conderino2024,Chen2022} has employed a variant within the superpopulation modeling framework known as multilevel regression and poststratification (MRP) that was popularized in political science \citep{Little1993,Gelman1997,Wang2015}. MRP proceeds by partitioning the data into cross-classified cells defined by known characteristics (e.g., demographics), using multilevel regression to generate cell-level estimates by inducing shrinkage in cells with a small sample size, and applying a poststratification step to adjust the cell-level estimates to the target population. \citet{Chen2022} and \citet{Conderino2024} implemented MRP to generate population-based estimates of chronic diseases and health behaviors, leveraging parallel survey-based estimates for external validation. Despite the popularity of MRP, simulation studies have shown that its performance is variable, contingent on factors like completeness of predictors, predictor strength, and sample size \citep{Buttice2013,Conderino2024}. 

In this paper, we propose and evaluate three different superpopulation models for estimating diabetes prevalence in children and adolescents, overall and by demographic- and geographic-based subgroups. We adopt a Bayesian approach because it allows us to naturally account for uncertainty in the estimation process, and Bayesian methods are especially conducive to missing data problems \citep{Best1996}. The different Bayesian superpopulation models are mainly distinguished by how they account for two important features of the DiCAYA data: unobserved heterogeneity and data sparsity. First, in contrast to previous EHR-based surveillance that was limited in geographic scope \citep{Flood2015,Chen2022,Klompas2017,Thorpe2016,McVeigh2016,Conderino2024}, DiCAYA comprises EHRs from multiple health system centers across the US that may exhibit marked heterogeneity not only from underlying population health status, but also from center-level attributes like penetrance of the EHR network and level of care represented. Moreover, while in DiCAYA, the EHR sample size may be large, covariates include only basic demographics, leaving the potential for significant heterogeneity from unmeasured sources. Second, the nonrandom selection process into the EHR sample may result in sparsely observed subgroups, particularly for combinations of demographic characteristics and geography. Therefore, techniques for regularizing estimation, such as approaches to borrowing information that are commonly used in small area estimation, are critical to prevalence estimation in DiCAYA.

The first Bayesian superpopulation model is based on analysis of variance (ANOVA) using Bayesian hierarchical regression \citep{Gelman2005} through which we group predictors, such as main effects and their interactions, into batches with each batch having its own variance component. Although some combinations of predictors may be very sparsely observed, by assigning a variance component, we induce borrowing of information that stabilizes estimation of their effects. We call this model Bayesian Hierarchical Analysis of Variance (BHAV). In the second Bayesian superpopulation model, which we refer to as Partitioning Unobserved Geographic Variations (PUGV), we adapt a previous model of mammogram utilization rates built in a complex survey sample that allowed the joint effects of different demographic variables to vary locally \citep{Malec2008}. In DiCAYA, we hypothesized that unobserved heterogeneity in diabetes prevalence may be explained by geographic-based effects that vary according to demographic group. Third, we apply MRP to DiCAYA. As the only one of the three methods previously applied in EHR-based surveillance, we consider MRP the reference model. In addition, we include prevalence estimates based on the proportion of patients with diabetes in the DiCAYA data, which we refer to as the Na\"ive estimation method.

We proceed by describing DiCAYA in Section \ref{s:MotivatingData} while emphasizing data features that may be important to model building. In Section \ref{s:StatisticalMethods}, we explicate the different superpopulation models and their corresponding prevalence estimators. Before applying the different superpopulation models to DiCAYA, in Section \ref{s:HeuristicExample} we use a heuristic Bayesian setting to provide intuition for why the three Bayesian superpopulation models may differ from Na\"ive estimation. Section \ref{s:DataApplication} presents the application of the superpopulation models and the Na\"ive method to estimate diabetes prevalence in DiCAYA. We highlight how BHAV and PUGV differ from MRP as the reference model and how the three superpopulation models differ from the Na\"ive method. In Section \ref{s:Simulation}, in the absence of a parallel gold standard dataset, we conduct a plasmode simulation study to evaluate the estimation performance of the superpopulation models. In Section \ref{s:Discussion}, we conclude with some discussion.


\section{Motivating DiCAYA Data}
\label{s:MotivatingData}





The study design in DiCAYA has been described in detail in \citet{Hirsch2024}. The DiCAYA Network comprises health system centers, which represent healthcare delivery systems in locales where there are also other non-study health systems (patients may not receive their care from a single system), and geographical-based centers. These centers  are designed to cover an entire state and include EHR data streams from all major health systems in the state plus augmentation of records from medical claims data \citep{Hirsch2024}. We focus on estimating type 1 diabetes (T1D) prevalence in children and adolescents in health system centers. Three DiCAYA health system centers focus on children and adolescents, namely, the Ann \& Robert H. Lurie Children’s Hospital of Chicago (LUR), PEDSNet (PED), and OneFlorida+ (UFL). LUR, in collaboration with the National Patient-Centered Clinical Research Network (PCORnet{\tiny\CircledTop{R}}) coordinating center, assembled data from multiple health systems across the country; PED is a clinical research network composed of several pediatric medical centers across the country; and UFL covers most of Florida \citep{Hirsch2024}. We defined the target population for inference as children and adolescents from the 135 counties in the catchment areas of these three centers for 2018 and 2022.



\begin{figure}[htbp!]
\centering
\includegraphics[scale = 0.40]{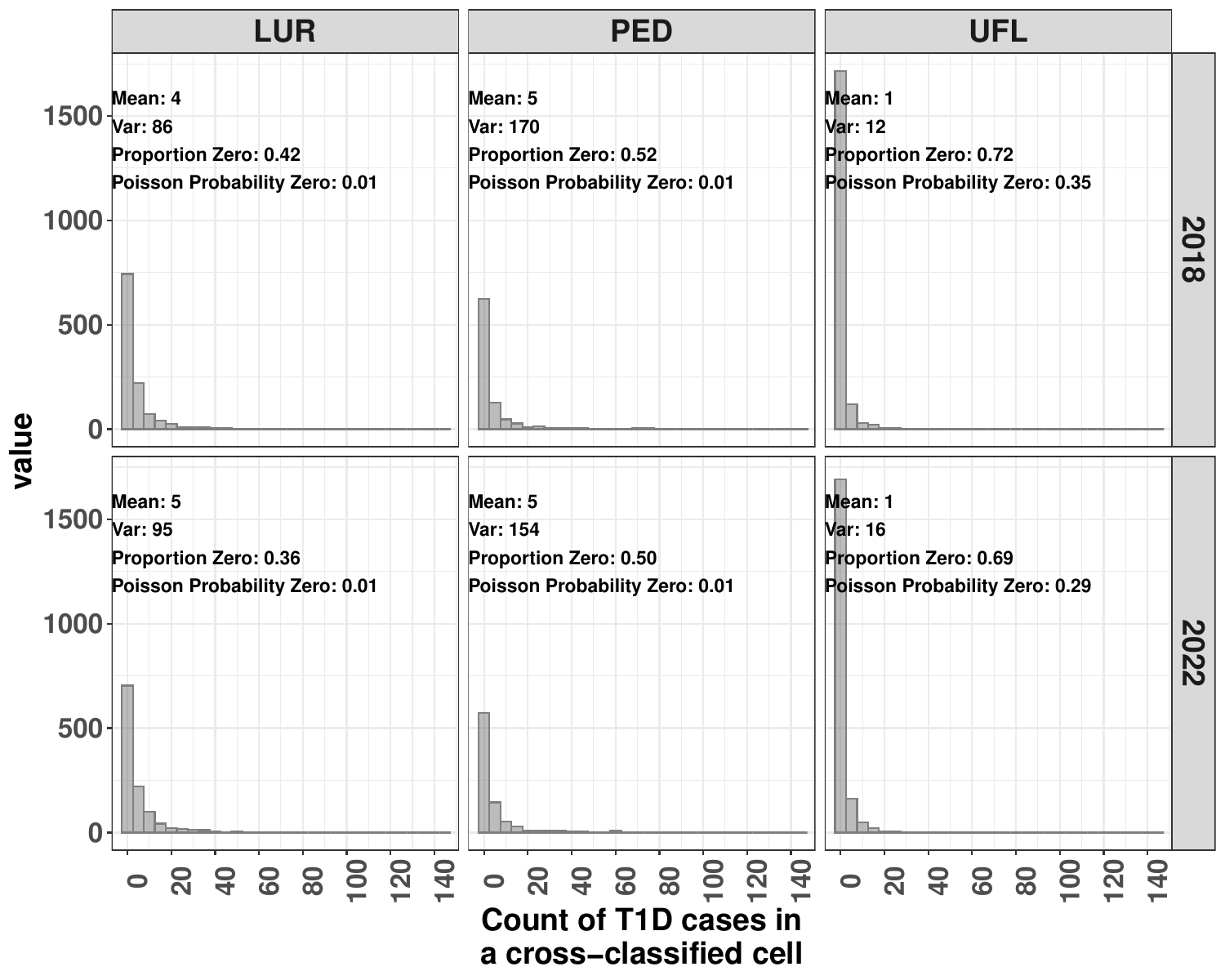}
\caption{Distribution of counts of type 1 diabetes (T1D) cases by health system center and year. LUR refers to Lurie Children's Hospital; PED, PEDSNet; and UFL, OneFlorida$+$.}
\label{f:diabetes_heterogeneity}
\end{figure}

The DiCAYA data are in the format of counts in cross-classified cells defined by county, year, and prespecified categories for age groups (0-5, 5-10, 10-15, and 15-18 years), sex (female and male), and race and ethnicity (non-Hispanic Asian, non-Hispanic Black, Hispanic, and non-Hispanic White). The utilization denominator in a given cell was defined as the total number of unique patients who interacted with the health system center at least once in the index year or in the previous two years. The outcome is the count of T1D cases in a cross-classified cell. Figure \ref{f:diabetes_heterogeneity} shows that the distributions of the T1D counts exhibited heterogeneity by center. Overdispersion was greatest for PED and smallest for UFL. Reflecting the rarity of T1D, case counts in each center showed a higher proportion of zeros than would be expected under a Poisson distribution with the center-specific sample mean, especially for LUR and PED. 

To define the finite population for surveillance, we obtained the Census population count in each cross-classified cell for the 135 counties in the target population from the US Census. Table \ref{t:nonrepresentativeness} compares characteristics of the EHR sample to the target population for inference overall and by health system center in 2018. Overall, the EHR sample totaled 3,674,328 patients, representing 25\% of individuals in the target population. While LUR and PED each had EHR sample sizes of more than one million patients, UFL had the smallest sample size with just over 600,000 patients. By center, the percent of the population included in the EHR sample was 21\% in LUR, 61\% in PED, and 15\% in UFL. The degree of nonrepresentativeness according to demographic characteristics varied by center, especially for race and ethnicity and age. For example, the EHR sample in LUR overrepresented non-Hispanic Whites (42\% vs. 37\%), while the EHR sample in UFL underrepresented them (33\% vs. 45\%). Based on sociodemographic variables from the American Community Survey (ACS) linked at the county-year level, the EHR sample in UFL exhibited notably higher percentage of foreign-born individuals than the corresponding population (25\% vs. 21\%). Patterns of EHR sample inclusion and nonrepresentativeness were consistent in 2022.


\begin{table}
\footnotesize
\caption{Comparison of the electronic health record (EHR) sample and Census population overall and by health system center in 2018. To apply county-level American Community Survey (ACS) variables to the target (Census) population, we used weighted means, with weights defined by each county’s share of the population. For the EHR sample, the weight was replaced with the proportion of the EHR sample included in a county. SNAP refers to the Supplemental Nutrition Assistance Program. LUR refers to Lurie Children's Hospital; PED, PEDSNet; and UFL, OneFlorida$+$.}
\label{t:nonrepresentativeness}
\begin{tabular}{p{4.5cm}cccccccc}
  \hline
 & \multicolumn{2}{c}{Overall} & \multicolumn{2}{c}{LUR} & \multicolumn{2}{c}{PED} & \multicolumn{2}{c}{UFL} \\ 
  \hline
 & \multicolumn{1}{c}{\underline{EHR}} & \multicolumn{1}{c}{\underline{Census}} & \multicolumn{1}{c}{\underline{EHR}} & \multicolumn{1}{c}{\underline{Census}} & \multicolumn{1}{c}{\underline{EHR}} & \multicolumn{1}{c}{\underline{Census}} & \multicolumn{1}{c}{\underline{EHR}} & \multicolumn{1}{c}{\underline{Census}} \\  
 Size & 3674328 & 14489962 & 1666530 & 8024381 & 1365902 & 2243995 & 641896 & 4221586 \\ 
  \% of target population included in the EHR sample  & 25 &  & 21 &  & 61 &  & 15 &  \\ 
  \underline{EHR variables} &  \underline{\%} & \underline{\%} & \underline{\%} & \underline{\%} & \underline{\%} & \underline{\%} & \underline{\%} & \underline{\%} \\ 
  Race and ethnicity: &  &  &  &  &  &  &  &  \\ 
   \hspace{0.1cm}Non-Hispanic Asian & 6 & 9 & 7 & 12 & 7 & 11 & 1 & 4 \\ 
   \hspace{0.1cm}Non-Hispanic Black & 21 & 19 & 19 & 18 & 23 & 19 & 25 & 22 \\ 
   \hspace{0.1cm}Hispanic & 26 & 29 & 32 & 33 & 11 & 13 & 40 & 30 \\ 
   \hspace{0.1cm}Non-Hispanic White & 47 & 42 & 42 & 37 & 59 & 57 & 33 & 45 \\ 
  Sex: &  &  &  &  &  &  &  &  \\ 
   \hspace{0.1cm}Female & 48 & 49 & 48 & 49 & 48 & 49 & 48 & 49 \\ 
   \hspace{0.1cm}Male & 52 & 51 & 52 & 51 & 52 & 51 & 52 & 51 \\ 
  Age, years: &  &  &  &  &  &  &  &  \\ 
   \hspace{0.1cm}0-5 & 32 & 27 & 33 & 27 & 29 & 29 & 34 & 27 \\ 
   \hspace{0.1cm}5-10 & 26 & 27 & 25 & 27 & 26 & 25 & 26 & 27 \\ 
   \hspace{0.1cm}10-15 & 26 & 29 & 25 & 29 & 27 & 28 & 24 & 29 \\ 
   \hspace{0.1cm}15-18 & 17 & 17 & 16 & 17 & 18 & 19 & 15 & 17 \\ 
  Health system center: &  &  &  &  &  &  &  &  \\ 
   \hspace{0.1cm}LUR & 45 & 55 & - & - & - & - & - & - \\ 
   \hspace{0.1cm}PED & 37 & 15 & - & - & - & - & - & - \\ 
   \hspace{0.1cm}UFL & 17 & 29 & - & - & - & - & - & - \\ 
  \underline{ACS county-level variables} &  &  &  &  &  &  &  &  \\ 
    Median household income & 65943 & 66179 & 67738 & 71088 & 70386 & 72149 & 51829 & 53676 \\ 
 \% Bachelor's degree or more & 36 & 35 & 37 & 38 & 37 & 38 & 30 & 29 \\ 
  \% Foreign born & 19 & 21 & 23 & 24 & 12 & 13 & 25 & 21 \\ 
  \% Poverty & 15 & 14 & 15 & 14 & 13 & 12 & 17 & 15 \\ 
  \% Unemployment & 6 & 6 & 7 & 6 & 6 & 6 & 6 & 6 \\ 
  \% SNAP & 14 & 13 & 15 & 13 & 12 & 12 & 18 & 15 \\ 
  \% Female-headed household & 26 & 24 & 26 & 24 & 24 & 23 & 27 & 26 \\ 
  \% Black & 17 & 16 & 18 & 16 & 16 & 15 & 17 & 16 \\ 
  \% Insurance & 91 & 89 & 91 & 90 & 94 & 94 & 86 & 86 \\ 
  \% Private insurance & 67 & 66 & 65 & 67 & 73 & 73 & 61 & 62 \\ 
  \% Public insurance & 34 & 34 & 35 & 33 & 33 & 32 & 34 & 36 \\ 
 \hline
\end{tabular}
\end{table}

The distributions of the utilization denominator in cross-classified cells by center revealed that sparsely observed cross-classified cells were particularly problematic in UFL and in racial and ethnic minorities (SM Table 1). While the median of the utilization denominator by categories of race and ethnicity, age, and sex was generally in the hundreds of patients for LUR and PED, each demographic subgroup except non-Hispanic White in UFL had a median utilization denominator below 50. Across centers, compared to non-Hispanic Whites, the utilization denominator for racial and ethnic minorities was smaller. In PED and UFL, the 25$^{th}$ percentile of the utilization denominator ranged from 7--14 for non-Hispanic Blacks and Hispanics versus 47--782 for non-Hispanic Whites. In the most extreme case, among non-Hispanic Asians, the interquartile range of the utilization denominator was 0--8 in UFL, compared to 54--490 in LUR and 5--659 in PED.

\section{Bayesian Superpopulation Models}
\label{s:StatisticalMethods}

\subsection{Setting}

Suppose there are $G$ health system centers ($g=1,\dots,G$). Each center comprises $C_g$ unique counties with varying degrees of coverage in the EHRs. The target population for inference is derived from a total of $C$ counties where $C = \sum_{g=1}^G C_g$ ($c = 1,\dots,C$). The target population can be cross-classified by demographic characteristics available in the EHRs, namely, $A$ age groups; $S$ sex categories; and, $R$ racial and ethnic groups; ($a = 1,\dot,A$, $s = 1,\dots,S$, $r = 1,\dots,R$). In year $t$ ($t \in 2018,2022$), define $S_{tc}$, $R_{tc}$, and $A_{tc}$ as the number of sex, racial and ethnic, and age groups, respectively, from county $c$ in the population. For a finite population of size $N_t$, let $J_t$ ($j = 1,\dots,J_t$) be the total number of cells in the population obtained by cross-classifying county and these demographic variables, with $J_t=\sum_{c=1}^C S_{tc} \times R_{tc} \times A_{tc}$. For T1D surveillance in children and adolescents in DiCAYA, there are $G=3$ health system centers, $C=135$ counties, and $A=4$, $S=2$, and $R=4$ categories for age, sex, and race and ethnicity, respectively. 

For year $t$ in cell $j$ of population size $N_{tj}$ ($N_t=\sum_{j=1}^{J_{t}} N_{tj}$ ), let $Y_{tj}$ be a random variable for the count of T1D cases that can be partitioned between observed cases $y_{tj}$ from individuals included in the EHR sample and unobserved cases $Y_{tj}^{exc}$ from individuals who did not select into the EHR sample. Define $n_{tj}$ as the utilization denominator, defined as the count of patients in the EHR sample from which the observed cases $y_{tj}$ arise. Note that in cross-classified cells in the target population that do not appear in the EHR sample, $n_{tj}=0$.

In each super population model, in year $t$, for cross-classified cell $j$ ($j \text{:} \, \, n_{tj}>0$), we assume $y_{tj} \sim Poisson(\theta_{tj})$, where 
\begin{gather}
\log(\theta_{tj}) = \log(n_{tj}) + \eta_{tj}. \label{e:poisson_model} 
\end{gather} 
In equation (\ref{e:poisson_model}), we use a log link function to link the Poisson mean $\theta_{tj}$ with the offset $n_{tj}$ and linear predictor $\eta_{tj}$ in cell $j$ of year $t$. BHAV, PUGV, and MRP are differentiated in terms of specification of their linear predictor, as explicated in subsections \ref{s:BHAV}--\ref{s:MRP}.

\subsection{Bayesian Hierarchical Analysis of Variance (BHAV)}
\label{s:BHAV}

In the spirit of ANOVA, and more broadly, the variance components literature, in BHAV, we group the coefficients into batches where each batch represents a source of variation in the data \citep{Gelman2005}. Assuming that the coefficients within each batch are exchangeable, we assign them a common prior distribution with unknown variance. For example, in DiCAYA, we may group the main effects of race and ethnicity into a batch, along with a batch for the interaction terms of race and ethnicity and age. Through the variance component, coefficient estimates in a batch account for not only the available information in the data, but also the shrinkage effect from the prior distribution, which stabilizes estimation in sparsely observed cells. When for a given batch the effect of shrinkage is larger, the variance will be closer to zero, thus indicating little contribution of the batch to variation in the data. By flexibly enabling the inclusion of many predictors, the hierarchical approach facilitates modeling of complex data structures while alleviating concern about model misspecification.

To specify BHAV, in year $t$, for cross-classified cell $j$ ($j \text{:} \, \, n_{tj}>0$), the linear predictor from equation (\ref{e:poisson_model}) is given by
\begin{eqnarray}
\eta_{tj} & = & b^{county}_{gc[j]} \label{e:lp_bhav} \\
& + & \mathbf{x}_{tj}^{race'} \mathbf{\beta}^{race} + \mathbf{x}_{tj}^{age'} \mathbf{\beta}^{age} + x_{tj}^{sex} \beta^{sex} + \mathbf{x}_{g[j]}^{center'} \mathbf{\beta}^{center} + {x}_{tj}^{year} \beta^{year}  \nonumber   \\
& + & \mathbf{x}_{tj}^{age,race'} \mathbf{\beta}^{age,race} + \mathbf{x}_{tj}^{age,sex'} \mathbf{\beta}^{age,sex} + \mathbf{x}_{tj}^{race,sex'} \mathbf{\beta}^{race,sex}  \nonumber \\
& + &  \mathbf{x}_{tg[j]}^{center,year'} \mathbf{\beta}^{center,year}  \nonumber  \\
& + & \mathbf{x}_{tg[j]}^{center,race'} \mathbf{\beta}^{center,race}  + \mathbf{x}_{tg[j]}^{center,age'} \mathbf{\beta}^{center,age} + \mathbf{x}_{tg[j]}^{center,sex'} \mathbf{\beta}^{center,sex} \nonumber   \\
& + & \mathbf{x}_{tg[j]}^{year,race'} \mathbf{\beta}^{year,race}  + \mathbf{x}_{tg[j]}^{year,age'} \mathbf{\beta}^{year,age} + \mathbf{x}_{tg[j]}^{year,sex'} \mathbf{\beta}^{year,sex} \nonumber  \\
& + &  \mathbf{x}_{tc[j]}^{auxiliary'} \mathbf{\beta}^{auxiliary} + \alpha. \nonumber 
\end{eqnarray}
In equation (\ref{e:lp_bhav}), $b_{gc[j]}$ is a county-level random intercept for county $c$ in center $g$ to account for unobserved county-level variations. The bracket notation in the subscript indicates the nesting of cross-classified cell $j$. 
Next, line 2 of equation (\ref{e:lp_bhav}) presents the dummy vectors for the demographic variables in $\mathbf{x}_{tj}^{demographic}$ and health system center in $\mathbf{x}_{g[j]}^{center}$, with one of the levels excluded as the reference. $x_{tj}^{year}$ is a binary indicator for study year 2022 versus 2018. Lines 3--6  present the effects for the two-way interactions among the demographic variables, and between center and year, center and demographics, and year and demographics. The last line includes the auxiliary predictors from the ACS in $\mathbf{x}_{tc[j]}^{auxiliary}$, and $\alpha$ is the intercept term. Corresponding regression coefficients for the different covariate vectors are given by $\beta^{var}$ where $var$ indicates the covariates involved.


We assume that for county $c$ in center $g$, \begin{gather}
b_{gc} \sim N(0, \sigma^{2,county})  \label{e:bgc_distribution} \\ 
\sigma^{county} \sim t_{\text{student}}(df = 3, \mu = 0, \sigma^2 = 10), \ \sigma^{county} \ge 0 \nonumber
\end{gather} For each batch of coefficients for the main effects of predictors involving more than two categories (race, age, and center), we use a prior distribution for the coefficients and hyperprior for the unknown variance of the same form as (\ref{e:bgc_distribution}). The prior structure for all of the two-way interactions is the same except that we set the variance to 2.5 in the half-t distribution.

\subsection{Partition Unobserved Geographic Variations (PUGV)}
\label{s:PUGV}

In DiCAYA, the hypothesis that unobserved heterogeneity in T1D prevalence may be driven by geographic effects that differ by demographic group is reasonable especially because the available geographic area unit, county, is relatively coarse. Our proposed approach to partitioning unobserved geographic heterogeneity differs from \citet{Malec2008} in order to meet the scientific objectives of DiCAYA. While \citet{Malec2008} were able to define demographic variables to obtain a total of six subgroups that vary over more than 3,000 counties, for a given year in DiCAYA, we have 32 joint effects (4 age groups $\times$ 4 race and ethnicities $\times$ 2 sexes) to vary over only 135 counties. Therefore, rather than joint effects, we partition geographic variations with separate effects for each demographic characteristic. In PUGV, we specify the linear predictor $\eta_{tj}$ from equation (\ref{e:poisson_model}) as \begin{eqnarray}
      \eta_{tj} & = & b_{cr[j]}^{race} +b_{ca[j]}^{age} +b_{cs[j]}^{sex} +  
     \mathbf{x}_{tj}^{demographic'} \mathbf{\beta}^{demographic} \label{e:pugv_lp} \\
     & + & \mathbf{x}_{g[j]}^{center'} \mathbf{\beta}^{center}  + {x}_{tj}^{year} \beta^{year} 
      + \mathbf{x}_{tg[j]}^{center,year'} \mathbf{\beta}^{center,year}  +  
      \mathbf{x}_{tc[j]}^{auxiliary'} \mathbf{\beta}^{auxiliary}  + \alpha \nonumber
\end{eqnarray} \noindent 
In equation (\ref{e:pugv_lp}), $b_{cr[j]}^{race}$, $b_{ca[j]}^{age}$, and $b_{cs[j]}^{sex}$ are random effects to partition county-level unobserved heterogeneity according to race, age, and sex, respectively. For example, $b_{cr[j]}^{race}$ is the effect of racial and ethnic group $r$ in county $c$, where $cr[j]$ indicates that cell $j$ is nested in racial and ethnic group $r$ and county $c$. In addition to reflecting demographic group-specific patterns of unobserved heterogeneity, the random effects induce borrowing of information for counties with smaller sample size of a given demographic group. For example, $b_{cr}^{race}$ induces a county with a smaller sample size of, say, non-Hispanic Asians to borrow strength from counties with a larger sample of this racial and ethnic group. Next, as in BHAV, equation (\ref{e:pugv_lp}) contains terms for different fixed effects, including demographic effects not accounted for by the random effects structure in $\mathbf{x}_{tj}^{demographic}$, such as hypothesized interactions; effects for center and year and their interaction; and, sociodemographic variables from the ACS.

We model each of $b_{cr}^{race}$, $b_{ca}^{age}$, and $b_{cs}^{sex}$ hierarchically. To illustrate with racial and ethnic group-specific county effects $b_{cr}^{race}$ ($r = 1,\dots,R$; $c = 1,\dots,C$), we specify \begin{gather}
 b_{cr}^{race} \sim Normal( \beta_r^{race}, \sigma_r^{2,race} )   \label{e:bcr_distribution}    \\    
 \beta_r^{race} \sim Normal(0, \phi^{2,race} )     \label{e:race_means}         \\     
\sigma_r^{2, race } \sim Scale-inv- \chi^2 (\nu_0^{race} ,\sigma_0^{2,race} )   \label{e:bcr_variances}          \\
\phi^{race} \sim t_{student}  (df=3,\mu=0,\sigma^2=2.5),\phi^{race} \geq 0  \label{e:race_variances}     \end{gather}
In equation (\ref{e:bcr_distribution}), $b_{cr}^{race}$ are flexibly modeled to capture racial and ethnic group specific patterns in unobserved heterogeneity across counties. For each racial and ethnic group $r$, $b_{cr}^{race}$  ($c = 1,\dots,C$) are centered around racial and ethnic group-specific means $\beta_r^{race}$  and assigned racial and ethnic group-specific variance $\sigma_r^{2,race}$. In equation (\ref{e:race_means}), we assign the racial and ethnic group-specific means $\beta_r^{race}$ a hyperprior normal distribution with mean zero and variance $\phi^{2,race}$. For identifiability, we use a sum-to-zero constraint $\sum_{r=1}^R \beta_r^{race} =0$. In equation (\ref{e:bcr_variances}), the $\sigma_r^{2,race}$ are modeled according to a scaled inverse chi-squared distribution with $\nu_0^{race}$  as the prior degrees of freedom and $\sigma_0^{2,race}$ as the prior scale. This is a conjugate prior distribution that yields a posterior degrees of freedom equal to the sum of the prior and data-based degrees of freedom, and a posterior scale equal to a weighted average of the prior and data-based scales with weights equal to the degrees of freedom. In other words, we shrink the racial and ethnic group-specific variances toward a common prior variance $\sigma_0^{2,race}$. Using an empirical Bayes approach, we estimate $\sigma_0^{2,race}$ in the EHR sample. We obtain $\hat{\sigma}_0^{2,race}$ from a Poisson mixed model with a random intercept for race and ethnicity. We assign a small prior degrees of freedom to reduce the influence of prior information. Finally, in equation (\ref{e:race_variances}), we assign a hyperprior distribution for the standard deviation $\phi^{race}$ of the racial and ethnic group-specific means of the form of a half-t distribution. The specification of the county-level age and sex-specific random effects, $b_{ca}^{age}$  and $b_{cs}^{sex}$, respectively, follows suit.

\subsection{Multilevel Regression and Poststratification (MRP)}
\label{s:MRP}


To apply MRP in DiCAYA, we first build a multilevel regression model by specifying the linear predictor $\eta_{tj}$ from equation (\ref{e:poisson_model}) as
\begin{eqnarray}
\eta_{tj}  &=&  b_{r[j]}^{race} + b_{a[j]}^{age} + b_{gc[j]}^{county}  + \mathbf{x}_{tj}^{demographic'} \mathbf{\beta}^{demographic}  \label{e:mrp_lp} \\
& + & \mathbf{x}_{g[j]}^{center'} \mathbf{\beta}^{center} + {x}_{tj}^{year} \beta^{year}   + \mathbf{x}_{tg[j]}^{center,year'} \mathbf{\beta}^{center,year}  \mathbf{x}_{tc[j]}^{auxiliary'} \mathbf{\beta}^{auxiliary}  + \alpha  \nonumber 
\end{eqnarray}

\noindent where the $b^{county}_{gc[j]}$, $b^{race}_{r[j]}$, $b^{age}_{a[j]}$ are random effects at the levels of the county, racial and ethnic group, and age group, respectively. The prior distributions for each random effect are the same as in (\ref{e:bgc_distribution}), and the fixed effects are the same as in PUGV.

\subsection{Prevalence estimators}

We seek to make inference about T1D prevalence in the target population in year $t$, which we denote as $P_t$. Our estimand is given by
\begin{equation} 
P_t=\frac{1}{N_t} \bigg(\sum_{j:\,\, n_{tj} > 0} y_{tj} + \sum_{{ j:\,\, 0 \leq n_{tj} < N_{tj}}} Y_{tj}^{exc} \bigg), \label{e:superpopulation_estimand}
\end{equation} where the first term sums observed T1D cases in the EHR sample. The second summation in (\ref{e:superpopulation_estimand}) is for missing T1D cases from individuals who were not included in the EHR sample, either from cross-classified cells in the target population where $0 < n_{tj} < N_{tj}$ or $n_{tj} = 0$.

Let $\mathbf{\Theta}$ be a container for model parameters and $\tilde{\mathbf{x}}_{tj}$ include covariate information (e.g., $\mathbf{x}_{tj}^{demographic}$) to define cross-classified cells in which to predict $Y_{tj}^{exc}$. For BHAV and PUGV,  leveraging the posterior predictive distribution to predict $Y_{tj}^{exc}$, we construct the estimator of $P_t$ as
\begin{eqnarray}
    E[P_t \, | \, \mathbf{y}_t]   & = & \frac{1}{N_t} \bigg(\sum_{j:\,\, n_{tj} > 0} y_{tj} + \sum_{{ j:\,\, 0 \leq n_{tj} < N_{tj}}} E[\tilde{Y}^{exc}_{tj} \, | \, \mathbf{y}_t] \bigg). \label{e:estimator_pugv_bhav}
\end{eqnarray} The second term is the sum of the expected posterior predictive count of T1D cases among individuals who were not included in the EHR sample, where $E[\tilde{Y}^{exc}_{tj} \, | \, \mathbf{y}]$ is obtained from the posterior predictive distribution of $\tilde{Y}^{exc}_{tj}$ as \\
$ E[\tilde{Y}^{exc}_{tj} \, | \, \mathbf{y}] = \iint p(\tilde{Y}^{exc}_{tj} \, | \, \mathbf{\Theta}) p(\mathbf{\Theta} \, | \, \mathbf{y}, \tilde{\mathbf{x}}_{tj}) \tilde{Y}^{exc}_{tj} \partial \mathbf{\Theta} \partial \tilde{Y}^{exc}$. Using equation (\ref{e:estimator_pugv_bhav}), we can estimate the T1D prevalence in the population for any subdomain of interest (e.g, age and year; or, age, year, and center).

The prevalence estimator for MRP differs from BHAV and PUGV because rather than using the posterior predictive distribution to predict missing T1D for individuals not included in the EHR sample, we project from the sample to population with a poststratification step. In year $t$, based on the $J_t$ cross-classified cells defined by county and demographic characteristics, the MRP estimator is constructed with a weighted average of the cell-specific T1D prevalence where the weights are the cell's relative proportion in population, given by
\begin{equation} E[P_t \, | \, \mathbf{y}_t] = \frac{\sum_{j = 1}^{J_{t}} N_{tj} \pi_{tj}}{\sum_{j = 1}^{J_{t}} N_{tj}}. \label{e:estimator_mrp} \end{equation}
In (\ref{e:estimator_mrp}), when $n_{tj} > 0 $ for cell $j$ in year $t$, the cell-specific T1D prevalence is $\pi_{tj} = \frac{\theta_{tj}}{n_{tj}}$. For cross-classified cells in the target population but with $n_{tj} = 0$, we obtain the cell prevalence using the covariate information in $\tilde{\mathbf{x}}_{tj}$ and population size $N_{tj}$. The MRP estimator may be used to estimate prevalence for any subgroup of interest. 

\section{A Heuristic Comparison of Bayesian Superpopulation Versus Na\"ive Estimation}
\label{s:HeuristicExample}

 
 

To provide intuition for why the three Bayesian superpopulation models may differ from Na\"ive estimation (i.e., EHR-based sample means), we consider a simplified heuristic framework with a Poisson likelihood and conjugate gamma prior distribution. Although the heuristic framework is not identical to the proposed models, it captures the general shrinkage mechanism while yielding an analytically tractable expression for the superpopulation-based difference from Na\"ive estimation. Suppose our goal is to estimate $P_g$, which is the T1D prevalence in the target population from health system center $g$ ($g = 1,\dots,G$). Define $J_g$ as the number of cross-classified cells in center $g$. For $j = 1,\dots,J_g$, we assume that $y_j \sim Poisson(\theta_j)$ with $\theta_j = n_j \mu_g$, where $\mu_g$ is the underlying center-specific prevalence for cells in center $g$, and $n_j$ is the utilization denominator as previously defined. We assign $\mu_g$ a conjugate gamma prior distribution with shape $d_1$ and rate $d_2$ ($d_1,d_2 > 0$), and let $y_0$ denote the prior T1D prevalence with $y_0 = \frac{d_1}{d_2}$. We set $y_0$ equal to the overall T1D prevalence in the EHR sample $\hat{y}$ with $d_2 = m_0$ and $d_1 = \hat{y} m_0$, where $m_0$ is the amount of prior information. $m_0$ is typically a small value to reduce the influence of the prior distribution.



We can show that the posterior distribution of $\mu_g$ is gamma with posterior mean $E[\mu_g \, | \, \mathbf{y}] = \frac{\sum_{j = 1}^{J_{g}} y_j + d_1}{\sum_{j = 1}^{J_{g}} n_j + d_2}$. Let the Na\"ive estimate of the T1D prevalence in health system center $g$ be given by 
$\hat{y}_g = \frac{\sum_{j = 1}^{J_{g}} y_j}{\sum_{j = 1}^{J_{g}} n_j}$. Then, we can re-write $E[\mu_g \, | \, \mathbf{y}]$ as
\begin{equation}
    E[\mu_g \, | \, \mathbf{y}] = w_g \hat{y}_g + (1- w_g) y_0, \label{e:posterior_mean_mug}
\end{equation} where $w_g$ is a weight with $w_g = \frac{\sum_{j = 1}^{J_{g}} n_j}{\sum_{j = 1}^{J_{g}} n_j + d_2}$ and $(1 - w_g) = \frac{d_2}{\sum_{j = 1}^{J_{g}} n_j + d_2}$. Thus, $E[\mu_g \, | \, \mathbf{y}]$ is a weighted average of the Na\"ive estimate of the T1D prevalence in center $g$ and the prior prevalence $y_0$. The weighting with $w_g$ indicates that as the utilization denominator in health system center $g$ increases, the Na\"ive estimate receives relatively more weight than the prior prevalence, and vice versa. Then, the difference between the Na\"ive estimate and $E[\mu_g \, | \, \mathbf{y}]$ is given by 
\begin{equation}
\hat{y}_g -  E[\mu_g \, | \, \mathbf{y}] = (\hat{y}_g - y_0) (1 - w_g),    \end{equation} where as the utilization denominator grows, the difference between the Na\"ive estimate and the posterior mean of $\mu_g$ decreases. Depending on whether $\hat{y}_g$ is less than or greater than $y_0$, the difference will be positive or negative.

Using equations (\ref{e:superpopulation_estimand}) and (\ref{e:posterior_mean_mug}), we can express the posterior mean estimator for $P_g$ as
\begin{equation} \label{e:posterior_mean_pg}
E[P_g \, | \, \mathbf{y}] = p_g \hat{y}_g + (1 - p_g)  E[\tilde{\mu}_g \, | \, \mathbf{y}].
\end{equation} In equation (\ref{e:posterior_mean_pg}), $p_g$ is the proportion of the population included in the EHR sample, defined as $p_g = \frac{\sum_{j = 1}^{J_{g}} n_j}  {\sum_{j = 1}^{J_{g}} N_j}$. The second term represents the posterior expected T1D prevalence among individuals in the population who were not included in the EHR sample. As the proportion of the population included in the EHR sample increases, the Na\"ive estimate will dominate $E[P_g \, | \, \mathbf{y}]$.

Finally, let $\hat{E}[P_g \, | \, \mathbf{y}]$ be the superpopulation-based estimate of $P_g$. Then, the difference between the Na\"ive and Bayesian superpopulation-based estimates is given by 
\begin{equation} \label{e:diff_naive_posterior_mean_pg}
\hat{y}_g - \hat{E}[P_g \, | \, \mathbf{y}] =
(\hat{y}_g - y_0) \bigg((1 - p_g) (1 - w_g) \bigg).    
\end{equation} The absolute magnitude of the difference increases as the proportion of the population included in the EHR sample decreases and as the utilization denominator decreases. Equation (\ref{e:diff_naive_posterior_mean_pg}) implies that the difference between the three Bayesian superpopulation models and Na\"ive estimation is governed by (i) the extent to which the Na\"ive estimate differs from the prior prevalence; (ii) the proportion of the population not included in the EHR sample; and, (iii) the degree of Bayesian shrinkage induced by utilization denominators of limited sample size.

\section{Data Application}
\label{s:DataApplication}

We applied each superpopulation model to estimate T1D prevalence overall and by demographic and geographic-based subgroups. We compared the superpopulation-based estimates with estimates from the Na\"ive method in which we compute sample
means in the EHR data. 

To specify PUGV, we partitioned county-level variations by only race and ethnicity and age because of unstable estimation for the variance component for sex, which indicated that sex-specific patterning in county-level effects were minimal. In $\mathbf{x}_j^{demographic}$, we included the main effect for sex, and based on our hypothesis that differences in T1D by race and ethnicity and sex may become more apparent in older age groups as a function of timing of diagnosis, the interaction terms of age with sex and race and ethnicity. For the interaction terms with age, we used a cubic polynomial function of the midpoint of each age category. For MRP, $\mathbf{x}_j^{demographic}$ was similarly specified. For each superpopulation model, we specified $\mathbf{x}_{tc}^{auxiliary}$ with the following ACS sociodemographic variables: \% bachelor's degree or more, \% foreign born, \% Black, and \% public insurance.

To complete the specification of each Bayesian superpopulation model, we assigned prior distributions to remaining unknown parameters. We used independent normal prior distributions with mean zero and diffuse variance for the intercept and the regression coefficients in PUGV and MRP, in addition to the remaining coefficients in BHAV for covariates with only one category and the auxiliary variables. Bayesian models were run with three chains from dispersed initial values for 30,000 iterations with a burn-in of 25,000. Model convergence was assessed visually based on traceplots and using the Gelman-Rubin diagnostic with convergence indicated by values being below the threshold of 1.1 \citep{Gelman2013}. Posterior sampling was performed with \texttt{JAGS} software \citep{Plummer2017} via the \texttt{R} software package \texttt{R2jags} \citep{Su2024}.

\subsection{Model comparison among the Bayesian superpopulation models}

SM Tables 2--4 present the posterior means and 95\% credible intervals (CrIs) for parameters in BHAV, PUGV, and MRP. Here, we briefly summarize how each model explained unobserved heterogeneity in T1D. In BHAV (SM Table 2), among the main effects, the largest variance component was for age, followed by race and ethnicity, and center. Among the two-way interactions, center by age explained the most variation in T1D, while center by sex explained the least with a 95\% CrI bordering zero. The standard deviation of the county-level random intercept was of similar magnitude to the center main effect. For PUGV (SM Table 3), unobserved county-level effects exhibited racial and ethnic group-specific patterning, especially for Hispanic and non-Hispanic Black that exhibited non-overlapping CrIs. In contrast, the extent of unobserved county heterogeneity appeared similar by age group. In MRP (SM Table 4), the variance component for age showed the largest standard deviation, followed by race and ethnicity and county, with magnitudes similar to those in BHAV.

To compare the superpopulation models in terms of predictive accuracy, we use an information criterion known as the Widely Applicable Information Criterion (WAIC) \citep{Watanabe2010}. In contrast to the well-known Deviance Information Criterion (DIC) \citep{Spiegelhalter2002} that conditions on a point estimate (e.g., posterior mean) of model parameters, the WAIC is fully Bayesian in that it averages over their posterior distribution. Bayesian leave-out-one cross validation has been proven to be asymptotically equal to the WAIC \citep{Watanabe2010}. The WAIC is given by $\text{WAIC} =  -2 \text{lppd} + 2 p_{\text{WAIC2}}$ where $\text{lppd}$ is log pointwise predictive density, which is a measure of within-sample predictive accuracy. The second term, $p_{\text{WAIC2}}$, is the bias correction term for overfitting, which we calculated as the sum of the variance of the individual terms in the log predictive density. For comparison, we include the DIC which we compute using standard formulas \citep{Spiegelhalter2002}. Both the WAIC and DIC selected PUGV, followed by BHAV and MRP (SM Table 5). 

We conduct model checking with Bayesian posterior predictive p-values in which a discrepancy measure is computed using observed versus replicated T1D case counts \citep{Gelman1996}. The Bayesian predictive p-value denotes the probability that the discrepancy measure under the replicated counts is greater than that under the observed counts. A p-value near 0.5 indicates consistency between the observed and replicated discrepancy measures, while a p-value outside the range of 0.05 and 0.95 is considered to suggest a lack of model fit. First, we evaluate whether each superpopulation model is generally consistent with the observed data using three discrepancy measures, namely, the sample mean T1D count as $D_1 = \bar{y}$; a measure of overdispersion defined as the sample variance of the T1D count minus the mean count $D_2 = s^2_y - \bar{y}$; and, the proportion of zero counts, $D_3 = \frac{\sum_{t=1}^T \sum_{j = 1}^{J_{t}} \mathbf{1}_{y_{tj}=0}}{\sum_{t=1}^T J_t}$. We then consider whether each superpopulation model well-replicated subgroup-specific means for each racial and ethnic, age, and sex group, stratified by center and year. Each of BHAV, PUGV, and MRP well-replicated the mean, overdispersion, and proportion of zeros in the observed data with Bayesian posterior predictive p-values between 0.31--0.70 (SM Figure 1). However, for the subgroup-specific mean T1D counts, MRP showed posterior predictive p-values that were often too small or too large in racial and ethnic minority groups and in different age groups, especially for PED (SM Figures 2-7). While PUGV occasionally showed inconsistencies with the observed data, all of the posterior predictive p-values for BHAV were in the acceptable range. 

\subsection{T1D prevalence overall and by health system center}

\begin{table}[htbp!]
\centering
\footnotesize
\begin{tabular}{llcccc}
  \hline
\multicolumn{1}{c}{Center}  & \multicolumn{1}{c}{Year} & \multicolumn{1}{c}{BHAV} & \multicolumn{1}{c}{PUGV} & \multicolumn{1}{c}{MRP} & \multicolumn{1}{c}{Na\"ive} \\ 
  \hline
\multirow{2}{*}{All} & 2018 & 3.70 (0.05) [3.61, 3.80] & 3.72 (0.05) [3.63, 3.81] & 3.67 (0.05) [3.58, 3.77] & 3.11 (0.03) [3.05, 3.17] \\ 
   & 2022 & 3.58 (0.05) [3.50, 3.68] & 3.62 (0.05) [3.52, 3.71] & 3.58 (0.05) [3.49, 3.67] & 2.99 (0.03) [2.94, 3.04] \\ 
  \multirow{2}{*}{LUR} & 2018 & 3.24 (0.05) [3.14, 3.34] & 3.26 (0.05) [3.16, 3.36] & 3.23 (0.05) [3.12, 3.33] & 2.99 (0.04) [2.90, 3.07] \\ 
  & 2022 & 3.02 (0.04) [2.93, 3.10] & 3.02 (0.04) [2.94, 3.11] & 3.00 (0.04) [2.91, 3.09] & 2.89 (0.04) [2.81, 2.96] \\ 
 \multirow{2}{*}{PED} & 2018 & 3.47 (0.04) [3.40, 3.54] & 3.49 (0.04) [3.42, 3.56] & 3.46 (0.05) [3.36, 3.57] & 3.25 (0.05) [3.16, 3.35] \\ 
   & 2022 & 3.07 (0.03) [3.02, 3.13] & 3.11 (0.03) [3.05, 3.17] & 3.07 (0.05) [2.98, 3.17] & 3.02 (0.05) [2.93, 3.11] \\ 
  \multirow{2}{*}{UFL} & 2018 & 4.71 (0.13) [4.46, 4.98] & 4.72 (0.14) [4.44, 4.99] & 4.64 (0.13) [4.39, 4.91] & 3.11 (0.07) [2.97, 3.25] \\ 
   & 2022 & 4.84 (0.13) [4.60, 5.08] & 4.92 (0.13) [4.67, 5.18] & 4.86 (0.13) [4.62, 5.11] & 3.23 (0.07) [3.10, 3.36] \\    \hline
\end{tabular}
\caption{T1D cases per 1,000 persons overall and by health system center among children and adolescents. BHAV, PUGV, and MRP refer to Bayesian Hierarchical Analysis of Variance, Partitioning Unobserved Geographic Variations, and Multilevel Regression and Poststratification, respectively. LUR refers to Lurie Children's Hospital; PED, PEDSNet; and UFL, OneFlorida$+$. For BHAV, PUGV, and MRP, the posterior mean (posterior standard deviation) [95\% credible interval] is presented. The Na\"ive method uses the sample mean (standard error) [95\% confidence interval].} 
\label{t:table_prevalence_index_year_site_name1_index_year_T1DM_a}
\end{table}

Table \ref{t:table_prevalence_index_year_site_name1_index_year_T1DM_a} presents the estimated T1D prevalence among children and adolescents (expressed as T1D cases per 1,000 persons) overall and stratified by health system center based on the superpopulation models and Na\"ive estimation method. Among the superpopulation models, differences in the posterior mean T1D cases per 1,000 persons were less than one-tenth of a T1D case per 1,000 persons. BHAV and PUGV exhibited posterior standard deviations (SDs) less than or equal to those of MRP, with the exception of UFL in 2018, when the posterior SD of PUGV was larger than MRP. Considering all health system centers together, the superpopulation-based estimated T1D prevalences showed a consistent upward adjustment relative to the Na\"ive estimate. For example, in 2018, the PUGV-based T1D prevalence was 0.6 T1D cases per 1,000 persons higher than the Na\"ive estimate, representing a 20\% increase. The center-specific estimates suggest that the upward adjustment was primarily driven by UFL, where the superpopulation-based estimates were at least 1.5 T1D cases per 1,000 persons higher than the Na\"ive estimate, corresponding to a 49-52\% increase. Albeit to lesser extent, in LUR and PED the superpopulation-based estimates were also largely higher than their Na\"ive counterparts.


\subsubsection{T1D prevalence by health system center-specific demographic subgroups}

\begin{figure}[htbp!]
\centering
\includegraphics[scale = 0.65]{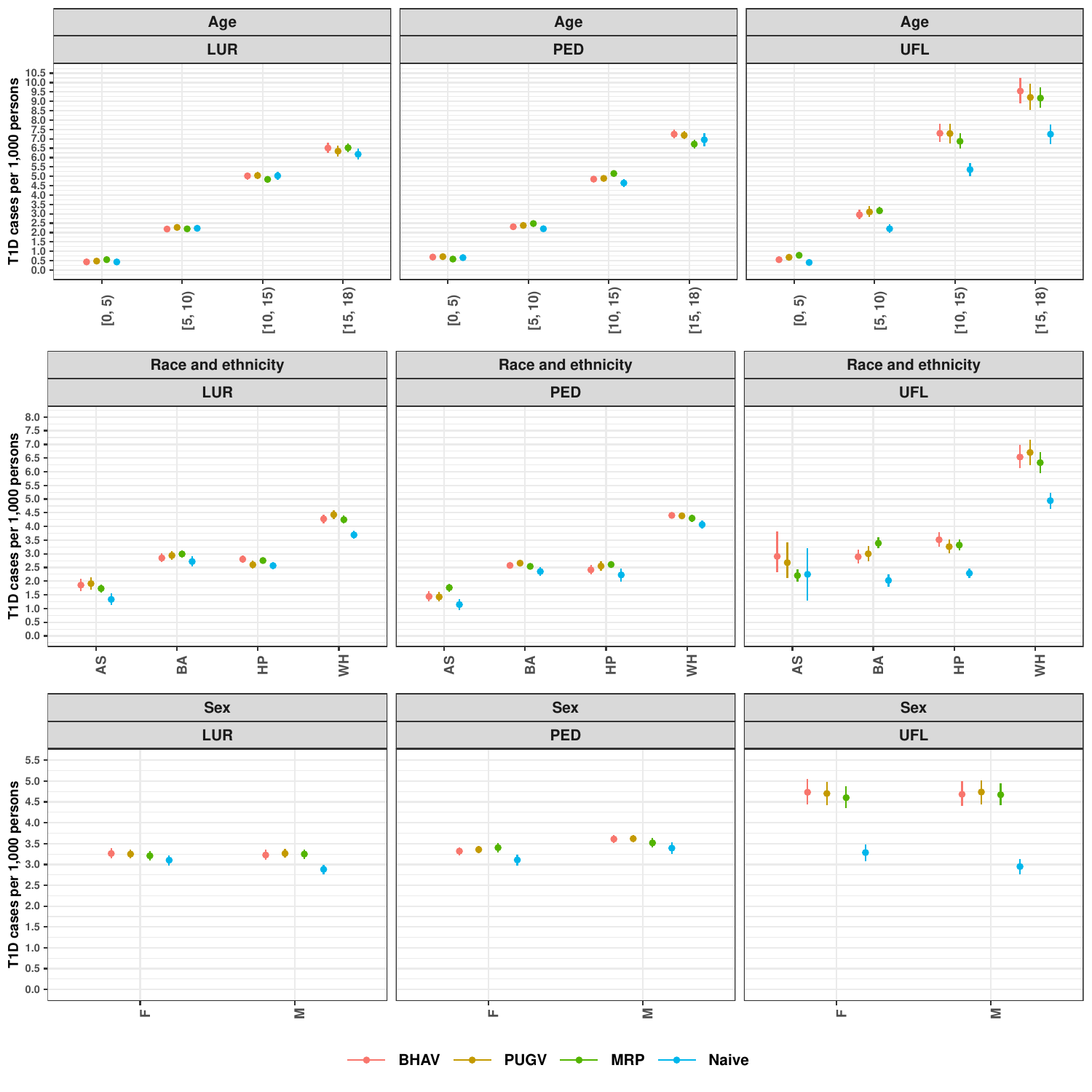}
\caption{T1D cases per 1,000 persons in 2018 by age in years, race and ethnicity, and sex among children and adolescents, stratified by health system center, color coded by estimation method. BHAV, PUGV, and MRP refer to Bayesian Hierarchical Analysis of Variance, Partitioning Unobserved Geographic Variations, and Multilevel Regression and Poststratification, respectively. LUR refers to Lurie Children's Hospital; PED, PEDSNet; and UFL, OneFlorida$+$. For BHAV, PUGV, and MRP, the points are posterior means, and the vertical bars are 95\% credible intervals. For Na\"ive, the points and vertical bars are sample means and 95\% confidence intervals, respectively. AS, BA, HP, and WH refer to non-Hispanic Asian, non-Hispanic Black, Hispanic, and non-Hispanic White, respectively. F and M refer to female and male, respectively.}
\label{f:Fig_prevalence_joined_age_race_sex_site_name1_index_year_T1DM2018_a}
\end{figure}

\begin{figure}[htbp!]
\centering
\includegraphics[scale = 0.5]{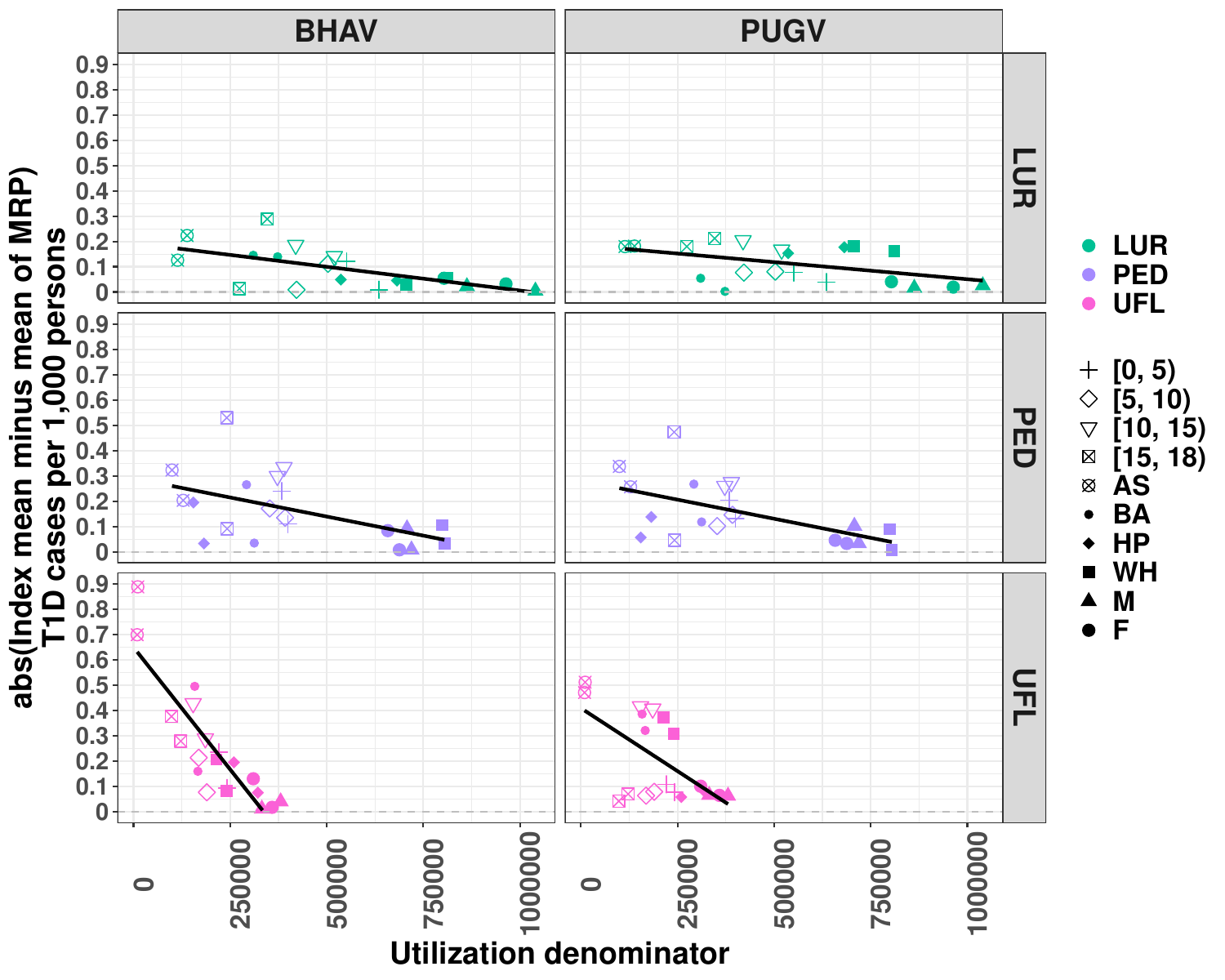}
\caption{Absolute difference in the T1D cases per 1,000 persons between the index posterior mean (BHAV or PUGV) and MRP for demographic subgroups in 2018 and 2022 according to the utilization denominator, stratified by health system center. BHAV, PUGV, and MRP refer to Bayesian Hierarchical Analysis of Variance, Partitioning Unobserved Geographic Variations, and Multilevel Regression and Poststratification, respectively. LUR refers to Lurie Children's Hospital; PED, PEDSNet; and UFL, OneFlorida$+$. AS, BA, HP, and WH refer to non-Hispanic Asian, non-Hispanic Black, Hispanic, and non-Hispanic White, respectively. M and F indicate male and female, respectively.}
\label{f:Fig_superpopulation_minus_MRP_utliz_den_site_name1_T1DM_a}
\end{figure}

\begin{figure}[htbp!]
\centering
\includegraphics[scale = 0.5]{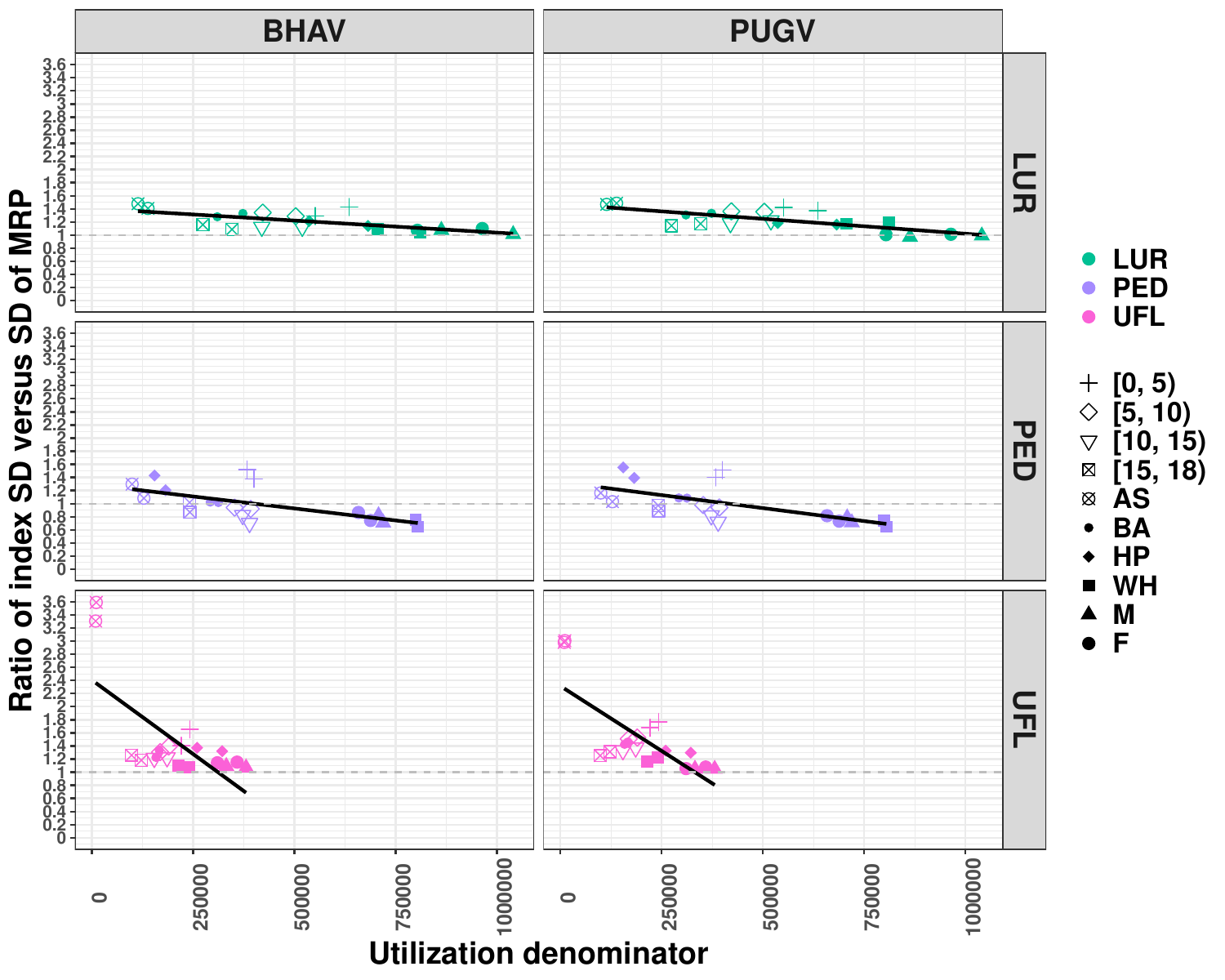}
\caption{Ratio of the index posterior standard deviation (SD) (BHAV or PUGV) of T1D cases per 1,000 persons and the posterior SD for MRP for demographic subgroups in 2018 and 2022 according to the utilization denominator, stratified by health system center. BHAV, PUGV, and MRP refer to Bayesian Hierarchical Analysis of Variance, Partitioning Unobserved Geographic Variations, and Multilevel Regression and Poststratification, respectively. LUR refers to Lurie Children's Hospital; PED, PEDSNet; and UFL, OneFlorida$+$. AS, BA, HP, and WH refer to non-Hispanic Asian, non-Hispanic Black, Hispanic, and non-Hispanic White, respectively. M and F indicate male and female, respectively.}
\label{f:Fig_superpopulation_sd_ratio_MRP_utliz_den_site_name1_T1DM_a}
\end{figure}

Within each health system center, superpopulation-based estimated prevalences were often comparable for age, race and ethnicity, and sex subgroups (Figure \ref{f:Fig_prevalence_joined_age_race_sex_site_name1_index_year_T1DM2018_a}). Mean absolute relative differences (\%) that compared BHAV and PUGV-based estimation versus MRP for the posterior mean and posterior SD of T1D prevalence are shown in SM Table 6. Mean absolute relative differences in the posterior mean were often greater in racial and ethnic minorities; and, more generally, in estimates from PED and UFL versus LUR. (Note that the large relative differences in age 0-5 years are driven by the very low prevalence in this subgroup.) Across demographic subgroups, mean absolute relative differences for the posterior SD were greater than those for the posterior mean. Figure \ref{f:Fig_superpopulation_minus_MRP_utliz_den_site_name1_T1DM_a} shows that on the absolute scale, differences in T1D prevalence comparing BHAV and PUGV versus MRP diminished with increasing utilization denominator. In Figure \ref{f:Fig_superpopulation_sd_ratio_MRP_utliz_den_site_name1_T1DM_a}, MRP exhibited greater precision, especially in more sparsely observed subgroups, and more generally at lower levels of the utilization denominator, where unlike BHAV and PUGV, MRP does not account for uncertainty from the posterior predictive distribution for missed T1D cases from individuals in the population who were not included in the EHR sample. Patterning was unchanged upon substituting the proportion of the population included in the EHR sample for the utilization denominator (SM Figures 8 and 9).

\begin{figure}[htbp!]
\centering
\includegraphics[scale = 0.5]{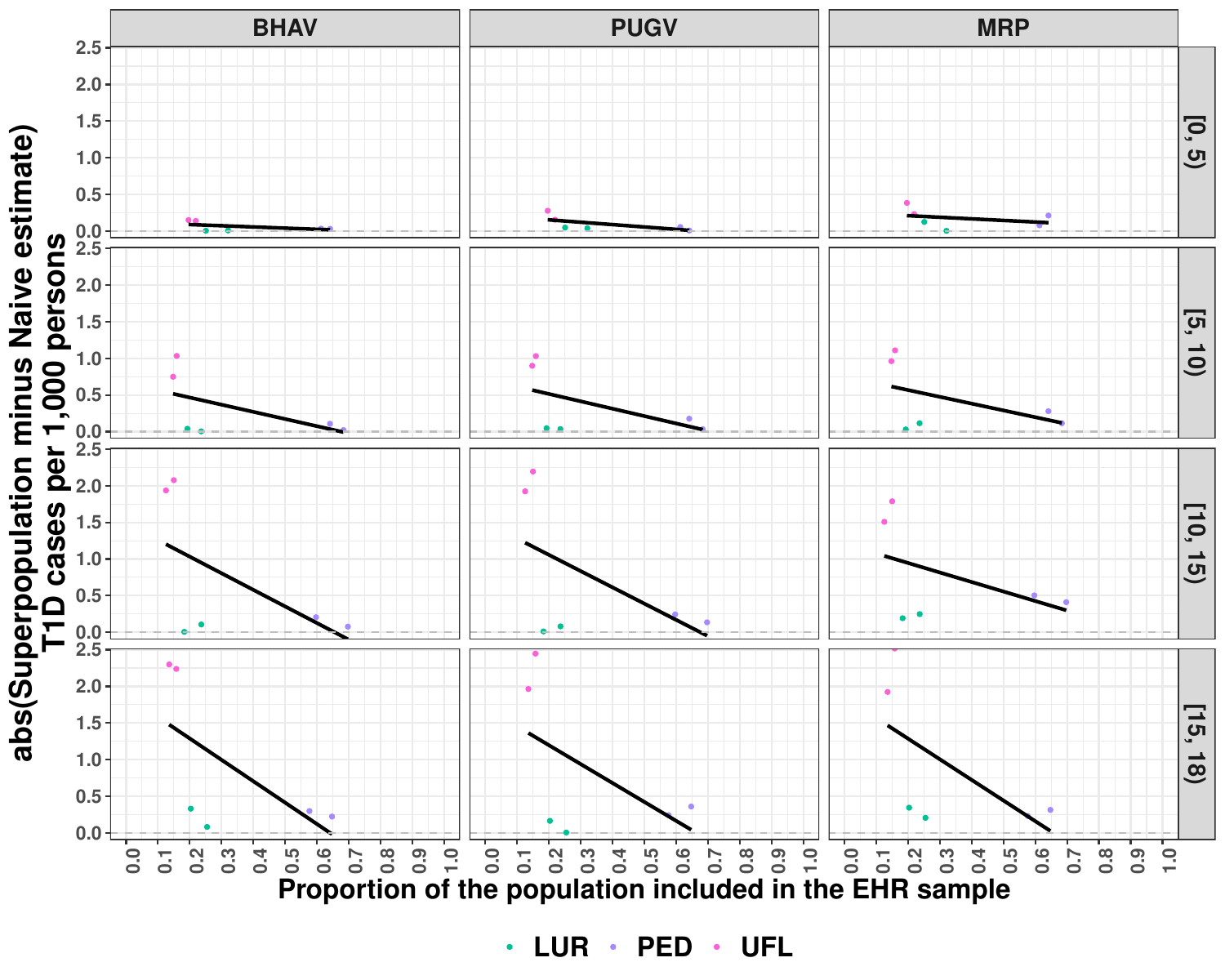}
\caption{Absolute difference in the T1D cases per 1,000 persons between the superpopulation-based and Na\"ive estimates for age subgroups (years) in 2018 and 2022 by superpopulation model according to the proportion of the population included in the EHR sample, color coded by health system center. BHAV, PUGV, and MRP refer to Bayesian Hierarchical Analysis of Variance, Partitioning Unobserved Geographic Variations, and Multilevel Regression and Poststratification, respectively. LUR refers to Lurie Children's Hospital; PED, PEDSNet; and UFL, OneFlorida$+$.}
\label{f:Fig_superpopulation_minus_naive_proppop_age_T1DM_a}
\end{figure}

When estimated T1D prevalence in demographic subgroups differed between the superpopulation models and Na\"ive estimation method, the superpopulation-based estimated prevalence was often higher than the Na\"ive estimate (Figure \ref{f:Fig_prevalence_joined_age_race_sex_site_name1_index_year_T1DM2018_a}). Figure \ref{f:Fig_prevalence_joined_age_race_sex_site_name1_index_year_T1DM2018_a} shows that compared to LUR and PED, UFL exhibited the largest differences between the superpopulation-based and Na\"ive prevalence estimates. For example, in UFL, the magnitude of the positive adjustment in the age-based estimates from the superpopulation models increased in a dose-response fashion with increasing age. By age 15-18 years, the superpopulation-based estimates were about 2 T1D cases per 1,000 persons higher than the Na\"ive estimate, reflecting a 27\% increase. For race and ethnicity and sex, UFL exhibited superpopulation-based estimates that were up to 67\% higher than those based on Na\"ive estimation. Consistent with equation (\ref{e:diff_naive_posterior_mean_pg}), Figure \ref{f:Fig_superpopulation_minus_naive_proppop_age_T1DM_a} demonstrates the inverse relationship between the absolute differences for age-based estimates in each year and health system center and the proportion of the population included in the EHR sample. Race and ethnicity and sex showed consistent findings except for some deviation in non-Hispanic Asians for MRP, which may be driven by the small EHR sample size in this subgroup (SM Figures 10 and 11). Patterning was unchanged with the utilization denominator instead of the proportion in the EHR sample.

\subsubsection{County-level T1D prevalence}

\begin{figure}[htbp!]
\centering
\includegraphics[scale = 0.5]{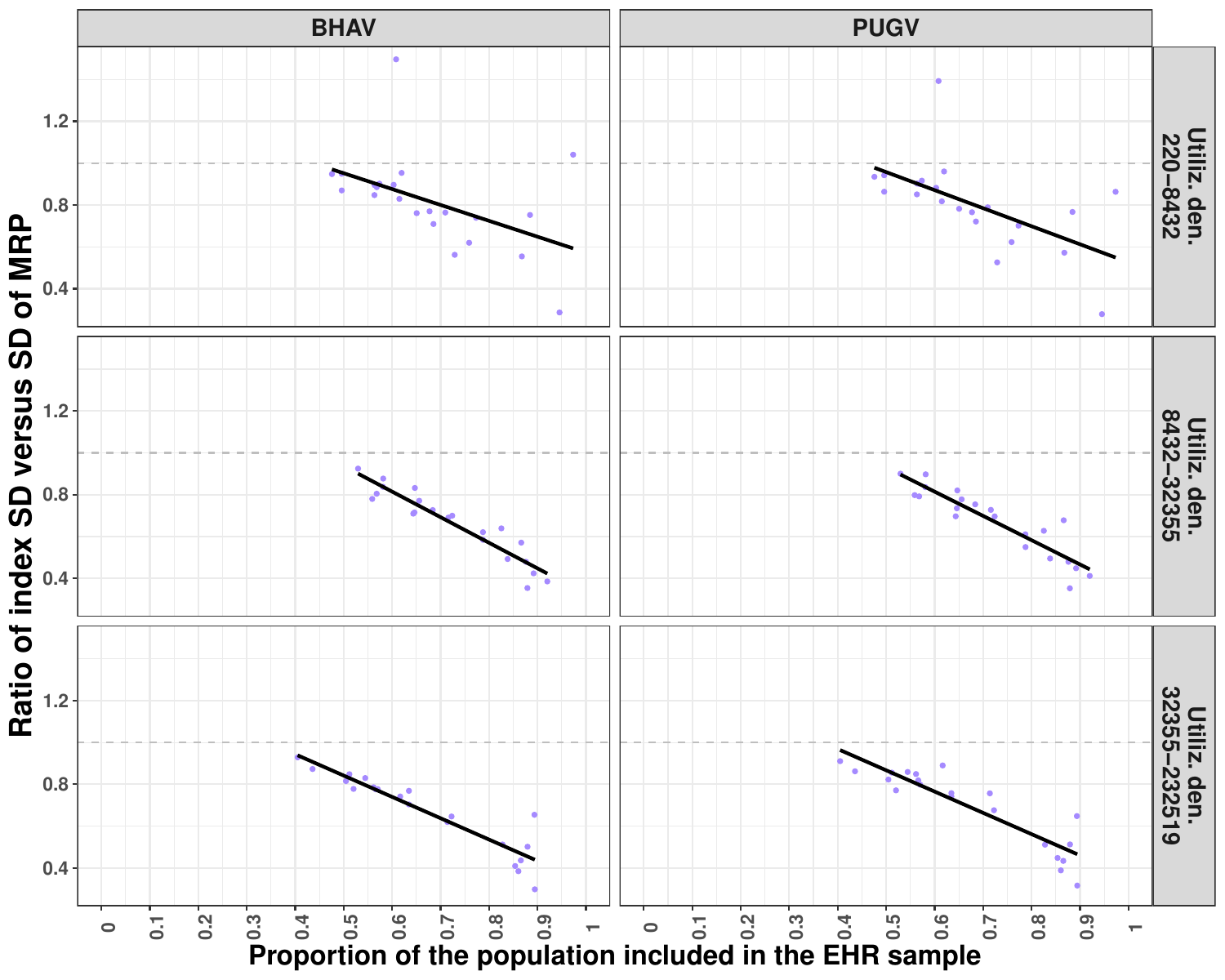}
\caption{Ratio of the index posterior standard deviation (SD) (BHAV or PUGV) of T1D cases per 1,000 persons and the posterior SD for MRP for counties in PEDSNet (PED) during 2018 and 2022, according to the proportion of the population included in the EHR sample and tertiles of the utilization denominator. BHAV, PUGV, and MRP refer to Bayesian Hierarchical Analysis of Variance, Partitioning Unobserved Geographic Variations, and Multilevel Regression and Poststratification, respectively.}
\label{f:Fig_superpopulation_sd_ratio_mrp_county_proppop_utliz_denPED_T1DM_a(Outlier)}
\end{figure}

SM Table 7 presents descriptive statistics of the distributions of the posterior means and posterior SDs for county-level T1D prevalence estimation among children and adolescents, in addition to the corresponding summaries for Na\"ive estimation. The median absolute relative differences (\%) in the posterior mean T1D prevalence comparing BHAV and PUGV versus MRP were less than 10\% (SM Table 8). For the posterior SD, absolute relative differences in PED were notably larger than in other health system centers. Figure \ref{f:Fig_superpopulation_sd_ratio_mrp_county_proppop_utliz_denPED_T1DM_a(Outlier)} shows that in PED, which had the highest proportion of the population included in the EHR sample among the health system centers, BHAV and PUGV exhibited a precision advantage over MRP across the distribution of proportion included in the EHR sample. Even in the lowest tertile of the utilization denominator, the precision advantage increased with increasing proportion included. In LUR and UFL, BHAV and PUGV demonstrated a precision advantage when the proportion of the population included in the EHR sample was higher and the utilization denominator was larger.

\begin{figure}[htbp!]
\centering
\includegraphics[scale = 0.5]{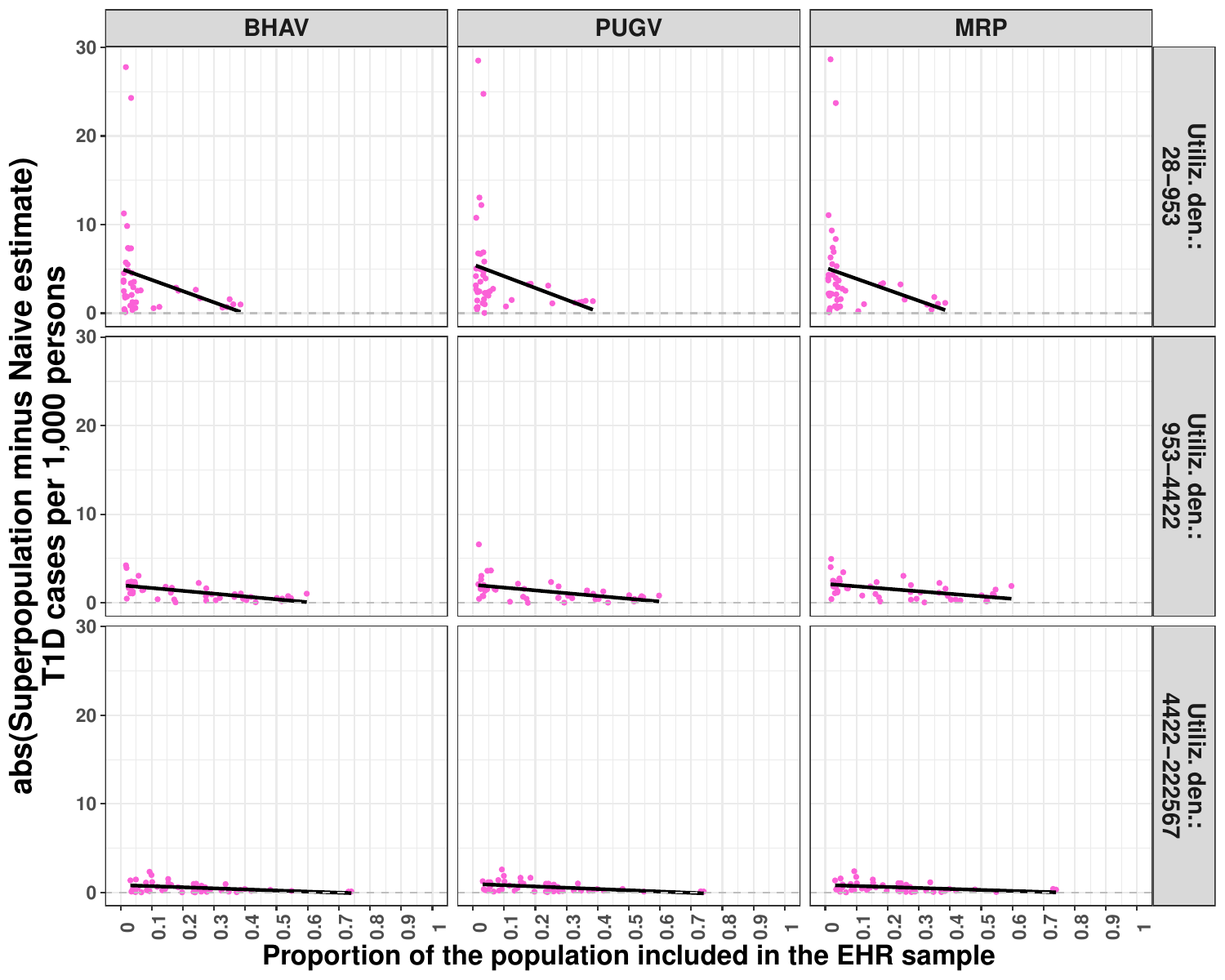}
\caption{Absolute difference in the T1D cases per 1,000 persons between the superpopulation-based and Na\"ive estimates for counties in OneFlorida$+$ (UFL) during 2018 and 2022 by superpopulation model, according to the proportion of the population included in the EHR sample and tertiles of the utilization denominator. BHAV, PUGV, and MRP refer to Bayesian Hierarchical Analysis of Variance, Partitioning Unobserved Geographic Variations, and Multilevel Regression and Poststratification, respectively.}
\label{f:Fig_superpopulation_minus_naive_county_proppop_utliz_denUFL_T1DM_a}
\end{figure}

Although in LUR and PED the distributions of the superpopulation-based and Na\"ive county-level T1D prevalence estimates were mostly similar, in UFL, the Na\"ive distribution differed due to estimated prevalences of zero in eight counties with no recorded T1D cases despite the utilization denominator ranging from 66 to 696 patients (SM Table 7). Figure \ref{f:Fig_superpopulation_minus_naive_county_proppop_utliz_denUFL_T1DM_a} uses UFL to depict the interplay of the utilization denominator and the proportion of the population included in the EHR sample on the absolute difference between the superpopulation-based and Na\"ive estimates. In the lowest tertile of the utilization denominator, the difference in the estimates diminishes as the proportion of the population included in the EHR sample increases. In higher tertiles of the utilization denominator, however, the effect of increasing the proportion of the population included in the EHR sample is attenuated. Indeed, the highest tertile shows little difference between the superpopulation and Na\"ive estimates even at low proportions of the population included in the EHR sample. The mitigating effect of the size of the utilization denominator was also present in LUR and PED (despite an outlying county in PED with a very small utilization denominator and very high proportion of the population included in the EHR sample).

\section{Plasmode Simulation Study}
\label{s:Simulation}

We designed a plasmode simulation study to examine the statistical properties of the proposed estimators under different assumptions for the selection mechanism. While ordinary simulation techniques based on fully synthetic simulated data may not be informative in the setting of EHR data because of their complexity \citep{Franklin2014}, a plasmode simulation design is based on the real data but with some aspect of the data-generating model controlled by the researcher, sometimes referred to as a ``spike-in'' \citep{Franklin2014,Vaughan2009}. In order to assess estimation performance for T1D prevalence in the population, our spike-in is the assumed relationship between T1D status and predictors among individuals in the population who were not included in the EHRs. 

The plasmode simulations proceeded in the three steps. First, to prepare to spike-in missing T1D status among individuals not included in the EHR sample, we used the population counts in each cross-classified cell to form an individual-level dataset in which each row refers to an individual in the population. Individuals who were not included in the EHR sample have their T1D status set to missing.  Second, to predict missing T1D status, we estimated a logistic regression model of T1D status as a function of covariate information in the EHR sample. To specify the model, in scenario 1 (S1), we included categorical race, age, and sex, and their two-way interactions, in addition to study year and county. In scenario 2 (S2), we additionally incorporated two-way interactions between health system center and each demographic variable and between center and year. Third, we ran the plasmode simulations by repeating the following steps: 1) Sample with replacement from each center with probability proportional to size; 2) Predict missing T1D status based on empirical associations in the EHR sample and in turn, compute the true T1D prevalence; and, 3) Apply the estimation methods in the EHR sample to estimate T1D prevalence in the population.

We fixed $N = 5,000,000$ individuals in the simulated universe. We ran 150 simulations. For each scenario, we verified that the plasmode and real DiCAYA data matched in terms of the proportion of the population included in the EHR sample and sample T1D prevalence. We evaluated the BHAV, PUGV, MRP, and Na\"ive estimation methods according to average bias, mean squared error (MSE), 95\% coverage probability, and average length of the 95\% CrI. The Na\"ive method replaced the CrI with the confidence interval. We considered estimands of T1D prevalence at a coarser scale, including T1D prevalence by year and by demographic subgroup and year; at a more resolved level, by center and year and by demographic subgroup in each center and year; and, at the county-level. Herein, for the purpose of brevity, we summarize the simulation results, particularly as they relate to findings in the real data analysis while providing details on estimation performance in SM Tables 9--19.

\begin{figure}[htbp!]
\centering
\includegraphics[scale = 0.5]{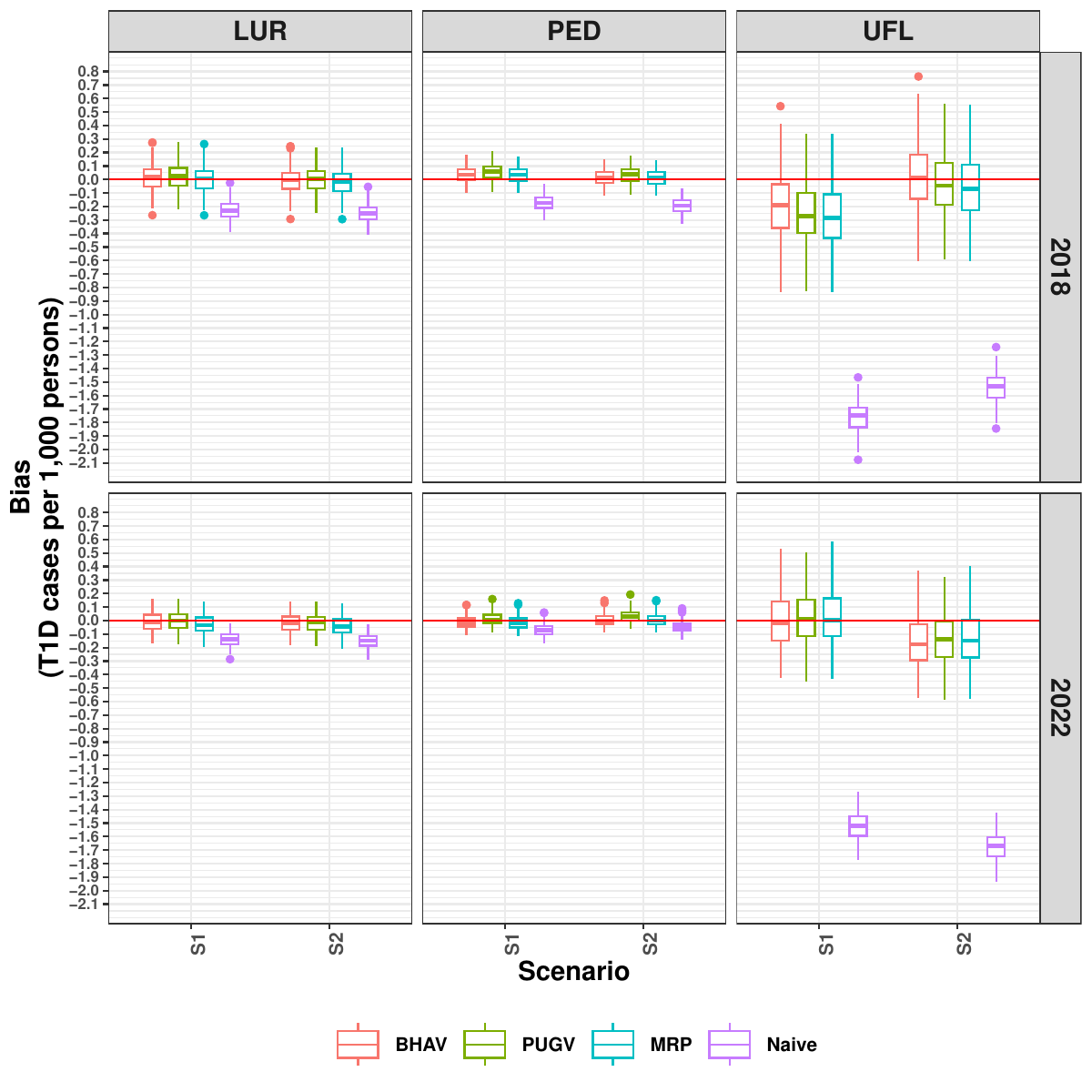}
\caption{Distribution of bias in T1D prevalence by health system center and year according to the superpopulation models and Na\"ive estimation method by scenario S1 and S2 in the plasmode simulations, color coded by estimation method. BHAV, PUGV, and MRP refer to Bayesian Hierarchical Analysis of Variance, Partitioning Unobserved Geographic Variations, and Multilevel Regression and Poststratification, respectively. LUR refers to Lurie Children's Hospital; PED, PEDSNet; and UFL, OneFlorida$+$.}
\label{f:Fig_plasmode_sim_boxplotbias_index_year_site_name1_a}
\end{figure}

The superpopulation models largely reduced bias in T1D prevalence estimates. 
By year, the average bias in the superpopulation-based estimates was less than 0.1 T1D cases per 1,000 persons. By year in each health system center, although the superpopulation models sometimes exhibited greater bias in UFL than LUR and PED, the bias did not exceed about 0.25 T1D cases per 1,000 persons (Figure \ref{f:Fig_plasmode_sim_boxplotbias_index_year_site_name1_a}, SM Table 13). For more granular estimands in demographic subgroups stratified by center and year, superpopulation-based estimates for subgroups with smaller utilization denominators (e.g., non-Hispanic Asian, non-Hispanic Black, and older age groups), and more generally for UFL were more likely to have some residual bias. However, the magnitude of the bias was often small, exceeding 1 T1D case per 1,000 persons only in one instance, for non-Hispanic Asians in UFL. In PED where the proportion of the population included in the EHR sample was highest (unlike BHAV and PUGV), the 95\% coverage probability for MRP twice reached 1 when it was the least biased estimator. Moreover, for racial and ethnic minorities in PED and UFL that had limited utilization denominators, MRP was only once the least bias estimator with 95\% coverage probability near the nominal level. Consistent with the real data analysis, the average bias from the Na\"ive method was consistently negative and especially pronounced in UFL.

Considering all estimands together (except those at the county level), BHAV performed best in terms of most often exhibiting the least average bias with nominal coverage, followed by PUGV (SM Tables 20--23). However, the best performing method varied by estimand type, scenario, and health system center. At a coarser level of granularity, PUGV performed best for estimands by year and those by demographic subgroup and year. MRP won under the simpler selection mechanism of S1, except for center-specific estimands in PED and UFL where BHAV prevailed. For estimands involving racial and ethnic minorities in PED and UFL where utilization denominators were limited, BHAV and PUGV outperformed MRP. 

For county-level T1D prevalence, among the superpopulation models, BHAV exhibited the least bias, followed by PUGV and MRP (SM Figure 12), and most often showed coverage probabilities closer to the nominal 0.95 level (SM Figure 13). Across scenarios and health system centers, BHAV demonstrated the best estimation performance in terms of being the least bias estimator with nominal coverage (SM Table 24). In Figure \ref{f:Fig_plasmode_sim_ave_bias_fips_county_code_index_year_a_proppoputilization_den}, in the lowest tertile of the proportion of the population included in the EHR sample, the average bias of the superpopulation models and the Na\"ive method diminished with increasing tertile of the utilization denominator. Conversely, in the lowest tertile of the utilization denominator, the average bias diminished with an increasing proportion of the population included in the EHR sample. Figure \ref{f:Fig_plasmode_sim_coverage_fips_county_code_index_year_a_proppoputilization_den} shows that across superpopulation models, for a given level of proportion of the population included in the EHR sample, the 95\% coverage probabilities improved with higher utilization denominators. Patterning by increasing proportion included within a tertile of the utilization denominator, however, was less consistent.

\begin{figure}[htbp!]
\centering
\includegraphics[scale = 0.5]{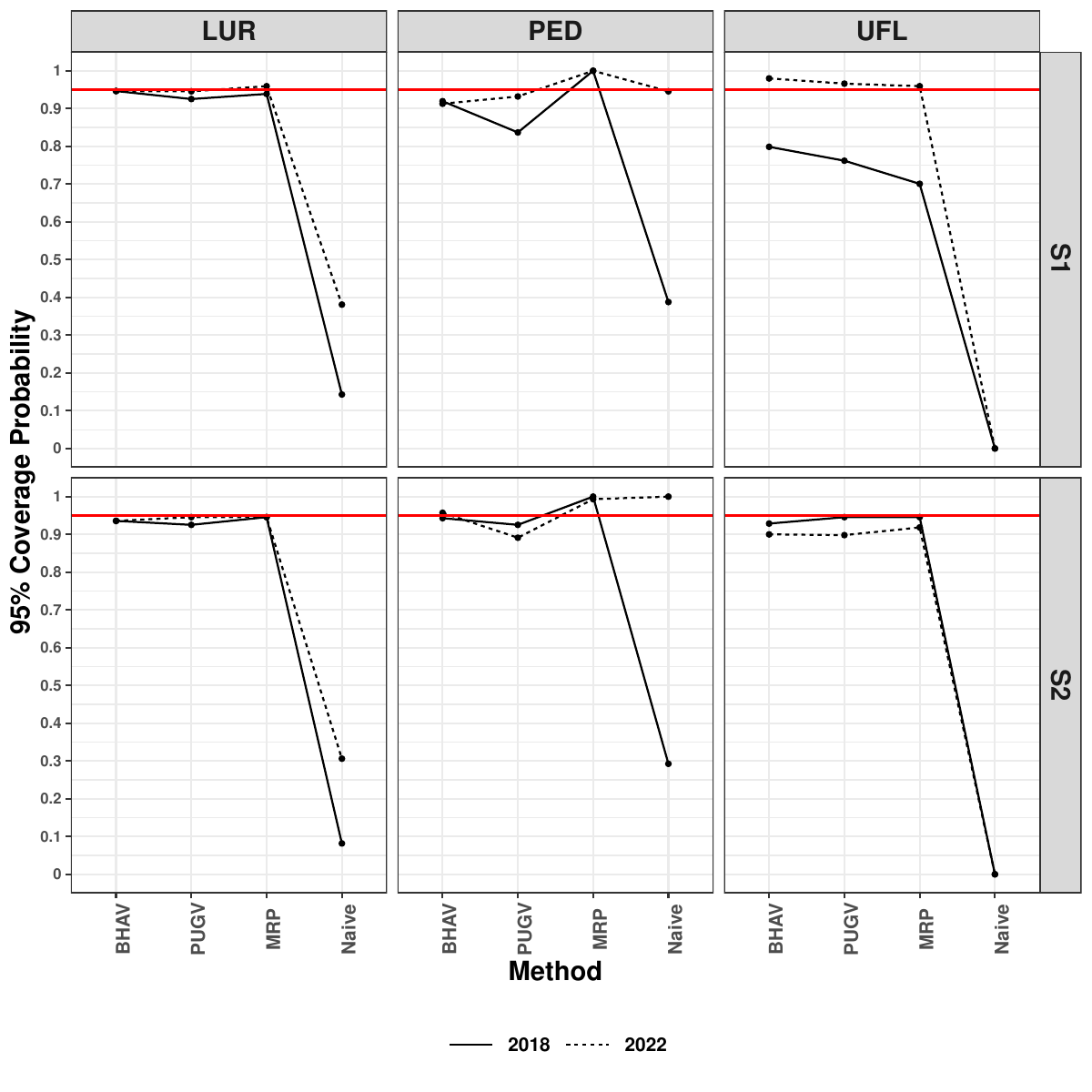}
\caption{95\% coverage probability for T1D prevalence by health system center and year according to the superpopulation models and Na\"ive estimation method by scenario S1 and S2 in the plasmode simulations. BHAV, PUGV, and MRP refer to Bayesian Hierarchical Analysis of Variance, Partitioning Unobserved Geographic Variations, and Multilevel Regression and Poststratification, respectively. LUR refers to Lurie Children's Hospital; PED, PEDSNet; and UFL, OneFlorida$+$.}
\label{f:Fig_plasmode_sim_lineplotcoverage_index_year_site_name1_a}
\end{figure}

\begin{figure}[htbp!]
\centering
\includegraphics[scale = 0.43]{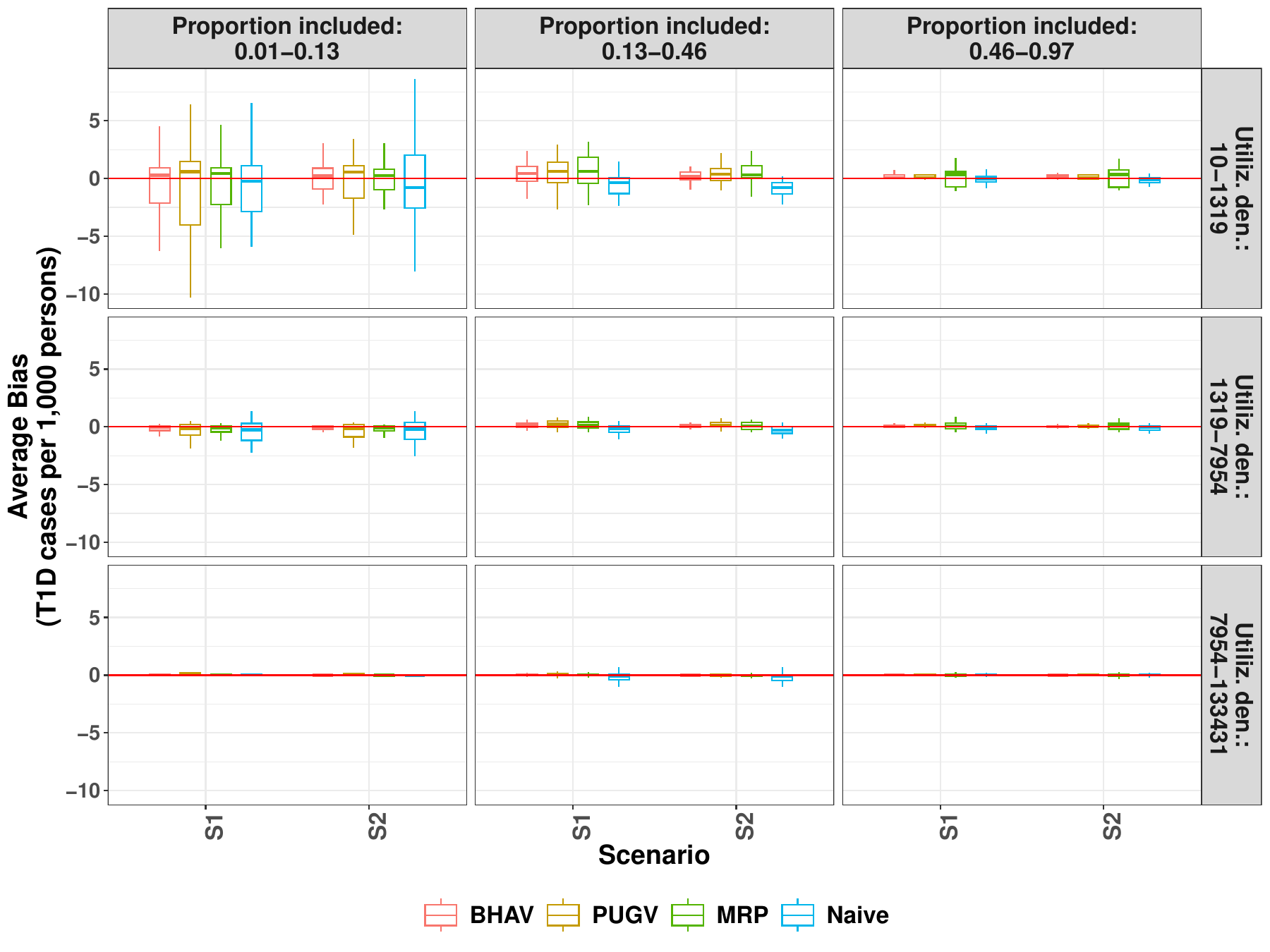}
\caption{Box plots of the average bias in T1D prevalence in counties for 2018 and 2022 according to the superpopulation models and Na\"ive estimation method by scenario S1 and S2, stratified by tertiles of proportion of the population included in the EHR sample and utilization denominator. BHAV, PUGV, and MRP refer to Bayesian Hierarchical Analysis of Variance, Partitioning Unobserved Geographic Variations, and Multilevel Regression and Poststratification, respectively.}
\label{f:Fig_plasmode_sim_ave_bias_fips_county_code_index_year_a_proppoputilization_den}
\end{figure}

\begin{figure}[htbp!]
\centering
\includegraphics[scale = 0.43]{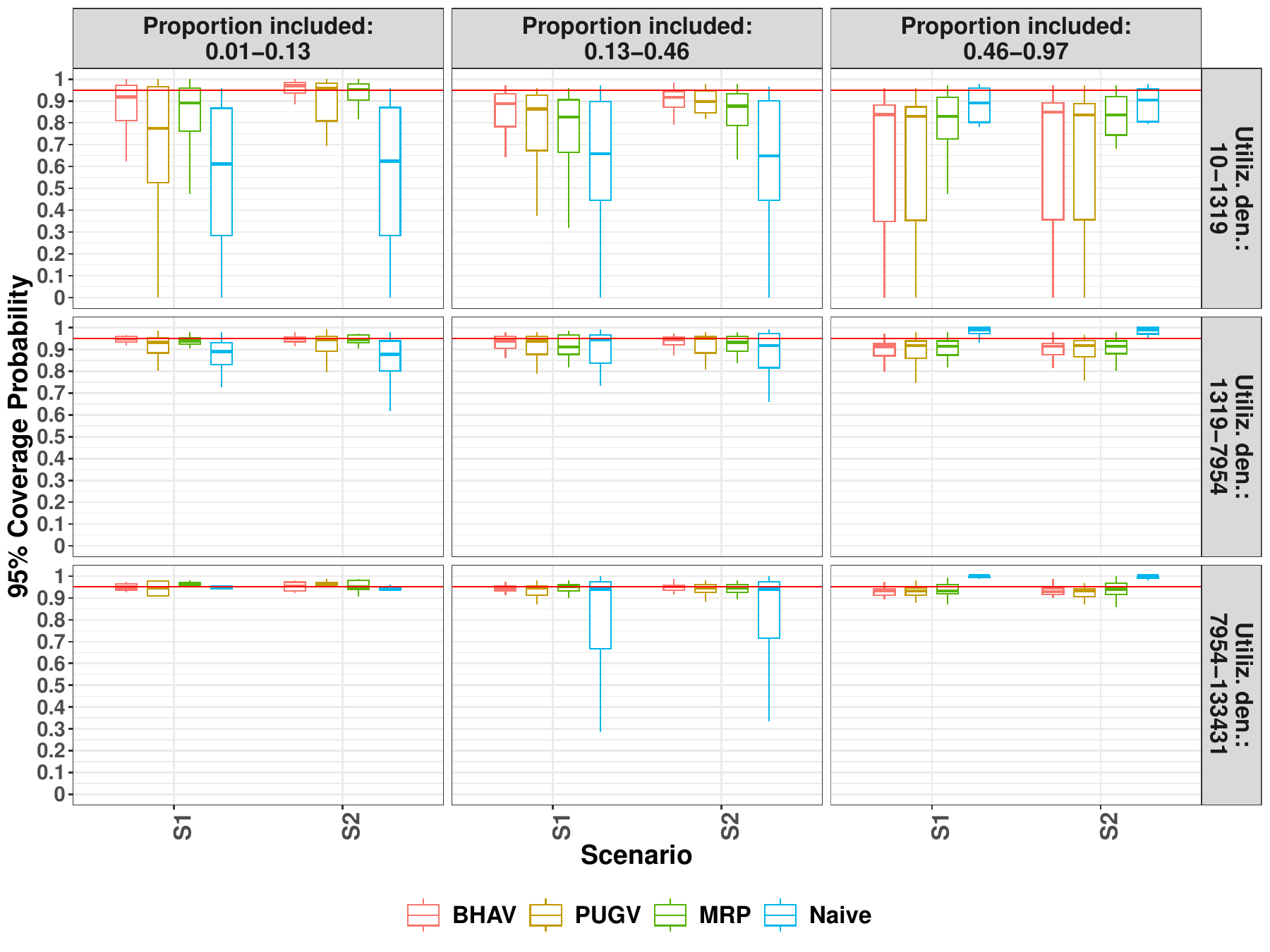}
\caption{Box plots of the 95\% coverage probabilities for T1D prevalence in counties for 2018 and 2022 according to the superpopulation models and Na\"ive estimation method by scenario S1 and S2, stratified by tertiles of proportion of the population included in the EHR sample and utilization denominator. BHAV, PUGV, and MRP refer to Bayesian Hierarchical Analysis of Variance, Partitioning Unobserved Geographic Variations, and Multilevel Regression and Poststratification, respectively.}
\label{f:Fig_plasmode_sim_coverage_fips_county_code_index_year_a_proppoputilization_den}
\end{figure}

\clearpage

\section{Discussion}
\label{s:Discussion}

In this study, we proposed and evaluated three Bayesian superpopulation models for population-based T1D surveillance in children and adolescents using data collected in EHRs. The superpopulation models flexibly enable estimating T1D prevalence at varying levels of granularity from estimation overall to demographic- and geographic-based subgroups. In the setting of multi-site EHR data from across the US, the superpopulation models allowed center-specific patterns of response compared to Na\"ive estimation. Among the superpopulation models, BHAV demonstrated its robustness to estimating prevalence for alternative estimands, under varying sizes of the utilization denominator and proportions of the population included in the EHR sample, and under different assumptions for the selection mechanism. In practice, BHAV is advantageous because it reduces the emphasis on correctly specifying the model while characterizing uncertainty well in prevalence estimates.

The influence of the utilization denominator's size and the proportion of the population included in the EHR sample on the difference between the superpopulation-based and Na\"ive prevalence estimates in equation (\ref{e:diff_naive_posterior_mean_pg}) was borne out by our findings. As the health system center with the smallest utilization denominator and the lowest proportion of the population included in the EHR sample, UFL exhibited the largest differences between the superpopulation-based and Na\"ive estimates. While the proportion in the EHR sample was similar in LUR, its utilization denominator was more than double that of UFL, thus having an attenuating effect on the differences. With the highest proportion in the EHR sample and a utilization denominator of intermediate size, PED showed differences between superpopulation and Na\"ive estimation in between those of LUR and UFL. More generally, we observed across superpopulation models that larger utilization denominators may offset the adverse effects on estimation of lower proportions of the population included in the EHR sample, and vice versa. For example, even in very low proportions included in the EHR sample, a large utilization denominator will help ensure accurate point and interval estimation. Thus, rather than operating in isolation, the utilization denominator and the proportion of the population included in the EHR sample together impact the quality of population-based prevalence estimates derived from EHR data.

The validity of the T1D estimates depends on whether the implicit assumption about the selection mechanism in a given estimation method holds. Consistent with previous research using EHRs to estimate childhood obesity prevalence  \citep{Flood2015}, we found that the Na\"ive method underestimated the true T1D prevalence in the population, with the superpopulation models largely correcting this negative bias. The Na\"ive method is unbiased if individuals included in EHRs are a random sample from the target population (termed selection completely at random). The prevailing notion that data collected in EHRs overrepresent sicker individuals from the target population implies that the Na\"ive method will overestimate the true T1D prevalence, contrary to our findings. By using a utilization denominator based on the index year with the two previous years, DiCAYA sought to mitigate selection bias related to varying patterns of health system interaction. However, the extent to which patterns of health system interaction underlie the difference between the EHR sample-based and population prevalence for a disease like T1D is unclear. For example, compared to type 2 diabetes, T1D often requires more continuous and specialized care for insulin management, thus leading to regular clinic visits over time. An alternative explanation is that youth with T1D may receive their care outside the health systems included in DiCAYA. In contrast to Na\"ive estimation, superpopulation-based estimates are unbiased under the less stringent assumption of selection at random, which means that for a given superpopulation model, conditional on the observed covariates and the latent variables, the probability of being included in the EHR sample is independent of missing diabetes status from individuals not included in the EHR sample. As the selection mechanism in EHRs may not be fully understood, flexibly accounting for unobserved sources of variation in estimation approaches is critical for valid inference.

Our work underscored the potential complexities of using data collected in EHRs for finite population quantity estimation. In the plasmode simulation study, while the estimation performance of BHAV was the best overall (in terms of bias and 95\% coverage probability), the performance of the superpopulation models tended to vary by health system center. We observed that MRP and PUGV were competitive methods in LUR, but in PED and UFL, BHAV prevailed. Moreover, the level of granularity of the estimand mattered. We found that for granular estimands involving demographic subgroups with smaller utilization denominators, the estimation performance of MRP in particular worsened, especially in terms of interval estimation. This is consistent with previous research showing that MRP may not appropriately propagate uncertainty in finite population quantity estimates \citep{Buttice2013,Valliant2020}. Finally, the superpopulation models appeared sensitive to the assumed selection mechanism. For example, the estimation performance of MRP was stronger under the simpler scenario of S1 versus S2. Among the superpopulation models, BHAV emerged as the best method overall due to its robust performance across different settings.

Future research may prioritize evaluating the generalizability of the proposed Bayesian superpopulation models to other settings. For example, although the Poisson random effects framework adequately accounted for features of the T1D count distribution in children and adolescents, alternative models like the zero-inflated negative binomial or Poisson mixture may be necessary for other outcomes like type 2 diabetes or different target populations like older adults. Comparative studies of superpopulation modeling with quasi-randomization and machine learning methods, especially when scientific interest is in estimates of varying granularity, are needed. Finally, although the superpopulation models in our work are premised on selection at random, selection not at random (akin to missing not at random) in data collected in EHRs is certainly plausible in light of the nonprescheduled nature of the data generating process. Extensions to selection not at random are warranted.

\clearpage



\subsection*{Acknowledgments}

The research reported in this publication was conducted using data from the DiCAYA (Assessing the Burden of Diabetes by Type in Children, Adolescents, and Young Adults) Network. DiCAYA is funded by the Centers for Disease Control and Prevention (CDC) (DP20-001) and the National Institute of Diabetes and Digestive and Kidney Diseases to modernize diabetes surveillance efforts using electronic health record data and advanced statistical analysis.
This study includes data from the following institutions: the University of South Carolina, the University of Colorado Denver, the University of Florida, Lurie Children’s Hospital, Kaiser Permanente Southern California, Geisinger, and Indiana University, in partnership with the Regenstrief Institute. The Regenstrief Institute is the honest data broker for the Indiana Network for Patient Care (INPC), which includes EHR data from 38 distinct health systems across Indiana, as well as a variety of outpatient, laboratory, long-term care, and public health sources managed by the Indiana Health Information Exchange (IHIE).

The research reported in this study was also conducted using PEDSnet and the OneFlorida+ Clinical Research Network and several other clinical research networks in the project led by Lurie Children’s Hospital. PEDSnet, a pediatric learning health system, includes data from the following member institutions: Children’s Hospital Colorado, Children’s Hospital of Philadelphia, Cincinnati Children’s Hospital Medical Center, Lurie Children’s Hospital, Nationwide Children’s Hospital, Nemours Children’s Health, and Seattle Children’s Hospital. PEDSnet is a partner network clinical data research network in PCORnet, the National Patient-Centered Clinical Research Network, an initiative funded by the Patient-Centered Outcomes Research Institute (PCORI). OneFlorida+ is a PCORnet-affiliated research network spanning Florida, Georgia, and Alabama that brings together academic health systems, community healthcare partners, and state agencies to support pragmatic clinical trials, comparative effectiveness and implementation research, observational studies, and cohort discovery.

The Lurie Children’s Hospital DiCAYA project is composed of health care institutions from multiple clinical research networks: The INSIGHT Clinical Research Network (CRN) is a collaborative initiative that integrates clinical and social determinants of health data from more than 18 million patients across New York City, Long Island, NY, and Houston, TX. This work is supported in-part by the PCORI-funded PCORnet award to the INSIGHT CRN (Grant No. RI-CORNELL-01-MC). The approach described in this manuscript was developed in partnership with Research Action for Health Network (REACHnet), funded by PCORI (PCORI Award RI-LPHI-01-MC). REACHnet is a partner network in PCORnet, which was developed with funding from PCORI. The Accelerating Data Value Across a National Community Health Center Network (ADVANCE) Clinical Research Network is a member of PCORnet. ADVANCE is a multicenter collaborative led by OCHIN (not an acronym) in partnership with Fenway Health, Health Choice Network, and Oregon Health \& Science University. The Chicago Area Patient-Centered Outcomes Research Network (CAPriCORN) is a partnership between healthcare and research institutions that provides data through a federated, harmonized common data model. We acknowledge CAPriCORN’s partners, the CAPriCORN Patient Community Advisory Board, and the Chicago Area Institutional Review Board (CHAIRb), which serves as the central IRB of record for CAPriCORN-supported research. PaTH is a Network Partner in PCORnet, the National Patient-Centered Clinical Research Network. PCORnet has been developed with funding from the Patient-Centered Outcomes Research Institute (PCORI). Johns Hopkins’ participation in PCORnet was funded through PCORI Award (RI-PITT-01-PS1).

\subsection*{Financial Disclosure}

None reported.

\subsection*{Conflicts of Interest}

The authors declare no conflicts of interest.

\bibliography{ref}

@Manual{Plummer2017,
    title = {JAGS version 4.3.0 user manual},
    author = {Martyn Plummer},
    year = {2017}
}

@Manual{Su2024,
    title = {R2jags: Using R to Run 'JAGS'},
    author = {Yu-Sung Su and Masanao Yajima},
    year = {2024},
    note = {R package version 0.8-5},
    url = {https://CRAN.R-project.org/package=R2jags},
}

@article{Gelman1996,
author = {Gelman, A. and Meng, X. and Stern, H.},
journal = {Statistica Sinica},
number = {4},
pages = {733--760},
title = {{Posterior Predictive Assessment of Model Fitness via Realized Discrepancies}},
volume = {6},
year = {1996}
}

@article{Conderino2024,
  title={Addressing selection biases within electronic health record data for estimation of diabetes prevalence among New York City young adults: a cross-sectional study},
  author={Conderino, Sarah and Thorpe, Lorna E and Divers, Jasmin and Albrecht, Sandra S and Farley, Shannon M and Lee, David C and Anthopolos, Rebecca},
  journal={BMJ public health},
  volume={2},
  number={2},
  year={2024},
  publisher={BMJ Publishing Group Ltd}
}

@article{McVeigh2016,
  title={Can electronic health records be used for population health surveillance? Validating population health metrics against established survey data},
  author={McVeigh, Katharine H and Newton-Dame, Remle and Chan, Pui Ying and Thorpe, Lorna E and Schreibstein, Lauren and Tatem, Kathleen S and Chernov, Claudia and Lurie-Moroni, Elizabeth and Perlman, Sharon E},
  journal={EGEMs},
  volume={4},
  number={1},
  pages={1267},
  year={2016}
}

@article{Thorpe2016,
  title={Monitoring prevalence, treatment, and control of metabolic conditions in New York City adults using 2013 primary care electronic health records: a surveillance validation study},
  author={Thorpe, Lorna E and McVeigh, Katharine H and Perlman, Sharon and Chan, Pui Ying and Bartley, Katherine and Schreibstein, Lauren and Rodriguez-Lopez, Jesica and Newton-Dame, Remle},
  journal={eGEMs},
  volume={4},
  number={1},
  pages={1266},
  year={2016}
}

@incollection{Malec2008,
  title={A Bayesian semi-parametric model for small area estimation},
  author={Malec, Donald and M{\"u}ller, Peter},
  booktitle={Pushing the Limits of Contemporary Statistics: Contributions in Honor of Jayanta K. Ghosh},
  volume={3},
  pages={223--237},
  year={2008},
  publisher={Institute of Mathematical Statistics}
}

@article{Spiegelhalter2002,
  title={Bayesian measures of model complexity and fit},
  author={Spiegelhalter, David J and Best, Nicola G and Carlin, Bradley P and Van Der Linde, Angelika},
  journal={Journal of the Royal Statistical Society: Series b (statistical methodology)},
  volume={64},
  number={4},
  pages={583--639},
  year={2002},
  publisher={Wiley Online Library}
}

@article{Watanabe2010,
  title={Asymptotic equivalence of Bayes cross validation and widely applicable information criterion in singular learning theory},
  author={Watanabe, Sumio},
  journal={Journal of machine learning research},
  volume={11},
  number={12},
  year={2010}
}

@article{Buttice2013,
  title={How does multilevel regression and poststratification perform with conventional national surveys?},
  author={Buttice, Matthew K and Highton, Benjamin},
  journal={Political Analysis},
  volume={21},
  number={04},
  pages={449--467},
  year={2013}
}

@article{Hirsch2024,
  title={Using electronic health records to enhance surveillance of diabetes in children, adolescents and young adults: a study protocol for the DiCAYA Network},
  author={Hirsch, Annemarie G and Conderino, Sarah and Crume, Tessa L and Liese, Angela D and Bellatorre, Anna and Bendik, Stefanie and Divers, Jasmin and Anthopolos, Rebecca and Dixon, Brian E and Guo, Yi and others},
  journal={BMJ open},
  volume={14},
  number={1},
  pages={e073791},
  year={2024},
  publisher={British Medical Journal Publishing Group}
}

@article{Chen2022,
  title={Small-area estimation for public health surveillance using electronic health record data: reducing the impact of underrepresentation},
  author={Chen, Tom and Li, Wenjun and Zambarano, Bob and Klompas, Michael},
  journal={BMC Public Health},
  volume={22},
  number={1},
  pages={1515},
  year={2022},
  publisher={Springer}
}

@article{Klompas2017,
   author = {M. Klompas and N. M. Cocoros and J. T. Menchaca and D. Erani and E. Hafer and B. Herrick and M. Josephson and M. Lee and M. D. P. Weiss and B. Zambarano and K. R. Eberhardt and J. Malenfant and L. Nasuti and T. Land},
   doi = {10.2105/AJPH.2017.303874},
   issn = {15410048},
   issue = {9},
   journal = {American Journal of Public Health},
   month = {9},
   pages = {1406-1412},
   pmid = {28727539},
   publisher = {American Public Health Association Inc.},
   title = {State and Local Chronic Disease Surveillance Using Electronic Health Record Systems},
   volume = {107},
   year = {2017},
}

@article{Flood2015,
  title={Electronic health records and community health surveillance of childhood obesity},
  author={Flood, Tracy L and Zhao, Ying-Qi and Tomayko, Emily J and Tandias, Aman and Carrel, Aaron L and Hanrahan, Lawrence P},
  journal={American journal of preventive medicine},
  volume={48},
  number={2},
  pages={234--240},
  year={2015},
  publisher={Elsevier}
}

@article{Daley2020,
  title={Estimating childhood obesity prevalence in communities through multi-institutional data sharing},
  author={Daley, Matthew F and Barrow, Jennifer C and Tabano, David C and Reifler, Liza M and Kraus, Emily M and Davies, Sara Deakyne and Williford, Devon L and White, Benjamin and Shupe, Alyson and Davidson, Arthur J},
  journal={Journal of Public Health Management and Practice},
  volume={26},
  number={4},
  pages={E1--E10},
  year={2020},
  publisher={LWW}
}

@article{Cheng2019,
  title={Prevalence of diabetes by race and ethnicity in the United States, 2011-2016},
  author={Cheng, Yiling J and Kanaya, Alka M and Araneta, Maria Rosario G and Saydah, Sharon H and Kahn, Henry S and Gregg, Edward W and Fujimoto, Wilfred Y and Imperatore, Giuseppina},
  journal={Jama},
  volume={322},
  number={24},
  pages={2389--2398},
  year={2019},
  publisher={American Medical Association}
}

@article{Vaughan2009,
  title={The use of plasmodes as a supplement to simulations: a simple example evaluating individual admixture estimation methodologies},
  author={Vaughan, Laura K and Divers, Jasmin and Padilla, Miguel A and Redden, David T and Tiwari, Hemant K and Pomp, Daniel and Allison, David B},
  journal={Computational statistics \& data analysis},
  volume={53},
  number={5},
  pages={1755--1766},
  year={2009},
  publisher={Elsevier}
}

@article{Franklin2014,
  title={Plasmode simulation for the evaluation of pharmacoepidemiologic methods in complex healthcare databases},
  author={Franklin, Jessica M and Schneeweiss, Sebastian and Polinski, Jennifer M and Rassen, Jeremy A},
  journal={Computational statistics \& data analysis},
  volume={72},
  pages={219--226},
  year={2014},
  publisher={Elsevier}
}

@article{Wang2015,
  title={Forecasting elections with non-representative polls},
  author={Wang, Wei and Rothschild, David and Goel, Sharad and Gelman, Andrew},
  journal={International Journal of Forecasting},
  volume={31},
  number={3},
  pages={980--991},
  year={2015},
  publisher={Elsevier}
}

@article{Gelman1997,
  title={Poststratification into many categories using hierarchical logistic regression},
  author={Gelman, Andrew},
  journal={Survey Methodology},
  volume={23},
  pages={127},
  year={1997}
}

@article{Little1993,
  title={Post-stratification: a modeler's perspective},
  author={Little, Roderick JA},
  journal={Journal of the American Statistical Association},
  volume={88},
  number={423},
  pages={1001--1012},
  year={1993},
  publisher={Taylor \& Francis}
}

@book{Gelman2013,
  title={Bayesian Data Analysis},
  author={Gelman, Andrew and Carlin, John B and Stern, Hal S and Dunson, David B and Vehtari, Aki and Rubin, Donald B},
  year={2013},
  publisher={Chapman and Hall/CRC}
}

@article{Gelman2005,
  title={Analysis of variance--why it is more important than ever},
  author={Gelman, Andrew},
 journal={The Annals of Statistics},
volume = {33},
number = {1},
pages = {1--53},
  year={2005}
}

@article{Best1996,
   author = {N. G. Best and D. J. Spiegelhalter and A. Thomas and C. E. G. Brayne},
   doi = {10.2307/2983178},
   issn = {09641998},
   issue = {2},
   journal = {Journal of the Royal Statistical Society. Series A (Statistics in Society)},
   pages = {323},
   publisher = {JSTOR},
   title = {Bayesian Analysis of Realistically Complex Models},
   volume = {159},
   year = {1996},
}

@article{Bower2017,
   author = {J. K. Bower and S. Patel and J. E. Rudy and A. S. Felix},
   doi = {10.1007/s40471-017-0130-z},
   issue = {4},
   journal = {Current Epidemiology Reports},
   month = {12},
   pages = {346-352},
   publisher = {Springer Nature},
   title = {Addressing Bias in Electronic Health Record-based Surveillance of Cardiovascular Disease Risk: Finding the Signal Through the Noise},
   volume = {4},
   year = {2017},
}

@article{Brick2013,
   author = {J. M. Brick and D. Williams},
   doi = {10.1177/0002716212456834},
   issn = {00027162},
   issue = {1},
   journal = {Annals of the American Academy of Political and Social Science},
   month = {1},
   pages = {36-59},
   title = {Explaining Rising Nonresponse Rates in Cross-Sectional Surveys},
   volume = {645},
   year = {2013},
}

@article{Elliott2017,
   author = {M. R. Elliott and R. Valliant},
   doi = {10.1214/16-STS598},
   issn = {08834237},
   issue = {2},
   journal = {Statistical Science},
   month = {5},
   pages = {249-264},
   publisher = {Institute of Mathematical Statistics},
   title = {Inference for Nonprobability Samples},
   volume = {32},
   year = {2017},
}

@report{Kohut2012,
   author = {A. Kohut and S. Keeter and C. Doherty and M. Dimock and L. Christian},
   institution = {The Pew Research Center For the People & The Press},
   pages = {1-52},
   title = {Assessing the Representativeness of Public Opinion Surveys},
   volume = {15},
   url = {www.peoplepress.org},
   year = {2012},
}

@article{Paul2015,
   author = {M. M. Paul and C. M. Greene and R. Newton-Dame and L. E. Thorpe and S. E. Perlman and K. H. McVeigh and M. N. Gourevitch},
   doi = {10.1089/pop.2014.0093},
   issn = {19427905},
   issue = {3},
   journal = {Population Health Management},
   month = {6},
   pages = {209-216},
   pmid = {25608033},
   publisher = {Mary Ann Liebert Inc.},
   title = {The State of Population Health Surveillance Using Electronic Health Records: A Narrative Review},
   volume = {18},
   year = {2015},
}

@article{Romo2016,
   author = {M. L. Romo and P. Y. Chan and E. Lurie-Moroni and S. E. Perlman and R. Newton-Dame and L. E. Thorpe and K. H. McVeigh},
   doi = {10.5888/pcd13.150500},
   issn = {21665435},
   issue = {E56},
   journal = {Preventing Chronic Disease},
   month = {4},
   pages = {1-15},
   pmid = {27126554},
   publisher = {Centers for Disease Control and Prevention (CDC)},
   title = {Characterizing Adults Receiving Primary Medical Care in New York City: Implications for Using Electronic Health Records for Chronic Disease Surveillance},
   volume = {13},
   year = {2016},
}

@article{Valliant2020,
   author = {Richard Valliant},
   doi = {10.1093/jssam/smz003},
   issn = {23250992},
   issue = {2},
   journal = {Journal of Survey Statistics and Methodology},
   month = {4},
   pages = {231-263},
   publisher = {American Association for Public Opinion Research},
   title = {Comparing Alternatives for Estimation from Nonprobability Samples},
   volume = {8},
   year = {2020},
}

\clearpage

\appendix

\begin{center}
\huge
Supplementary Materials
\end{center}

\setcounter{figure}{0}
\setcounter{table}{0}

\section{Motivating Data}
\label{sm:MotivatingData}

\begin{table}[htbp!]
\centering
\begin{tabular}{lccccccc}
  \hline
 & \multicolumn{7}{c}{Utilization Denominator} \\ \cline{2-8}
\underline{Demographic subgroup} & \multicolumn{1}{c}{\underline{min}} & \multicolumn{1}{c}{\underline{q10}} & \multicolumn{1}{c}{\underline{q25}} & \multicolumn{1}{c}{\underline{q50}} & \multicolumn{1}{c}{\underline{q75}} & \multicolumn{1}{c}{\underline{q90}} & \multicolumn{1}{c}{\underline{max}} \\ 
& \multicolumn{7}{c}{LUR} \\ \hline
Age, years: & &  &  &  &  &  & \\
  \hspace{0.1cm}0-5 & 1 & 82 & 258 & 900 & 2494 & 4749 & 27487 \\ 
  \hspace{0.1cm}5-10 & 1 & 71 & 217 & 694 & 1862 & 4010 & 19560 \\ 
  \hspace{0.1cm}10-15 & 1 & 70 & 224 & 752 & 1890 & 4304 & 18410 \\ 
  \hspace{0.1cm}15-18 & 0 & 41 & 156 & 500 & 1282 & 3070 & 11238 \\ 
Race and ethnicity: & &  &  &  &  &  & \\
\hspace{0.1cm}Non-Hispanic Asian & 0 & 12 & 54 & 166 & 490 & 1232 & 5153 \\ 
  \hspace{0.1cm}Non-Hispanic Black & 3 & 81 & 206 & 528 & 1112 & 3374 & 13121 \\ 
  \hspace{0.1cm}Hispanic & 16 & 179 & 374 & 989 & 2588 & 4797 & 24549 \\ 
  \hspace{0.1cm}Non-Hispanic White & 37 & 274 & 765 & 1560 & 3221 & 5917 & 27487 \\ 
Sex: & &  &  &  &  &  & \\
  \hspace{0.1cm}Female & 1 & 63 & 202 & 669 & 1814 & 3925 & 23806 \\ 
  \hspace{0.1cm}Male & 0 & 67 & 206 & 708 & 1910 & 4084 & 27487 \\ 
& \multicolumn{7}{c}{PED} \\ \hline
Age, years: & &  &  &  &  &  & \\
  \hspace{0.1cm}0-5 & 0 & 5 & 14 & 495 & 1480 & 5160 & 18832 \\ 
  \hspace{0.1cm}5-10 & 0 & 6 & 18 & 462 & 1514 & 4945 & 21545 \\ 
  \hspace{0.1cm}10-15 & 0 & 6 & 17 & 467 & 1448 & 4904 & 21600 \\ 
  \hspace{0.1cm}15-18 & 0 & 4 & 10 & 282 & 888 & 3124 & 11192 \\ 
Race and ethnicity: & &  &  &  &  &  & \\
  \hspace{0.1cm}Non-Hispanic Asian & 0 & 2 & 5 & 35 & 659 & 1595 & 7299 \\ 
  \hspace{0.1cm}Non-Hispanic Black & 0 & 6 & 14 & 60 & 822 & 3143 & 21600 \\ 
  \hspace{0.1cm}Hispanic & 0 & 6 & 13 & 105 & 772 & 2562 & 7092 \\ 
  \hspace{0.1cm}Non-Hispanic White & 263 & 555 & 782 & 1503 & 5224 & 9601 & 19996 \\ 
  Sex: & &  &  &  &  &  & \\
  \hspace{0.1cm}Female & 0 & 5 & 14 & 375 & 1313 & 4408 & 20460 \\ 
  \hspace{0.1cm}Male & 0 & 4 & 16 & 378 & 1418 & 4762 & 21600 \\ 
& \multicolumn{7}{c}{UFL} \\ \hline
  Age, years: & &  &  &  &  &  & \\
  \hspace{0.1cm}0-5 & 0 & 1 & 6 & 48 & 195 & 907 & 26858 \\ 
  \hspace{0.1cm}5-10 & 0 & 0 & 4 & 32 & 154 & 642 & 24311 \\ 
  \hspace{0.1cm}10-15 & 0 & 1 & 4 & 32 & 148 & 631 & 23199 \\ 
  \hspace{0.1cm}15-18 & 0 & 0 & 3 & 23 & 100 & 438 & 12637 \\ 
Race and ethnicity: & &  &  &  &  &  & \\
  \hspace{0.1cm}Non-Hispanic Asian & 0 & 0 & 0 & 1 & 8 & 54 & 524 \\ 
  \hspace{0.1cm}Non-Hispanic Black & 0 & 3 & 10 & 35 & 146 & 683 & 5424 \\ 
  \hspace{0.1cm}Hispanic & 0 & 2 & 7 & 36 & 137 & 795 & 26858 \\ 
  \hspace{0.1cm}Non-Hispanic White & 0 & 14 & 47 & 130 & 484 & 1224 & 4835 \\ 
Sex: & &  &  &  &  &  & \\
  \hspace{0.1cm}Female & 0 & 1 & 4 & 30 & 142 & 623 & 24449 \\ 
  \hspace{0.1cm}Male & 0 & 0 & 4 & 33 & 158 & 648 & 26858 \\ 
   \hline
\end{tabular}
\caption{Distribution of the utilization denominator in cross-classified cells by age, race and ethnicity, and sex, stratified by health system center. The utilization denominator in a given cell is the total number of unique patients who interacted with the health
system center at least once in the index year or in the previous two years. LUR refers to Lurie Children's Hospital; PED, PEDSNet; and UFL, OneFlorida$+$.} 
\end{table}

\clearpage

\section{Data Application}
\label{sm:DataApplication}

\begin{table}[htbp!]
\centering
\tiny
\begin{tabular}{ccc}
  \hline
 \multicolumn{1}{c}{Parameter} &  \multicolumn{1}{c}{Description} &  \multicolumn{1}{c}{Posterior Mean (95\% CrI)} \\ 
  \hline
  $\sigma^{county}$ & Sd for county & 0.541 (0.468, 0.626) \\  \hline

  $\beta^{race}_1$ & Non-Hispanic Asian vs. non-Hispanic White & -0.909 (-1.072, -0.749) \\ 
  $\beta^{race}_2$ & Non-Hispanic Black vs. non-Hispanic White & -0.378 (-0.462, -0.291) \\ 
  $\beta^{race}_3$ & Hispanic vs. non-Hispanic White & -0.459 (-0.536, -0.383) \\ 
  $\sigma^{race}$ & Sd for race & 1.048 (0.383, 2.832) \\  \hline

  $\beta^{age}_1$ & Age 0-5 vs. 10-15 years   & -2.409 (-2.544, -2.277) \\ 
  $\beta^{age}_2$ & Age 5-10 vs. 10-15 years & -0.826 (-0.903, -0.746) \\ 
  $\beta^{age}_4$ & Age 15-18 vs. 10-15 years & 0.320 (0.254, 0.386) \\ 
   $\sigma^{age}$ & Sd for age & 2.135 (0.894, 5.074) \\  \hline
   
  $\beta^{sex}$ & Female vs. male & -0.008 (-0.066, 0.048) \\  \hline
 
  $\beta^{age,race}_1$ & Non-Hispanic Asian by age 0-5 years & -0.032 (-0.251, 0.154) \\ 
  $\beta^{age,race}_2$ & Non-Hispanic Asian by age 5-10 years & -0.074 (-0.243, 0.068) \\ 
  $\beta^{age,race}_3$ & Non-Hispanic Asian by age 15-18 years & 0.084 (-0.049, 0.230) \\  
  $\beta^{age,race}_4$ & Non-Hispanic Black by age 0-5 years & -0.118 (-0.272, 0.018) \\ 
  $\beta^{age,race}_5$ & Non-Hispanic Black by age 5-10 years & -0.157 (-0.255, -0.058) \\ 
  $\beta^{age,race}_6$ & Non-Hispanic Black by age 15-18 years & 0.063 (-0.013, 0.135) \\ 
  $\beta^{age,race}_7$ & Hispanic by age 0-5 years & 0.056 (-0.076, 0.183) \\ 
  $\beta^{age,race}_8$ & Hispanic by age 5-10 years & -0.091 (-0.176, -0.010) \\ 
  $\beta^{age,race}_9$ & Hispanic by age 15-18 years & 0.048 (-0.022, 0.119) \\ 
  $\sigma^{age,race}$ & Sd for age,race & 0.125 (0.054, 0.239) \\ \hline
  
  $\beta^{age,sex}_1$ & Female by age 0-5 years& -0.038 (-0.149, 0.074) \\ 
  $\beta^{age,sex}_2$ & Female by age 5-10 years & 0.102 (0.032, 0.173) \\ 
  $\beta^{age,sex}_3$ & Female by age 15-18 years & -0.204 (-0.262, -0.145) \\ 
   $\sigma^{age,sex}$ & Sd for age,sex & 0.258 (0.083, 0.831) \\  \hline
 
  $\beta^{race,sex}_1$ & Female by Non-Hispanic Asian & 0.160 (0.005, 0.320) \\   
  $\beta^{race,sex}_2$ & Female by Non-Hispanic Black & 0.144 (0.071, 0.211) \\ 
  $\beta^{race,sex}_3$ & Female by Hispanic & 0.120 (0.054, 0.185) \\ 
  $\sigma^{race,sex}$ & Sd for race,sex & 0.281 (0.076, 0.972) \\  \hline

   $\beta^{center}_1$ & PED & -0.187 (-0.499, 0.057) \\ 
  $\beta^{center}_2$ & UFL & 0.048 (-0.194, 0.289) \\ 
   $\sigma^{center}$ & Sd for center & 0.517 (0.027, 2.496) \\ \hline
 
  $\beta^{year}$ & Year 2022 & -0.172 (-0.246, -0.096) \\ \hline
  
  $\beta^{center,year}_1$ & PED by Year 2022 & -0.034 (-0.094, 0.026) \\ 
  $\beta^{center,year}_2$ & UFL by Year 2022 & 0.140 (0.063, 0.211) \\ 
$\sigma^{center,year}$ & Sd for center,year & 0.290 (0.055, 1.113) \\ \hline
  
  $\beta^{center,race}_1$ & PED by Non-Hispanic Asian & -0.265 (-0.425, -0.117) \\ 
  $\beta^{center,race}_2$ & UFL by Non-Hispanic Asian & 0.167 (-0.023, 0.454) \\ 
   $\beta^{center,race}_3$ & PED by Non-Hispanic Black & 0.063 (-0.019, 0.146) \\ 
  $\beta^{center,race}_4$ & UFL by Non-Hispanic Black & -0.120 (-0.215, -0.028) \\ 
  $\beta^{center,race}_5$ & PED by Hispanic & -0.055 (-0.145, 0.028) \\ 
  $\beta^{center,race}_6$ & UFL by Hispanic & 0.025 (-0.066, 0.112) \\ 
   $\sigma^{center,race}$ & Sd center,race & 0.197 (0.085, 0.421) \\ \hline
 
   $\beta^{center,age}_1$ & PED by age 0-5 years & 0.524 (0.397, 0.657) \\ 
  $\beta^{center,age}_2$ & UFL by age 0-5 years & -0.115 (-0.286, 0.055) \\ 
  $\beta^{center,age}_3$ & PED by age 5-10 years & 0.078 (-0.002, 0.158) \\ 
  $\beta^{center,age}_4$ & UFL by age 5-10 years & -0.083 (-0.179, 0.008) \\ 
  $\beta^{center,age}_5$ & PED by age 15-18 years & 0.163 (0.094, 0.230) \\ 
  $\beta^{center,age}_6$ & UFL by age 15-18 years & 0.017 (-0.063, 0.095) \\ 
  $\sigma^{center,age}$ & Sd for center,age & 0.305 (0.156, 0.626) \\  \hline
  
  $\beta^{center,sex}_1$ & PED by Female & -0.011 (-0.064, 0.035) \\ 
  $\beta^{center,sex}_2$ & UFL by Female & 0.022 (-0.028, 0.090) \\ 
  $\sigma^{center,sex}$ & Sd for center,sex & 0.100 (0.002, 0.588) \\ \hline
 
  $\beta^{year,race}_1$ & Year 2022 by Non-Hispanic Asian & 0.082 (-0.056, 0.231) \\ 
 $\beta^{year,race}_2$ & Year 2022 by Non-Hispanic Black & 0.135 (0.059, 0.211) \\ 
  $\beta^{year,race}_3$ & Year 2022 by Hispanic & 0.003 (-0.062, 0.070) \\ 
  $\sigma^{year,race}$ & Sd year,race & 0.184 (0.045, 0.603) \\ \hline

  $\beta^{year,age}_1$ & Year 2022 by age 0-5 years & 0.226 (0.113, 0.348) \\ 
  $\beta^{year,age}_2$ & Year 2022 by age 5-10 years & 0.077 (0.006, 0.146) \\ 
  $\beta^{year,age}_3$ & Year 2022 by age 15-18 years & -0.053 (-0.111, 0.007) \\ 
   $\sigma^{year,age}$ & Sd for year,age & 0.263 (0.076, 0.800) \\  \hline

  $\beta^{year,sex}$ & Year 2022 by Female & -0.015 (-0.066, 0.036) \\ \hline
  
  $\beta^{auxiliary}_1$ & \% Public insurance & 0.325 (0.215, 0.436) \\ 
  $\beta^{auxiliary}_2$ & \% Bachelor's degree or more & 0.101 (-0.033, 0.236) \\ 
  $\beta^{auxiliary}_3$ & \% Black & -0.201 (-0.299, -0.103) \\ 
  $\beta^{auxiliary}_4$ & \% Foreign born & -0.012 (-0.128, 0.104) \\ 
$\alpha$ & Intercept & -4.946 (-5.098, -4.768) \\ 
   \hline
\end{tabular}
\caption{Parameter estimates from Bayesian Hierarchical Analysis of Variance (BHAV). LUR refers to Lurie Children's Hospital; PED, PEDSNet; and UFL, OneFlorida$+$.} 
\label{t:parameter_estimates_bhav}
\end{table}

\begin{table}[htbp!]
\centering
\footnotesize
\begin{tabular}{ccc}
  \hline
 \multicolumn{1}{c}{Parameter} &  \multicolumn{1}{c}{Description} &  \multicolumn{1}{c}{Posterior Mean (95\% CrI)} \\ 
  \hline
  \multicolumn{3}{c}{\underline{Age group-specific county variance component}} \\
  \multicolumn{3}{c}{Age group-specific county Sd $\mathbf{\sigma}^{age}$}  \\
  $\sigma^{age}_1$ & 0-5 years  & 0.350 (0.278, 0.436) \\ 
  $\sigma^{age}_2$ & 5-10 years & 0.302 (0.245, 0.371) \\ 
  $\sigma^{age}_3$ & 10-15 years & 0.292 (0.235, 0.357) \\ 
  $\sigma^{age}_4$ & 15-18 years & 0.282 (0.227, 0.347) \\ 
   \multicolumn{3}{c}{Age group-specific mean $\mathbf{\beta}^{age}$} \\
  $\beta^{age}_1$ & 0-5 years & -1.476 (-1.587, -1.375) \\ 
  $\beta^{age}_2$ & 5-10 years & -0.146 (-0.217, -0.074) \\ 
  $\beta^{age}_3$ & 10-15 years & 0.637 (0.571, 0.705) \\ 
  $\beta^{age}_4$ & 15-18 years & 0.985 (0.917, 1.054) \\ 
  $\phi^{age}$ & Sd for $\mathbf{\beta}^{age}$ & 1.396 (0.651, 3.069) \\ 
  
   \multicolumn{3}{c}{\underline{Racial and ethnic group-specific county variance component}} \\
   \multicolumn{3}{c}{Racial and ethnic group-specific county Sd $\mathbf{\sigma}^{race}$}  \\
  $\sigma^{race}_1$ & Non-Hispanic Asian & 0.437 (0.278, 0.648) \\ 
  $\sigma^{race}_2$ & Non-Hispanic Black & 0.322 (0.245, 0.417) \\ 
  $\sigma^{race}_3$ & Hispanic & 0.556 (0.451, 0.686) \\ 
  $\sigma^{race}_4$ & Non-Hispanic White  & 0.415 (0.340, 0.503) \\  
   \multicolumn{3}{c}{Racial and ethnic group-specific mean $\mathbf{\beta}^{race}$}  \\
  $\beta^{race}_1$ & Non-Hispanic Asian & -0.389 (-0.533, -0.256) \\ 
  $\beta^{race}_2$ & Non-Hispanic Black & 0.103 (0.010, 0.193) \\ 
  $\beta^{race}_3$ & Hispanic & -0.117 (-0.237, -0.005) \\ 
  $\beta^{race}_4$ & Non-Hispanic White & 0.403 (0.316, 0.490) \\ 
  $\phi^{race}$ & Sd for $\mathbf{\beta}^{race}$ & 0.564 (0.201, 1.433) \\ \hline
 \multicolumn{3}{c}{\underline{Fixed effects}} \\
  $\beta^{demographic}_1$ & Female vs. male & 0.124 (0.080, 0.167) \\ 
  $\beta^{demographic}_2$ & Female by age & -0.100 (-0.184, -0.012) \\ 
  $\beta^{demographic}_3$ & Female by age, squared & -0.121 (-0.165, -0.078) \\ 
  $\beta^{demographic}_4$ & Female by age, cubed & 0.016 (-0.042, 0.070) \\ 
  $\beta^{demographic}_5$ & Non-Hispanic Black by age & 0.181 (0.060, 0.309) \\ 
  $\beta^{demographic}_{6}$ & Non-Hispanic Asian by age & 0.128 (-0.125, 0.374) \\ 
  $\beta^{demographic}_{7}$ & Non-Hispanic Asian by age, squared & 0.038 (-0.077, 0.165) \\ 
  $\beta^{demographic}_{8}$ & Non-Hispanic Asian by age, cubed & -0.037 (-0.185, 0.123) \\
  $\beta^{demographic}_9$ & Non-Hispanic Black by age, squared & -0.003 (-0.057, 0.052) \\ 
  $\beta^{demographic}_{10}$ & Non-Hispanic Black by age, cubed & -0.050 (-0.130, 0.028) \\ 
  $\beta^{demographic}_{11}$ & Hispanic by age & 0.089 (-0.034, 0.202) \\ 
  $\beta^{demographic}_{12}$ & Hispanic by age, squared & 0.044 (-0.010, 0.103) \\ 
  $\beta^{demographic}_{13}$ & Hispanic by age, cubed & -0.045 (-0.119, 0.031) \\ 
  $\beta^{center}_1$ & PED vs. LUR & -0.124 (-0.304, 0.052) \\ 
  $\beta^{center}_2$ & UFL vs. LUR & 0.059 (-0.111, 0.226) \\ 
  $\beta^{year}$ & 2022 vs. 2018 & -0.142 (-0.188, -0.096) \\ 
  $\beta^{center,year}_1$ & PED by 2022 & -0.020 (-0.074, 0.038) \\ 
  $\beta^{center,year}_2$ & UFL by 2022 & 0.145 (0.068, 0.220) \\ 
  $\beta^{auxiliary}_1$ & \% Public insurance & 0.283 (0.211, 0.354) \\ 
  $\beta^{auxiliary}_2$ & \% Bachelor's degree or more & 0.096 (0.008, 0.186) \\ 
  $\beta^{auxiliary}_3$ & \% Black & -0.248 (-0.308, -0.188) \\ 
  $\beta^{auxiliary}_4$ & \% Foreign born & -0.054 (-0.121, 0.015) \\ 
  $\alpha$ & Intercept & -6.048 (-6.172, -5.911) \\  \hline
   \hline
\end{tabular}
\caption{Parameter estimates from Partitioning Unobserved Geographic Variations (PUGV). LUR refers to Lurie Children's Hospital; PED, PEDSNet; and UFL, OneFlorida$+$.} 
\label{t:parameter_estimates_pugv}
\end{table}

\begin{table}[htbp!]
\centering
\begin{tabular}{ccc}
  \hline
\multicolumn{1}{c}{Parameter} &  \multicolumn{1}{c}{Description} &  \multicolumn{1}{c}{Posterior Mean (95\% CrI)} \\ 
  \hline
  $\sigma^{age}$ & Sd for age & 1.686 (0.671, 4.375) \\ 
  $\sigma^{county}$ & Sd for county & 0.538 (0.461, 0.625) \\ 
  $\sigma^{race}$ & Sd for race and ethnicity & 0.761 (0.249, 2.416) \\ \hline
  $\beta^{demographic}_1$ & Female vs. male & 0.126 (0.082, 0.168) \\ 
  $\beta^{demographic}_2$ & Female by age & -0.099 (-0.185, -0.015) \\ 
  $\beta^{demographic}_3$ & Female by age, squared & -0.121 (-0.163, -0.082) \\ 
  $\beta^{demographic}_4$ & Female by age, cubed & 0.014 (-0.041, 0.068) \\ 
  $\beta^{demographic}_{5}$ & Non-Hispanic Asian by age & 0.117 (-0.175, 0.396) \\ 
  $\beta^{demographic}_{6}$ & Non-Hispanic Asian by age, squared & 0.047 (-0.061, 0.159) \\ 
  $\beta^{demographic}_{7}$ & Non-Hispanic Asian by age, cubed & -0.014 (-0.193, 0.168) \\ 
 $\beta^{demographic}_8$ & Non-Hispanic Black by age & 0.234 (0.114, 0.349) \\ 
  $\beta^{demographic}_9$ & Non-Hispanic Black by age, squared & 0.009 (-0.050, 0.071) \\ 
  $\beta^{demographic}_{10}$ & Non-Hispanic Black by age, cubed & -0.077 (-0.149, -0.002) \\ 
  $\beta^{demographic}_{11}$ & Hispanic by age & 0.163 (0.059, 0.276) \\ 
  $\beta^{demographic}_{12}$ & Hispanic by age, squared & 0.015 (-0.035, 0.067) \\ 
  $\beta^{demographic}_{13}$ & Hispanic by age, cubed & -0.073 (-0.142, -0.006) \\ 
  $\beta^{center}_1$ & PED vs. LUR & -0.176 (-0.478, 0.131) \\ 
  $\beta^{center}_2$ & UFL vs. LUR & 0.016 (-0.256, 0.290) \\ 
  $\beta^{year}$ & 2022 vs. 2018 & -0.152 (-0.211, -0.093) \\ 
  $\beta^{center,year}_1$ & PED by 2022 & -0.029 (-0.090, 0.028) \\ 
  $\beta^{center,year}_2$ & UFL by 2022 & 0.153 (0.078, 0.230) \\ 
   $\beta^{auxiliary}_1$ & \% Public insurance & 0.337 (0.224, 0.448) \\ 
  $\beta^{auxiliary}_2$ & \% Bachelor's degree or more & 0.120 (-0.021, 0.256) \\ 
  $\beta^{auxiliary}_3$ & \% Black & -0.224 (-0.325, -0.123) \\ 
  $\beta^{auxiliary}_4$ & \% Foreign born & -0.028 (-0.149, 0.090) \\ 
  $\alpha$ & Intercept & -6.050 (-6.242, -5.864) \\ 
     \hline
\end{tabular}
\caption{Parameter estimates from Multilevel Regression and Poststratification (MRP). LUR refers to Lurie Children's Hospital; PED, PEDSNet; and UFL, OneFlorida$+$.} 
\label{t:parameter_estimates_MRP}
\end{table}

\begin{table}[htbp!]
\centering
\begin{tabular}{llccc}
  \hline
 &  & \multicolumn{1}{c}{BHAV} & \multicolumn{1}{c}{PUGV} & \multicolumn{1}{c}{MRP}\\ 
  \hline
\multirow{3}{*}{WAIC} & -2 $\times$ lppd & 16961 & 16197 & 17186 \\ 
  & $p_{\text{WAIC2}}$ & 382 & 1040 & 344 \\ 
  & WAIC & 17343 & 17238 & 17530 \\ 
  \multirow{3}{*}{DIC} & $\hat{D}$ & 16982 & 16162 & 17210 \\ 
   & $p_D$ & 159 & 498 & 139 \\ 
   & DIC & 17302 & 17156 & 17488 \\ 
   \hline
\end{tabular}
\caption{Information criteria. BHAV, PUGV, and MRP refer to Bayesian Hierarchical Analysis of Variance, Partitioning Unobserved Geographic Variations, and Multilevel Regression and Poststratification, respectively.} 
\label{t:model_comparison}
\end{table}

\begin{figure}[htbp!]
\centering
\includegraphics[scale = 0.5]{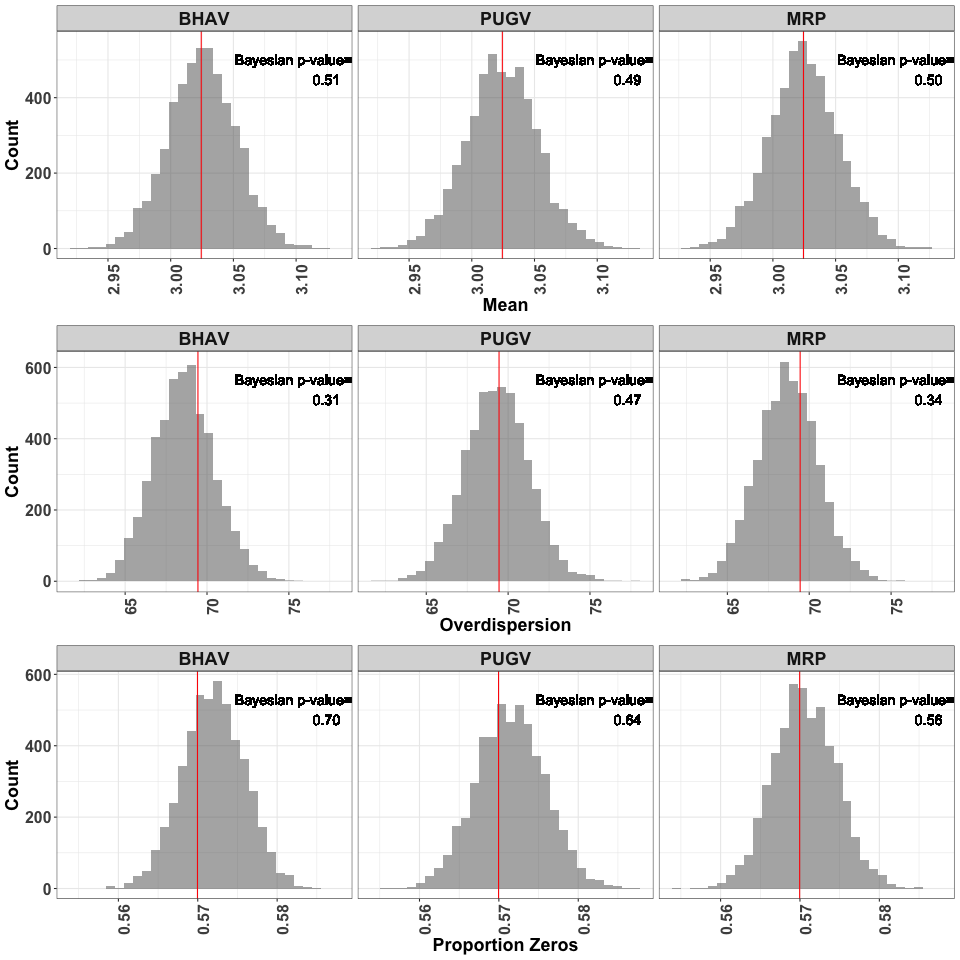}
\caption{Posterior predictive checks for BHAV, PUGV, and MRP using discrepancy measures including the mean T1D count, a measure of overdispersion equal to the variance minus the mean, and the proportion of zero counts. BHAV, PUGV, and MRP refer to Bayesian Hierarchical Analysis of Variance, Partitioning Unobserved Geographic Variations, and Multilevel Regression and Poststratification, respectively. The vertical red bar shows the discrepancy measure computed in the observed data. Each plot is annotated with the Bayesian posterior predictive p-value.}
\label{f:ppc__combined_T1DM_a}
\end{figure}

\clearpage

\begin{figure}[htbp!]
\centering
\includegraphics[scale = 0.5]{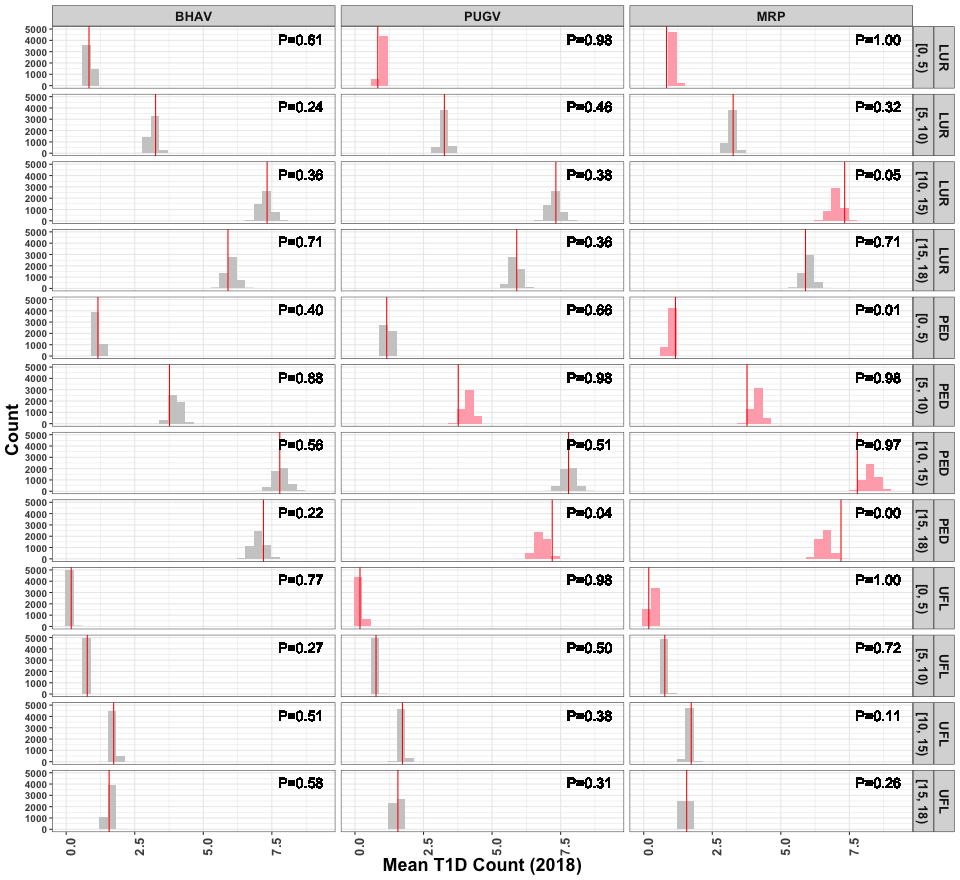}
\caption{Posterior predictive checks for BHAV, PUGV, and MRP using the mean T1D count in 2018 by health system center and age group (years). BHAV, PUGV, and MRP refer to Bayesian Hierarchical Analysis of Variance, Partitioning Unobserved Geographic Variations, and Multilevel Regression and Poststratification, respectively. LUR refers to Lurie Children's Hospital; PED, PEDSNet; and UFL, OneFlorida$+$. The vertical red bar shows the mean count computed in the observed data. Each plot is annotated with the Bayesian posterior predictive p-value. Pink shading indicates a Bayesian posterior predictive p-value outside the acceptable 0.05 to 0.95 range.}
\label{f:Fig_ppc_ageindex_yearsite_name1_mean_T1DM2018_a}
\end{figure}

\begin{figure}[htbp!]
\centering
\includegraphics[scale = 0.5]{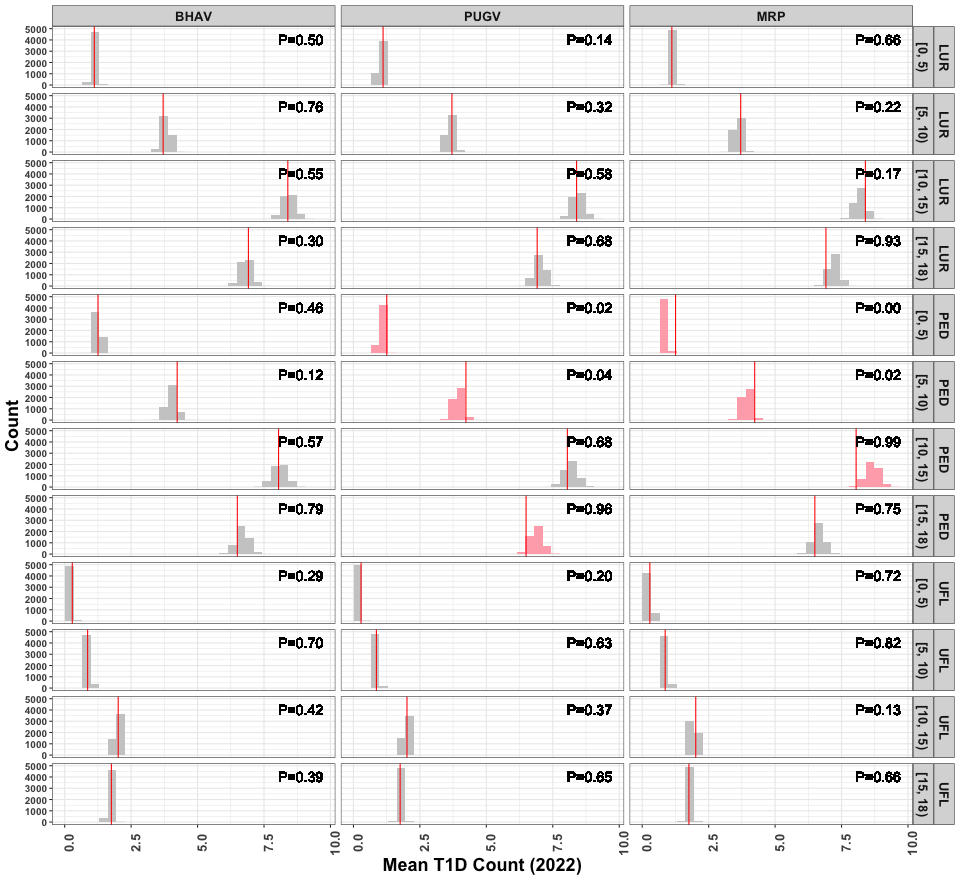}
\caption{Posterior predictive checks for BHAV, PUGV, and MRP using the mean T1D count in 2022 by health system center and age group (years). BHAV, PUGV, and MRP refer to Bayesian Hierarchical Analysis of Variance, Partitioning Unobserved Geographic Variations, and Multilevel Regression and Poststratification, respectively. LUR refers to Lurie Children's Hospital; PED, PEDSNet; and UFL, OneFlorida$+$. The vertical red bar shows the mean count computed in the observed data. Each plot is annotated with the Bayesian posterior predictive p-value. Pink shading indicates a Bayesian posterior predictive p-value outside the acceptable 0.05 to 0.95 range.}
\label{f:Fig_ppc_ageindex_yearsite_name1_mean_T1DM2022_a}
\end{figure}

\clearpage

\begin{figure}[htbp!]
\centering
\includegraphics[scale = 0.5]{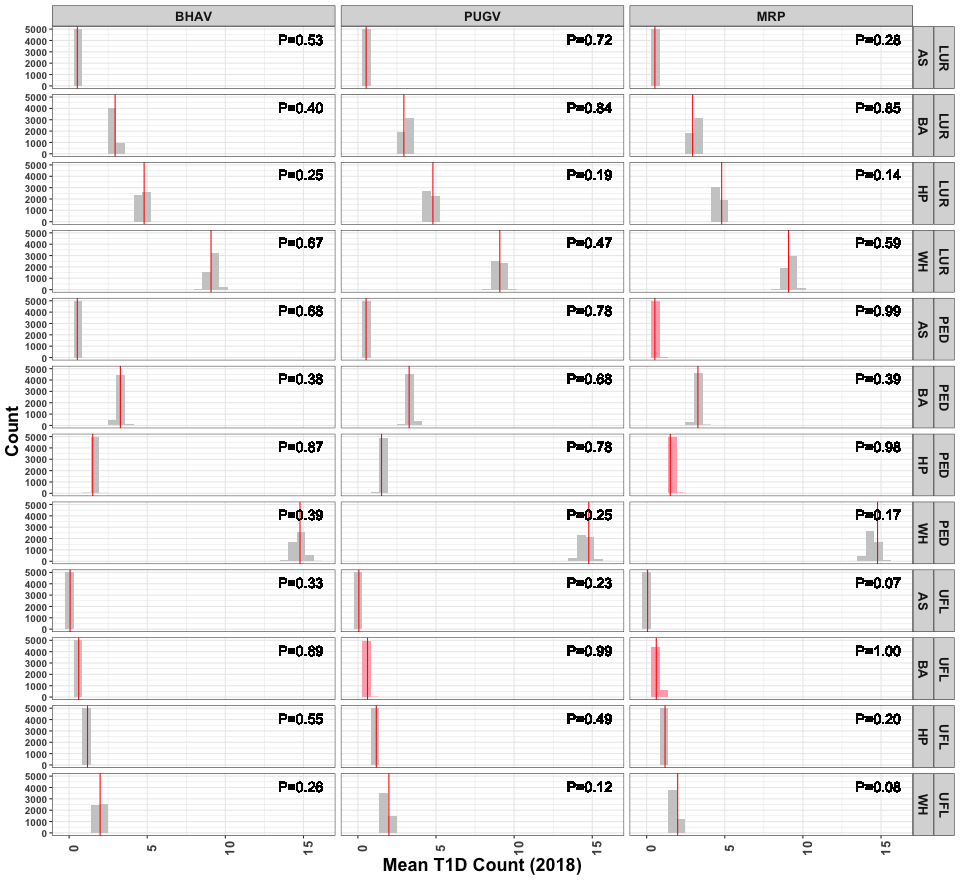}
\caption{Posterior predictive checks for BHAV, PUGV, and MRP using the mean T1D count in 2018 by health system center and race and ethnicity. BHAV, PUGV, and MRP refer to Bayesian Hierarchical Analysis of Variance, Partitioning Unobserved Geographic Variations, and Multilevel Regression and Poststratification, respectively. LUR refers to Lurie Children's Hospital; PED, PEDSNet; and UFL, OneFlorida$+$. AS, BA, HP, and WH refer to non-Hispanic Asian, non-Hispanic Black, Hispanic, and non-Hispanic White, respectively. The vertical red bar shows the mean count computed in the observed data. Each plot is annotated with the Bayesian posterior predictive p-value. Pink shading indicates a Bayesian posterior predictive p-value outside the acceptable 0.05 to 0.95 range.}
\label{f:Fig_ppc_raceindex_yearsite_name1_mean_T1DM2018_a}
\end{figure}

\begin{figure}[htbp!]
\centering
\includegraphics[scale = 0.5]{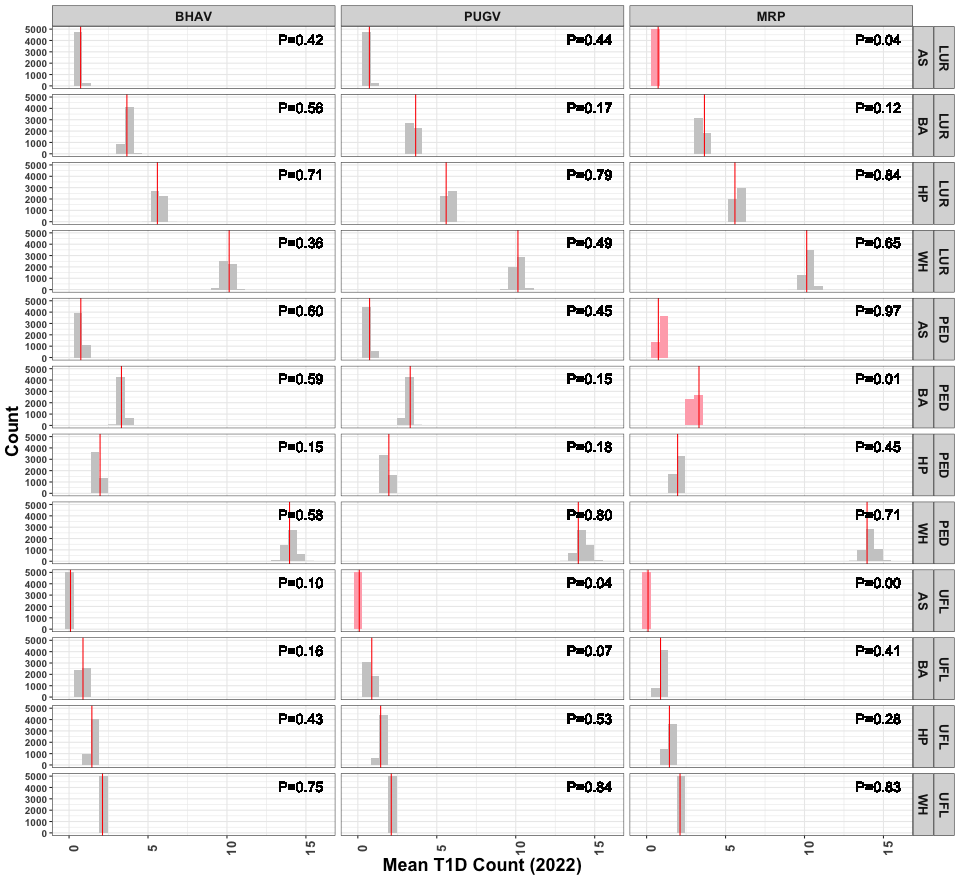}
\caption{Posterior predictive checks for BHAV, PUGV, and MRP using the mean T1D count in 2022 by health system center and race and ethnicity. BHAV, PUGV, and MRP refer to Bayesian Hierarchical Analysis of Variance, Partitioning Unobserved Geographic Variations, and Multilevel Regression and Poststratification, respectively. LUR refers to Lurie Children's Hospital; PED, PEDSNet; and UFL, OneFlorida$+$. AS, BA, HP, and WH refer to non-Hispanic Asian, non-Hispanic Black, Hispanic, and non-Hispanic White, respectively. The vertical red bar shows the mean count computed in the observed data. Each plot is annotated with the Bayesian posterior predictive p-value. Pink shading indicates a Bayesian posterior predictive p-value outside the acceptable 0.05 to 0.95 range.}
\label{f:Fig_ppc_raceindex_yearsite_name1_mean_T1DM2022_a}
\end{figure}

\clearpage

\begin{figure}[htbp!]
\centering
\includegraphics[scale = 0.5]{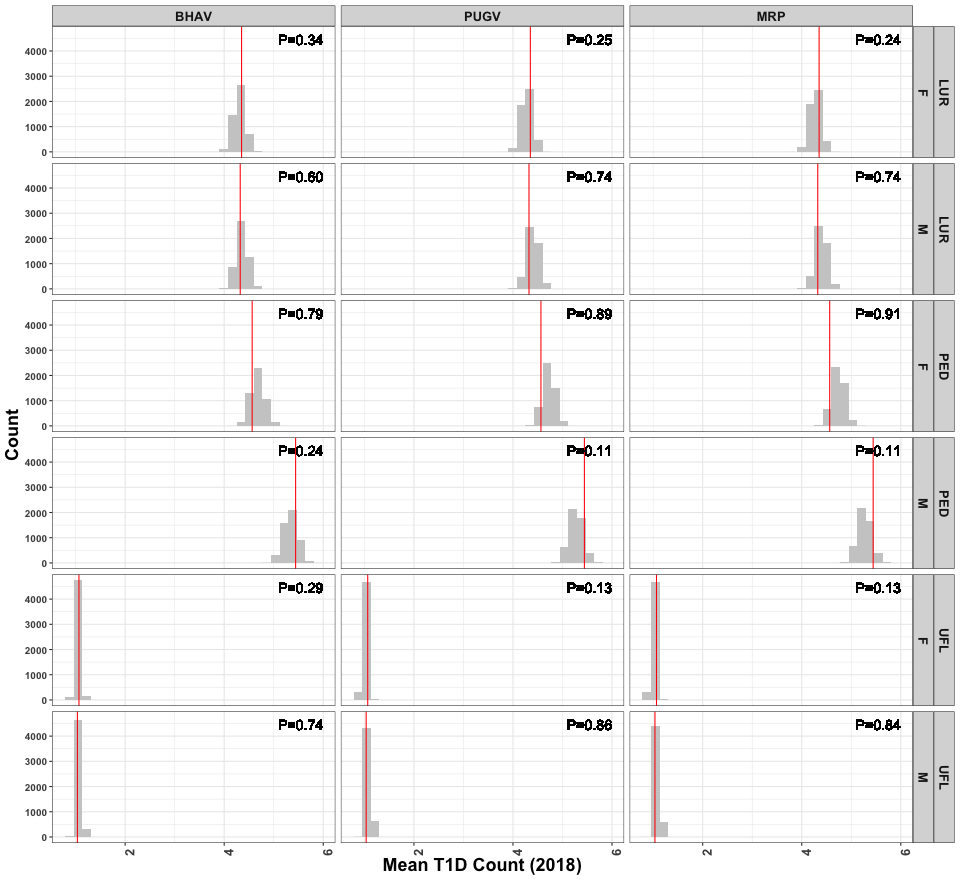}
\caption{Posterior predictive checks for BHAV, PUGV, and MRP using the mean T1D count in 2018 by health system center and sex. BHAV, PUGV, and MRP refer to Bayesian Hierarchical Analysis of Variance, Partitioning Unobserved Geographic Variations, and Multilevel Regression and Poststratification, respectively. LUR refers to Lurie Children's Hospital; PED, PEDSNet; and UFL, OneFlorida$+$. M and F indicate male and female, respectively. The vertical red bar shows the mean count computed in the observed data. Each plot is annotated with the Bayesian posterior predictive p-value. Pink shading indicates a Bayesian posterior predictive p-value outside the acceptable 0.05 to 0.95 range.}
\label{f:Fig_ppc_sexindex_yearsite_name1_mean_T1DM2018_a}
\end{figure}

\begin{figure}[htbp!]
\centering
\includegraphics[scale = 0.5]{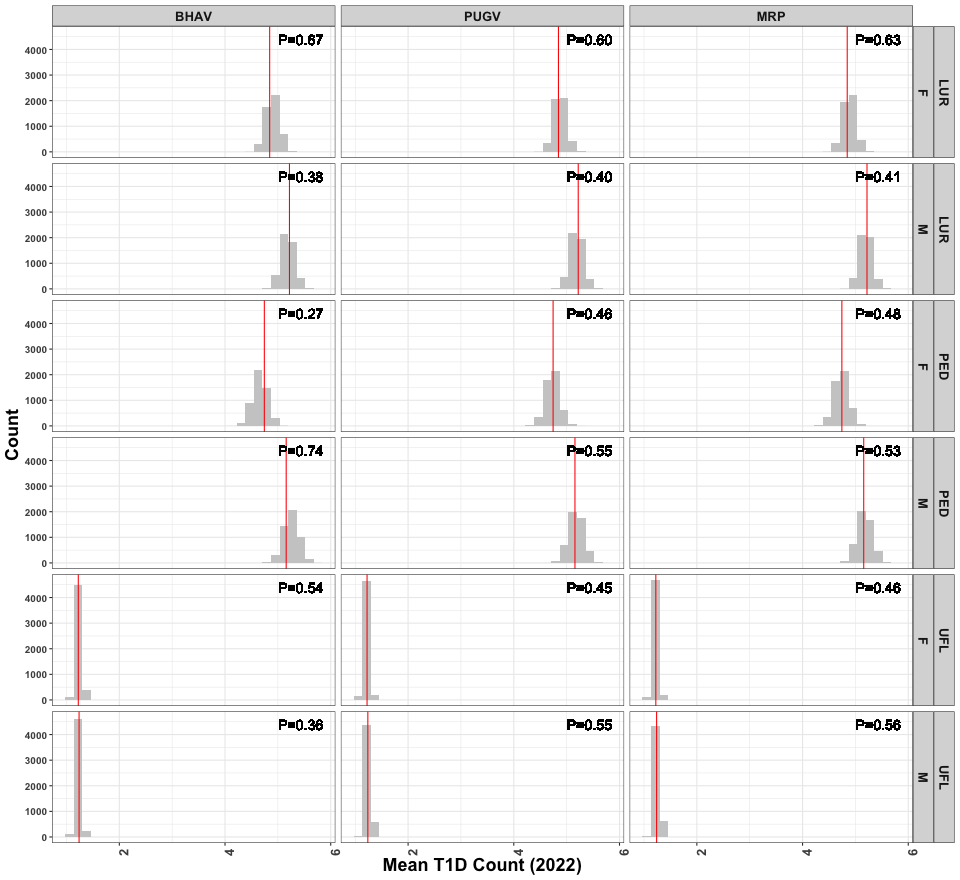}
\caption{Posterior predictive checks for BHAV, PUGV, and MRP using the mean T1D count in 2022 by health system center and sex. BHAV, PUGV, and MRP refer to Bayesian Hierarchical Analysis of Variance, Partitioning Unobserved Geographic Variations, and Multilevel Regression and Poststratification, respectively. LUR refers to Lurie Children's Hospital; PED, PEDSNet; and UFL, OneFlorida$+$. M and F indicate male and female, respectively. The vertical red bar shows the mean count computed in the observed data. Each plot is annotated with the Bayesian posterior predictive p-value. Pink shading indicates a Bayesian posterior predictive p-value outside the acceptable 0.05 to 0.95 range.}
\label{f:Fig_ppc_sexindex_yearsite_name1_mean_T1DM2022_a}
\end{figure}

\clearpage


\begin{table}[ht]
\footnotesize
\centering
\begin{tabular}{llcccc}
  \hline
 & &   \multicolumn{2}{c}{\underline{Posterior mean}} &  \multicolumn{2}{c}{\underline{Posterior SD}} \\ 
\multicolumn{1}{c}{Center} & \multicolumn{1}{c}{Subgroup} & \multicolumn{1}{c}{BHAV}  & \multicolumn{1}{c}{PUGV} & \multicolumn{1}{c}{BHAV} & \multicolumn{1}{c}{PUGV} \\ \hline
\multirow{13}{*}{LUR} & Age, years: &  & &  &  \\ 
 & \hspace{0.1cm}[0, 5) & 11.8 & 10.8 & 35.8 & 39.5 \\ 
   & \hspace{0.1cm}[5, 10) & 3.0 & 3.8 & 31.4 & 35.8 \\ 
   & \hspace{0.1cm}[10, 15) & 3.5 & 4.0 & 12.1 & 19.8 \\ 
  & \hspace{0.1cm}[15, 18) & 2.5 & 3.2 & 12.1 & 15.7 \\ 
   & Race and ethnicity: &  & &  &  \\ 
   & \hspace{0.1cm}AS & 10.4 & 10.7 & 44.1 & 47.5 \\ 
   & \hspace{0.1cm}BA & 5.0 & 1.0 & 30.4 & 31.9 \\ 
   & \hspace{0.1cm}HP & 1.8 & 6.2 & 17.5 & 17.1 \\ 
   & \hspace{0.1cm}WH & 1.1 & 4.2 & 5.9 & 17.9 \\ 
   & Sex: &  & &  &  \\ 
   & \hspace{0.1cm}M & 0.4 & 0.7 & 4.1 & 2.2 \\ 
   & \hspace{0.1cm}F & 1.4 & 1.0 & 8.5 & 0.8 \\ \hline
 \multirow{13}{*}{PED}  & Age, years: &  & &  &  \\ 
   & \hspace{0.1cm}[0, 5) & 32.4 & 30.9 & 44.9 & 45.6 \\ 
   & \hspace{0.1cm}[5, 10) & 6.6 & 5.4 & 7.1 & 4.1 \\ 
   & \hspace{0.1cm}[10, 15) & 6.4 & 5.4 & 23.9 & 23.3 \\ 
   & \hspace{0.1cm}[15, 18) & 4.7 & 3.9 & 7.5 & 7.1 \\
    & Race and ethnicity: &  & &  &  \\ 
   & \hspace{0.1cm}AS & 15.6 & 17.7 & 18.9 & 9.5 \\ 
   & \hspace{0.1cm}BA & 6.3 & 8.0 & 2.0 & 8.5 \\ 
   & \hspace{0.1cm}HP & 4.4 & 4.0 & 31.6 & 47.2 \\ 
   & \hspace{0.1cm}WH & 1.7 & 1.1 & 29.9 & 30.9 \\ 
     & Sex: &  & &  &  \\ 
   & \hspace{0.1cm}M & 1.4 & 2.0 & 23.7 & 25.3 \\ 
   & \hspace{0.1cm}F & 1.4 & 1.2 & 19.7 & 22.7 \\ \hline
  \multirow{13}{*}{UFL}  & Age, years: &  & &  &  \\ 
  & \hspace{0.1cm}[0, 5) & 20.7 & 11.6 & 52.9 & 72.1 \\ 
   & \hspace{0.1cm}[5, 10) & 4.5 & 2.2 & 34.9 & 51.9 \\ 
  & \hspace{0.1cm}[10, 15) & 5.1 & 5.9 & 21.9 & 35.2 \\ 
   & \hspace{0.1cm}[15, 18) & 3.5 & 0.6 & 22.0 & 28.2 \\ 
    & Race and ethnicity: &  & &  &  \\ 
   & \hspace{0.1cm}AS & 34.9 & 21.6 & 245.0 & 199.5 \\ 
   & \hspace{0.1cm}BA & 9.5 & 10.1 & 28.2 & 44.3 \\ 
   & \hspace{0.1cm}HP & 4.0 & 2.0 & 34.6 & 31.2 \\ 
   & \hspace{0.1cm}WH & 2.2 & 5.2 & 8.4 & 19.3 \\ 
     & Sex: &  & &  &  \\ 
   & \hspace{0.1cm}M & 0.5 & 1.3 & 8.0 & 4.8 \\ 
   & \hspace{0.1cm}F & 1.6 & 1.8 & 14.9 & 6.6 \\ 
   \hline
\end{tabular}
\caption{Mean absolute relative difference (\%) across years in the posterior mean and posterior standard deviation (SD) for T1D prevalence among children and adolescents comparing BHAV and PUGV to MRP by demographic subgroup. BHAV, PUGV, and MRP refer to Bayesian Hierarchical Analysis of Variance, Partitioning Unobserved Geographic Variations, and Multilevel Regression and Poststratification, respectively. LUR refers to Lurie Children's Hospital; PED, PEDSNet; and UFL, OneFlorida$+$.} 
\end{table}


\clearpage

\begin{figure}[htbp!]
\centering
\includegraphics[scale = 0.5]{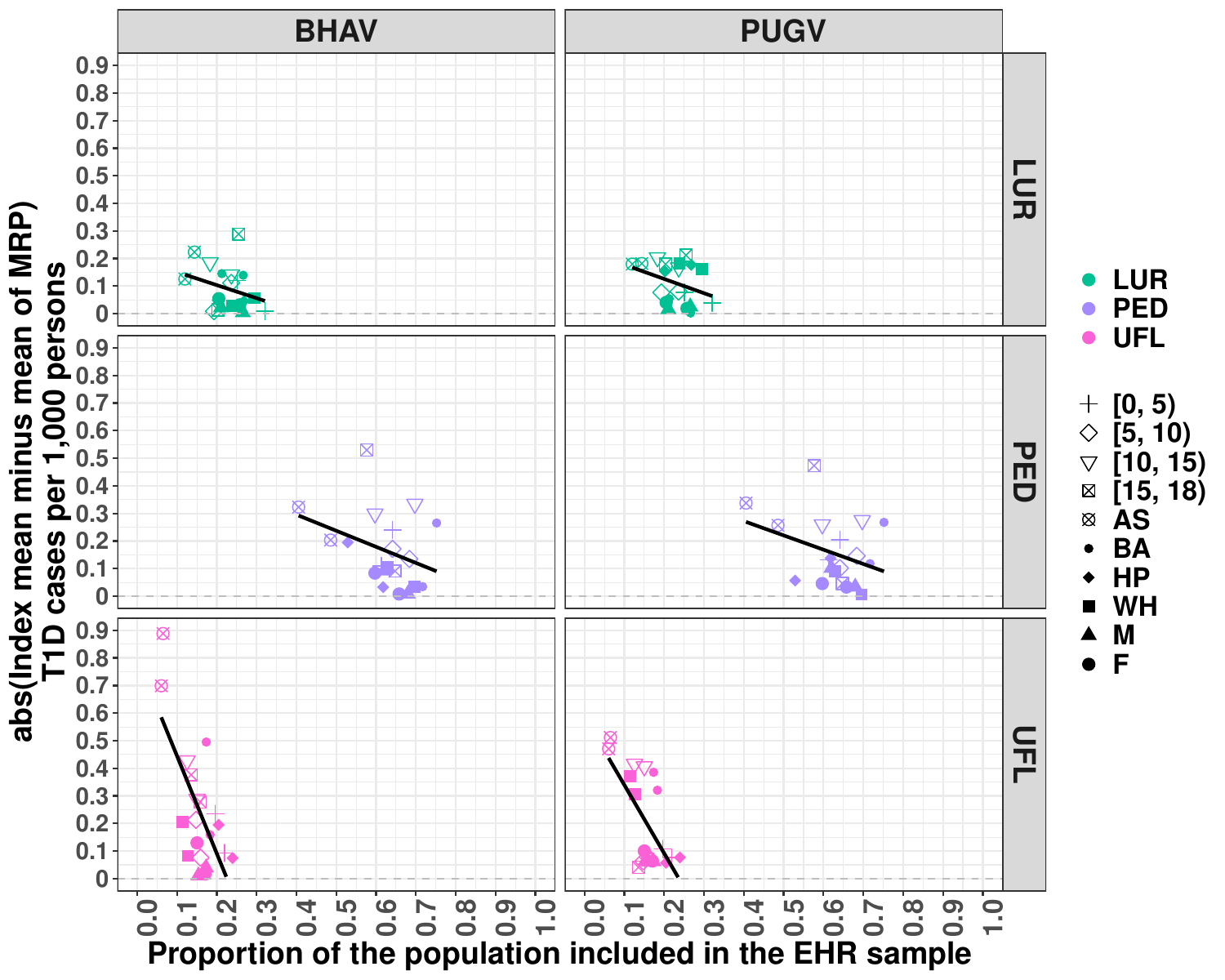}
\caption{Absolute difference in the T1D cases per 1,000 persons between the index posterior mean (BHAV or PUGV) and MRP for demographic subgroups in 2018 and 2022 according to the proportion of the population included in the EHR sample, stratified by health system center. BHAV, PUGV, and MRP refer to Bayesian Hierarchical Analysis of Variance, Partitioning Unobserved Geographic Variations, and Multilevel Regression and Poststratification, respectively. LUR refers to Lurie Children's Hospital; PED, PEDSNet; and UFL, OneFlorida$+$.  AS, BA, HP, and WH refer to non-Hispanic Asian, non-Hispanic Black, Hispanic, and non-Hispanic White, respectively. M and F indicate male and female, respectively.}
\label{f:Fig_superpopulation_minus_MRP_proppop_site_name1_T1DM_a}
\end{figure}


\begin{figure}[htbp!]
\centering
\includegraphics[scale = 0.5]{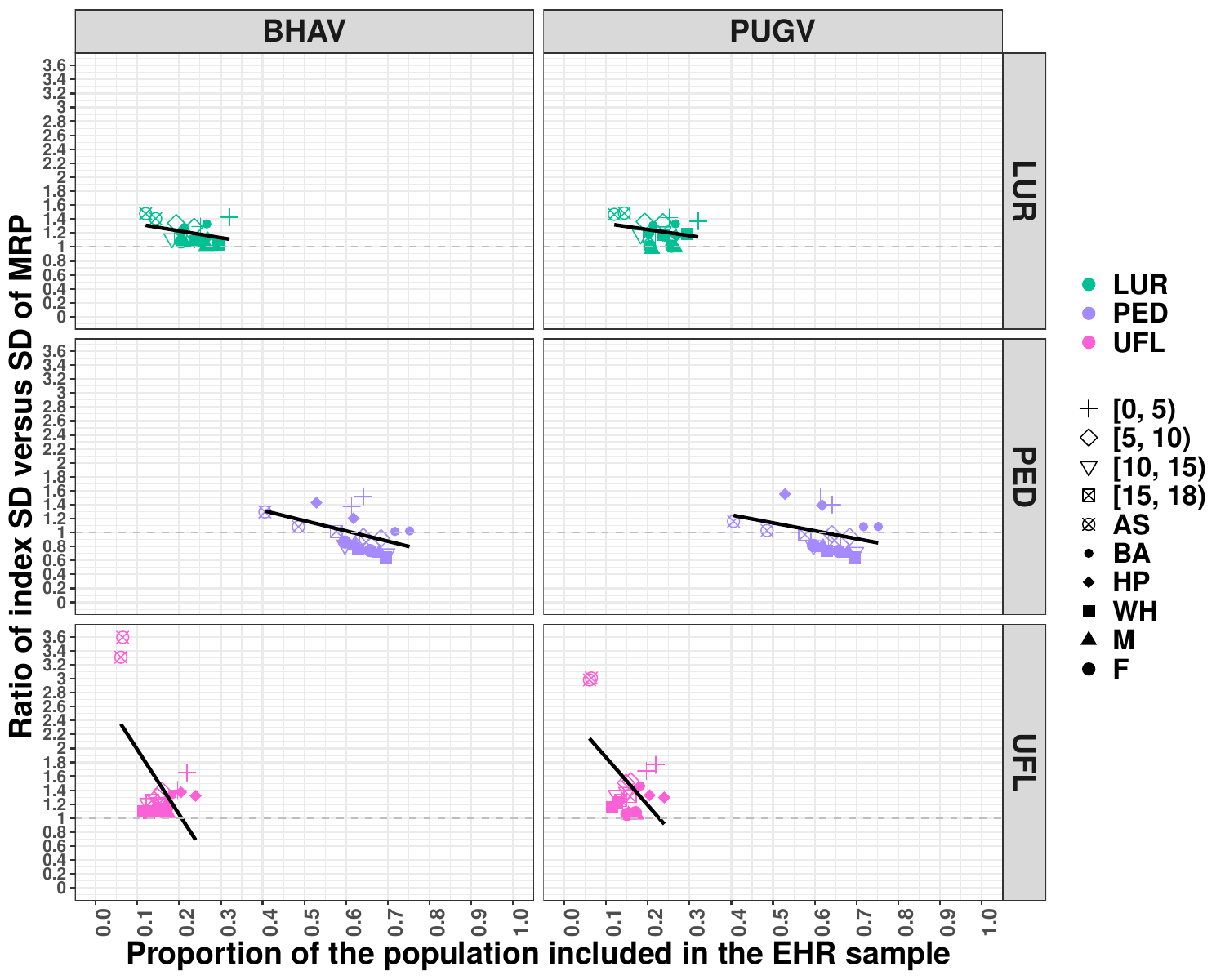}
\caption{Ratio of the index posterior standard deviation (SD) (BHAV or PUGV) of T1D cases per 1,000 persons and the posterior SD for MRP for demographic subgroups in 2018 and 2022 according to the proportion of the population included in the EHR sample, stratified by health system center. AS, BA, HP, and WH refer to non-Hispanic Asian, non-Hispanic Black, Hispanic, and non-Hispanic White, respectively. M and F indicate male and female, respectively.}
\label{f:Fig_superpopulation_sd_ratio_MRP_proppop_site_name1_T1DM_a}
\end{figure}

\clearpage

\begin{figure}[htbp!]
\centering
\includegraphics[scale = 0.5]{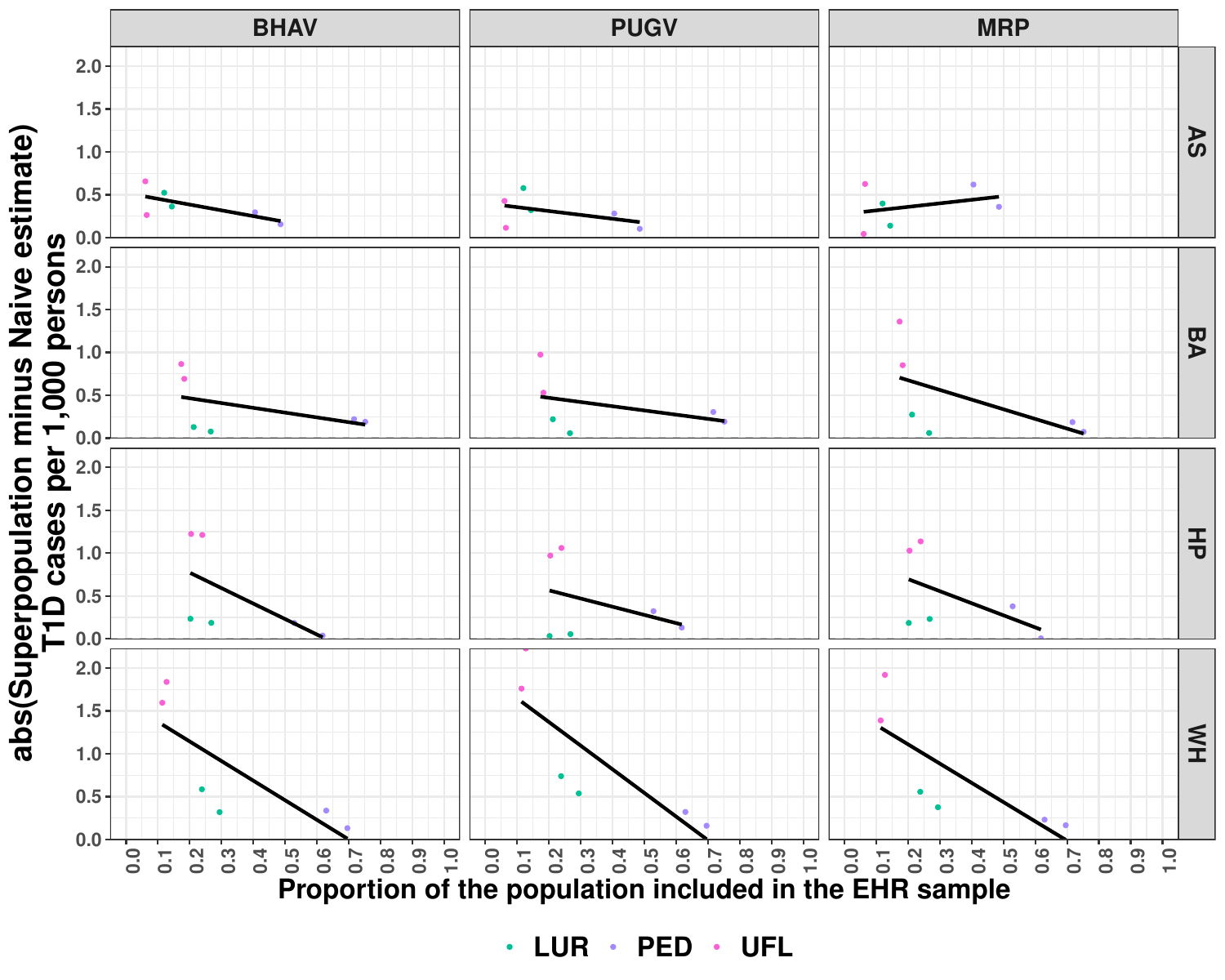}
\caption{Absolute difference in the T1D cases per 1,000 persons between the superpopulation-based and Na\"ive estimates for racial and ethnic subgroups in 2018 and 2022 by superpopulation model according to the proportion of the population included in the EHR sample, color coded by health system center. BHAV, PUGV, and MRP refer to Bayesian Hierarchical Analysis of Variance, Partitioning Unobserved Geographic Variations, and Multilevel Regression and Poststratification, respectively. LUR refers to Lurie Children's Hospital; PED, PEDSNet; and UFL, OneFlorida$+$. AS, BA, HP, and WH refer to non-Hispanic Asian, non-Hispanic Black, Hispanic, and non-Hispanic White, respectively.}
\label{f:Fig_superpopulation_minus_naive_proppop_race_T1DM_a}
\end{figure}

\begin{figure}[htbp!]
\centering
\includegraphics[scale = 0.5]{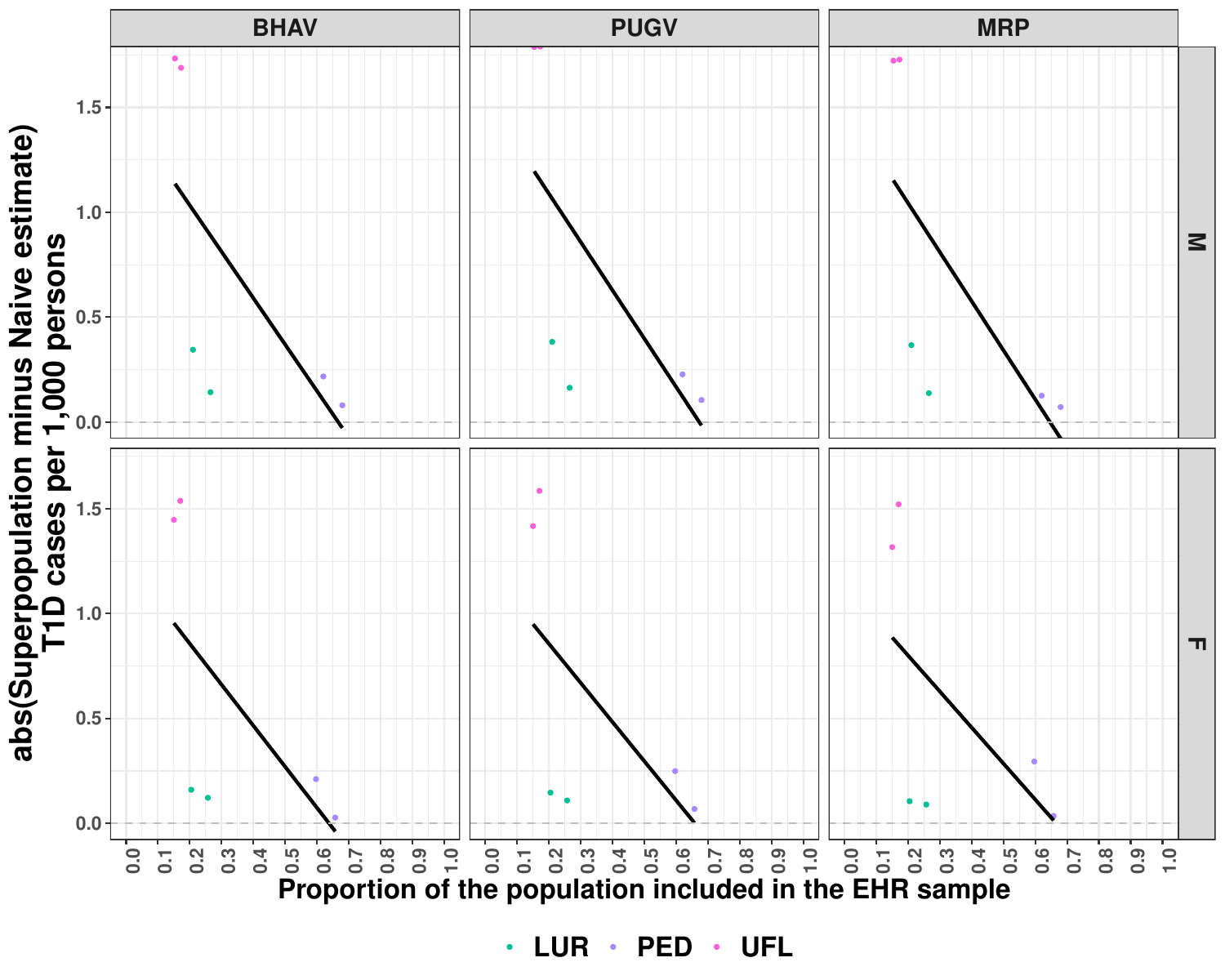}
\caption{Absolute difference in the T1D cases per 1,000 persons between the superpopulation-based and Na\"ive estimates for sex subgroups in 2018 and 2022 by superpopulation model according to the proportion of the population included in the EHR sample, color coded by health system center. BHAV, PUGV, and MRP refer to Bayesian Hierarchical Analysis of Variance, Partitioning Unobserved Geographic Variations, and Multilevel Regression and Poststratification, respectively. LUR refers to Lurie Children's Hospital; PED, PEDSNet; and UFL, OneFlorida$+$. M and F indicate male and female, respectively.}
\label{f:Fig_superpopulation_minus_naive_proppop_sex_T1DM_a}
\end{figure}

\clearpage

\begin{landscape}

\begin{table}[htbp!]
\tiny
\centering
\begin{tabular}{lcccccccccccccccc}
\hline
& \multicolumn{4}{c}{All Centers} & \multicolumn{4}{c}{LUR} & \multicolumn{4}{c}{PED} & \multicolumn{4}{c}{UFL} \\
No. counties & \multicolumn{4}{c}{($C=135$)} & \multicolumn{4}{c}{($C=36$)} & \multicolumn{4}{c}{($C=32$)} & \multicolumn{4}{c}{($C=67$)} \\ 
  \hline
 & \multicolumn{1}{c}{\underline{BHAV}} & \multicolumn{1}{c}{\underline{PUGV}} & \multicolumn{1}{c}{\underline{MRP}} & \multicolumn{1}{c}{\underline{Na\"ive}} & \multicolumn{1}{c}{\underline{BHAV}} & \multicolumn{1}{c}{\underline{PUGV}} & \multicolumn{1}{c}{\underline{MRP}} & \multicolumn{1}{c}{\underline{Na\"ive}} & \multicolumn{1}{c}{\underline{BHAV}} & \multicolumn{1}{c}{\underline{PUGV}} & \multicolumn{1}{c}{\underline{MRP}} & \multicolumn{1}{c}{\underline{Na\"ive}}& \multicolumn{1}{c}{\underline{BHAV}} & \multicolumn{1}{c}{\underline{PUGV}} & \multicolumn{1}{c}{\underline{MRP}} & \multicolumn{1}{c}{\underline{Na\"ive}} \\ 
  Posterior mean &  &  &  &  &  &  &  &  &  &  &  &  &  &  &  &  \\ 
  \hspace{0.05in}Minimum & 0.68 & 0.95 & 0.76 & 0.00 & 0.68 & 0.95 & 0.76 & 0.63 & 1.74 & 1.87 & 1.72 & 1.78 & 1.29 & 1.42 & 1.34 & 0.00 \\ 
  \hspace{0.05in}10\% & 1.94 & 2.05 & 2.00 & 1.15 & 1.39 & 1.60 & 1.59 & 1.24 & 2.34 & 2.33 & 2.37 & 2.23 & 2.07 & 2.16 & 2.12 & 0.00 \\ 
  \hspace{0.05in}25\% & 2.65 & 2.70 & 2.67 & 2.13 & 2.06 & 2.13 & 2.08 & 2.11 & 3.12 & 3.13 & 3.08 & 3.14 & 2.88 & 3.10 & 2.90 & 1.78 \\ 
  \hspace{0.05in}50\% & 4.03 & 4.17 & 3.97 & 3.56 & 2.84 & 2.82 & 2.87 & 2.85 & 4.11 & 4.14 & 4.13 & 3.97 & 4.55 & 4.67 & 4.43 & 3.56 \\ 
  \hspace{0.05in}75\% & 5.53 & 5.72 & 5.60 & 5.07 & 4.42 & 4.27 & 4.41 & 4.20 & 4.68 & 4.64 & 4.61 & 4.32 & 6.86 & 6.97 & 6.63 & 6.38 \\ 
  \hspace{0.05in}90\% & 8.44 & 8.28 & 8.39 & 8.56 & 6.04 & 6.16 & 5.88 & 6.80 & 5.79 & 5.84 & 5.46 & 5.68 & 10.87 & 10.72 & 10.43 & 14.83 \\ 
  \hspace{0.05in}Maximum & 20.31 & 17.64 & 19.29 & 35.71 & 11.99 & 11.31 & 11.74 & 12.05 & 7.47 & 7.44 & 7.74 & 7.43 & 20.31 & 17.64 & 19.29 & 35.71 \\ 
  Posterior SD &  &  &  &  &  &  &  &  &  &  &  &  &  &  &  &  \\ 
   \hspace{0.05in}Minimum & 0.04 & 0.04 & 0.07 & -0.00 & 0.08 & 0.08 & 0.09 & 0.10 & 0.04 & 0.04 & 0.07 & 0.09 & 0.08 & 0.08 & 0.08 & -0.00 \\ 
   \hspace{0.05in}10\% & 0.11 & 0.11 & 0.14 & 0.14 & 0.10 & 0.09 & 0.11 & 0.15 & 0.06 & 0.07 & 0.12 & 0.17 & 0.28 & 0.28 & 0.27 & -0.00 \\ 
   \hspace{0.05in}25\% & 0.19 & 0.19 & 0.21 & 0.26 & 0.14 & 0.15 & 0.15 & 0.20 & 0.15 & 0.15 & 0.20 & 0.27 & 0.60 & 0.61 & 0.58 & 0.42 \\ 
   \hspace{0.05in}50\% & 0.45 & 0.44 & 0.49 & 0.63 & 0.20 & 0.20 & 0.21 & 0.30 & 0.25 & 0.25 & 0.41 & 0.57 & 0.98 & 0.98 & 0.99 & 1.25 \\ 
   \hspace{0.05in}75\% & 1.01 & 1.00 & 1.02 & 1.40 & 0.31 & 0.34 & 0.31 & 0.45 & 0.49 & 0.49 & 0.57 & 0.76 & 1.70 & 1.52 & 1.57 & 2.11 \\ 
   \hspace{0.05in}90\% & 1.82 & 1.80 & 1.98 & 2.57 & 0.43 & 0.42 & 0.44 & 0.87 & 0.60 & 0.61 & 0.72 & 0.96 & 2.83 & 2.57 & 2.82 & 4.31 \\ 
   \hspace{0.05in}Maximum & 6.50 & 3.61 & 6.39 & 35.07 & 1.65 & 1.39 & 1.53 & 2.30 & 0.92 & 0.86 & 2.07 & 2.47 & 6.50 & 3.61 & 6.39 & 35.07 \\ 
   \hline
\end{tabular}
\caption{Posterior summaries of T1D cases per 1,000 persons at the county-level among children and adolescents in 2018, overall and by health system center. BHAV, PUGV, and MRP refer to Bayesian Hierarchical Analysis of Variance, Partitioning Unobserved Geographic Variations, and Multilevel Regression and Poststratification, respectively. LUR refers to Lurie Children's Hospital; PED, PEDSNet; and UFL, OneFlorida$+$. The Na\"ive method uses the sample mean and its standard error. For the  Na\"ive method, the summaries of the posterior mean and posterior standard deviation (SD) for all centers and for UFL have zeros due to eight counties in UFL with no recorded T1D cases in the EHRs despite the size of the utilization denominator ranging from 66 to 696 patients.} 
\end{table}

\end{landscape}

\clearpage

\begin{table}[ht]
\centering
\begin{tabular}{lcccccccc}
  \hline
& \multicolumn{2}{c}{All Centers} & \multicolumn{2}{c}{LUR} & \multicolumn{2}{c}{PED} & \multicolumn{2}{c}{UFL} \\
No. counties & \multicolumn{2}{c}{($C=135$)} & \multicolumn{2}{c}{($C=36$)} & \multicolumn{2}{c}{($C=32$)} & \multicolumn{2}{c}{($C=67$)} \\ 
  \hline
 & \multicolumn{1}{c}{BHAV}  & \multicolumn{1}{c}{PUGV} & \multicolumn{1}{c}{BHAV} & \multicolumn{1}{c}{PUGV} & \multicolumn{1}{c}{BHAV} & \multicolumn{1}{c}{PUGV}& \multicolumn{1}{c}{BHAV} & \multicolumn{1}{c}{PUGV} \\ \hline
  Posterior mean &  &  &  &  &  &  &  &  \\ 
   \hspace{0.05in}Minimum & 0.01 & 0.08 & 0.07 & 0.13 & 0.10 & 0.08 & 0.01 & 0.13 \\ 
   \hspace{0.05in}10\% & 0.59 & 0.75 & 0.28 & 0.52 & 0.78 & 0.42 & 0.64 & 1.35 \\ 
   \hspace{0.05in}25\% & 1.26 & 2.07 & 0.83 & 1.94 & 1.36 & 1.26 & 1.82 & 2.79 \\ 
   \hspace{0.05in}50\% & 3.30 & 4.89 & 2.09 & 3.74 & 3.77 & 3.94 & 3.94 & 6.08 \\ 
   \hspace{0.05in}75\% & 6.09 & 9.43 & 3.78 & 8.77 & 6.38 & 6.85 & 7.24 & 11.22 \\ 
   \hspace{0.05in}90\% & 10.37 & 17.70 & 4.90 & 14.01 & 9.68 & 11.83 & 11.94 & 19.97 \\ 
   \hspace{0.05in}Maximum & 403.63 & 403.13 & 22.39 & 41.89 & 403.63 & 403.13 & 26.47 & 65.19 \\ 
  Posterior SD &  &  &  &  &  &  &  &  \\ 
   \hspace{0.05in}Minimum & 0.04 & 0.24 & 0.11 & 0.24 & 4.07 & 3.95 & 0.04 & 0.29 \\ 
   \hspace{0.05in}10\% & 1.21 & 1.71 & 1.01 & 1.62 & 9.93 & 9.91 & 0.95 & 1.07 \\ 
   \hspace{0.05in}25\% & 3.35 & 4.03 & 2.22 & 4.12 & 16.67 & 14.78 & 2.46 & 2.57 \\ 
   \hspace{0.05in}50\% & 7.24 & 9.80 & 5.94 & 8.88 & 25.29 & 23.83 & 5.00 & 6.81 \\ 
   \hspace{0.05in}75\% & 17.04 & 20.95 & 11.20 & 14.32 & 41.92 & 39.03 & 9.17 & 13.26 \\ 
   \hspace{0.05in}90\% & 34.66 & 37.27 & 23.82 & 26.46 & 57.24 & 55.20 & 15.41 & 21.90 \\ 
   \hspace{0.05in}Maximum & 71.32 & 97.58 & 50.08 & 97.58 & 71.32 & 72.12 & 40.66 & 46.03 \\ 
   \hline
\end{tabular}
\caption{Mean absolute relative difference (\%) across years in the posterior mean and posterior standard deviation (SD) for T1D prevalence among children and adolescents comparing  BHAV and PUGV to MRP by county, overall and by health system center. BHAV, PUGV, and MRP refer to Bayesian Hierarchical Analysis of Variance, Partitioning Unobserved Geographic Variations, and Multilevel Regression and Poststratification, respectively. LUR refers to Lurie Children's Hospital; PED, PEDSNet; and UFL, OneFlorida$+$.} 
\end{table}


\clearpage

\section{Plasmode Simulations}

\begin{table}[htbp!]
\centering
\caption{Plasmode simulation results for estimation of T1D prevalence (T1D cases per 1,000 persons) by year and scenario under the \textbf{BHAV}, \textbf{PUGV}, \textbf{MRP}, and \textbf{Na\"ive} methods. BHAV, PUGV, and MRP refer to Bayesian Hierarchical Analysis of Variance, Partitioning Unobserved Geographic Variations, and Multilevel Regression and Poststratification, respectively. The method with the least bias is in bold. Coverage refers to the 95\% coverage probability. Length refers to the average length of the 95\% uncertainty interval.} 
\label{t:plasmode_simulation_year_scenario}
\begingroup\normalsize
\begin{tabular}{cccrrrrr}
  \hline
  \multicolumn{1}{c}{Scenario} & \multicolumn{1}{c}{Year} & \multicolumn{1}{c}{Truth} & \multicolumn{1}{c}{Method}& \multicolumn{1}{c}{Ave. Bias} & \multicolumn{1}{c}{MSE} & \multicolumn{1}{c}{Coverage} & \multicolumn{1}{c}{Ave. Length} \\
 \hline
S1 & 2018 & 3.727 & BHAV & \textbf{-0.041} & 0.009 & 0.899 & 0.306 \\ 
  S1 & 2018 & 3.727 & PUGV & -0.049 & 0.008 & 0.871 & 0.298 \\ 
  S1 & 2018 & 3.727 & MRP & -0.069 & 0.011 & 0.844 & 0.314 \\ 
  S1 & 2018 & 3.727 & Na\"ive & -0.618 & 0.384 & 0.000 & 0.194 \\ 
   \hline S1 & 2022 & 3.558 & BHAV & \textbf{-0.007} & 0.005 & 0.966 & 0.287 \\ 
  S1 & 2022 & 3.558 & PUGV & 0.008 & 0.005 & 0.973 & 0.283 \\ 
  S1 & 2022 & 3.558 & MRP & -0.008 & 0.005 & 0.959 & 0.304 \\ 
  S1 & 2022 & 3.558 & Na\"ive & -0.565 & 0.320 & 0.000 & 0.176 \\ 
   \hline S2 & 2018 & 3.681 & BHAV & 0.004 & 0.007 & 0.914 & 0.305 \\ 
  S2 & 2018 & 3.680 & PUGV & \textbf{-0.002} & 0.006 & 0.952 & 0.298 \\ 
  S2 & 2018 & 3.680 & MRP & -0.022 & 0.007 & 0.946 & 0.314 \\ 
  S2 & 2018 & 3.680 & Na\"ive & -0.572 & 0.329 & 0.000 & 0.194 \\ 
   \hline S2 & 2022 & 3.609 & BHAV & -0.057 & 0.008 & 0.900 & 0.286 \\ 
  S2 & 2022 & 3.609 & PUGV & \textbf{-0.044} & 0.006 & 0.918 & 0.283 \\ 
  S2 & 2022 & 3.610 & MRP & -0.059 & 0.009 & 0.898 & 0.304 \\ 
  S2 & 2022 & 3.610 & Na\"ive & -0.616 & 0.381 & 0.000 & 0.176 \\ 
   \hline \hline
\end{tabular}
\endgroup
\end{table}

\clearpage

\begin{table}[ht]
\centering
\caption{Plasmode simulation results for estimation of T1D prevalence (T1D cases per 1,000 persons) by year, age, and scenario under the \textbf{BHAV}, \textbf{PUGV}, \textbf{MRP}, and \textbf{Na\"ive} methods. BHAV, PUGV, and MRP refer to Bayesian Hierarchical Analysis of Variance, Partitioning Unobserved Geographic Variations, and Multilevel Regression and Poststratification, respectively. The method with the least bias is in bold. Coverage refers to the 95\% coverage probability. Length refers to the average length of the 95\% uncertainty interval.} 
\label{t:plasmode_simulation_year_age_S1}
\begingroup\tiny
\begin{tabular}{ccccrrrrr}
  \hline
  \multicolumn{1}{c}{Scenario} & \multicolumn{1}{c}{Year}  & \multicolumn{1}{c}{Age} & \multicolumn{1}{c}{Truth} & \multicolumn{1}{c}{Method}& \multicolumn{1}{c}{Ave. Bias} & \multicolumn{1}{c}{MSE} & \multicolumn{1}{c}{Coverage} & \multicolumn{1}{c}{Ave. Length} \\
 \hline
S1 & 2018 & [0, 5) & 0.613 & BHAV & -0.094 & 0.010 & 0.201 & 0.129 \\ 
  S1 & 2018 & [0, 5) & 0.614 & PUGV & -0.041 & 0.003 & 0.776 & 0.130 \\ 
  S1 & 2018 & [0, 5) & 0.613 & MRP & \textbf{0.010} & 0.001 & 0.871 & 0.121 \\ 
  S1 & 2018 & [0, 5) & 0.613 & Na\"ive & -0.113 & 0.014 & 0.088 & 0.138 \\ 
    \hline S1 & 2018 & [5, 10) & 2.539 & BHAV & -0.116 & 0.023 & 0.752 & 0.365 \\ 
  S1 & 2018 & [5, 10) & 2.539 & PUGV & -0.039 & 0.009 & 0.952 & 0.368 \\ 
  S1 & 2018 & [5, 10) & 2.538 & MRP & \textbf{-0.028} & 0.007 & 0.932 & 0.312 \\ 
  S1 & 2018 & [5, 10) & 2.538 & Na\"ive & -0.337 & 0.119 & 0.014 & 0.322 \\ 
    \hline S1 & 2018 & [10, 15) & 5.574 & BHAV & 0.064 & 0.035 & 0.899 & 0.626 \\ 
  S1 & 2018 & [10, 15) & 5.575 & PUGV & \textbf{0.042} & 0.029 & 0.925 & 0.619 \\ 
  S1 & 2018 & [10, 15) & 5.574 & MRP & -0.111 & 0.034 & 0.878 & 0.552 \\ 
  S1 & 2018 & [10, 15) & 5.574 & Na\"ive & -0.630 & 0.411 & 0.000 & 0.481 \\ 
    \hline S1 & 2018 & [15, 18) & 7.471 & BHAV & \textbf{-0.012} & 0.044 & 0.966 & 0.863 \\ 
  S1 & 2018 & [15, 18) & 7.470 & PUGV & -0.229 & 0.087 & 0.803 & 0.801 \\ 
  S1 & 2018 & [15, 18) & 7.471 & MRP & -0.188 & 0.074 & 0.844 & 0.756 \\ 
  S1 & 2018 & [15, 18) & 7.471 & Na\"ive & -0.816 & 0.687 & 0.000 & 0.693 \\ 
    \hline S1 & 2022 & [0, 5) & 0.616 & BHAV & \textbf{-0.015} & 0.002 & 0.913 & 0.132 \\ 
  S1 & 2022 & [0, 5) & 0.617 & PUGV & -0.029 & 0.002 & 0.871 & 0.125 \\ 
  S1 & 2022 & [0, 5) & 0.617 & MRP & -0.021 & 0.001 & 0.884 & 0.116 \\ 
  S1 & 2022 & [0, 5) & 0.617 & Na\"ive & -0.029 & 0.002 & 0.884 & 0.142 \\ 
    \hline S1 & 2022 & [5, 10) & 2.430 & BHAV & 0.023 & 0.008 & 0.973 & 0.349 \\ 
  S1 & 2022 & [5, 10) & 2.430 & PUGV & \textbf{-0.001} & 0.007 & 0.973 & 0.355 \\ 
  S1 & 2022 & [5, 10) & 2.430 & MRP & -0.027 & 0.005 & 0.966 & 0.298 \\ 
  S1 & 2022 & [5, 10) & 2.430 & Na\"ive & -0.242 & 0.064 & 0.109 & 0.295 \\ 
    \hline S1 & 2022 & [10, 15) & 5.253 & BHAV & 0.099 & 0.032 & 0.886 & 0.575 \\ 
  S1 & 2022 & [10, 15) & 5.252 & PUGV & 0.131 & 0.040 & 0.844 & 0.594 \\ 
  S1 & 2022 & [10, 15) & 5.254 & MRP & \textbf{-0.005} & 0.019 & 0.966 & 0.531 \\ 
  S1 & 2022 & [10, 15) & 5.254 & Na\"ive & -0.610 & 0.382 & 0.000 & 0.427 \\ 
    \hline S1 & 2022 & [15, 18) & 6.933 & BHAV & -0.209 & 0.077 & 0.812 & 0.754 \\ 
  S1 & 2022 & [15, 18) & 6.935 & PUGV & -0.122 & 0.043 & 0.905 & 0.759 \\ 
  S1 & 2022 & [15, 18) & 6.934 & MRP & \textbf{0.037} & 0.030 & 0.952 & 0.721 \\ 
  S1 & 2022 & [15, 18) & 6.934 & Na\"ive & -0.981 & 0.981 & 0.000 & 0.600 \\ \hline 
  S2 & 2018 & [0, 5) & 0.550 & BHAV & -0.031 & 0.002 & 0.843 & 0.129 \\ 
  S2 & 2018 & [0, 5) & 0.550 & PUGV & \textbf{0.022} & 0.002 & 0.878 & 0.129 \\ 
  S2 & 2018 & [0, 5) & 0.549 & MRP & 0.074 & 0.007 & 0.252 & 0.121 \\ 
  S2 & 2018 & [0, 5) & 0.549 & Na\"ive & -0.050 & 0.003 & 0.776 & 0.138 \\ 
    \hline S2 & 2018 & [5, 10) & 2.481 & BHAV & -0.061 & 0.013 & 0.893 & 0.364 \\ 
  S2 & 2018 & [5, 10) & 2.481 & PUGV & \textbf{0.019} & 0.008 & 0.966 & 0.368 \\ 
  S2 & 2018 & [5, 10) & 2.481 & MRP & 0.029 & 0.007 & 0.952 & 0.312 \\ 
  S2 & 2018 & [5, 10) & 2.481 & Na\"ive & -0.279 & 0.084 & 0.082 & 0.322 \\ 
    \hline S2 & 2018 & [10, 15) & 5.636 & BHAV & \textbf{-0.000} & 0.031 & 0.907 & 0.625 \\ 
  S2 & 2018 & [10, 15) & 5.634 & PUGV & -0.017 & 0.028 & 0.918 & 0.619 \\ 
  S2 & 2018 & [10, 15) & 5.634 & MRP & -0.171 & 0.051 & 0.755 & 0.552 \\ 
  S2 & 2018 & [10, 15) & 5.634 & Na\"ive & -0.690 & 0.490 & 0.000 & 0.481 \\ 
    \hline S2 & 2018 & [15, 18) & 7.292 & BHAV & 0.167 & 0.073 & 0.900 & 0.863 \\ 
  S2 & 2018 & [15, 18) & 7.290 & PUGV & -0.049 & 0.038 & 0.959 & 0.800 \\ 
  S2 & 2018 & [15, 18) & 7.292 & MRP & \textbf{-0.009} & 0.038 & 0.939 & 0.756 \\ 
  S2 & 2018 & [15, 18) & 7.292 & Na\"ive & -0.637 & 0.426 & 0.027 & 0.693 \\ 
    \hline S2 & 2022 & [0, 5) & 0.568 & BHAV & 0.035 & 0.002 & 0.829 & 0.132 \\ 
  S2 & 2022 & [0, 5) & 0.568 & PUGV & \textbf{0.018} & 0.001 & 0.932 & 0.125 \\ 
  S2 & 2022 & [0, 5) & 0.568 & MRP & 0.028 & 0.002 & 0.864 & 0.116 \\ 
  S2 & 2022 & [0, 5) & 0.568 & Na\"ive & 0.020 & 0.001 & 0.952 & 0.142 \\ 
    \hline S2 & 2022 & [5, 10) & 2.434 & BHAV & 0.022 & 0.008 & 0.971 & 0.350 \\ 
  S2 & 2022 & [5, 10) & 2.435 & PUGV & \textbf{-0.005} & 0.007 & 0.973 & 0.355 \\ 
  S2 & 2022 & [5, 10) & 2.434 & MRP & -0.031 & 0.006 & 0.952 & 0.298 \\ 
  S2 & 2022 & [5, 10) & 2.434 & Na\"ive & -0.247 & 0.066 & 0.102 & 0.295 \\ 
    \hline S2 & 2022 & [10, 15) & 5.465 & BHAV & -0.115 & 0.036 & 0.886 & 0.574 \\ 
  S2 & 2022 & [10, 15) & 5.463 & PUGV & \textbf{-0.080} & 0.030 & 0.918 & 0.596 \\ 
  S2 & 2022 & [10, 15) & 5.465 & MRP & -0.217 & 0.066 & 0.612 & 0.531 \\ 
  S2 & 2022 & [10, 15) & 5.465 & Na\"ive & -0.822 & 0.685 & 0.000 & 0.427 \\ 
    \hline S2 & 2022 & [15, 18) & 6.950 & BHAV & -0.223 & 0.085 & 0.786 & 0.754 \\ 
  S2 & 2022 & [15, 18) & 6.951 & PUGV & -0.139 & 0.048 & 0.891 & 0.757 \\ 
  S2 & 2022 & [15, 18) & 6.951 & MRP & \textbf{0.019} & 0.030 & 0.966 & 0.721 \\ 
  S2 & 2022 & [15, 18) & 6.951 & Na\"ive & -0.999 & 1.016 & 0.000 & 0.600 \\
    \hline  \hline
\end{tabular}
\endgroup
\end{table}

\clearpage

\begin{table}[ht]
\centering
\caption{Plasmode simulation results for estimation of T1D prevalence (T1D cases per 1,000 persons) by year, race and ethnicity, and scenario under the \textbf{BHAV}, \textbf{PUGV}, \textbf{MRP}, and \textbf{Na\"ive} methods. BHAV, PUGV, and MRP refer to Bayesian Hierarchical Analysis of Variance, Partitioning Unobserved Geographic Variations, and Multilevel Regression and Poststratification, respectively. The method with the least bias is in bold. Coverage refers to the 95\% coverage probability. Length refers to the average length of the 95\% uncertainty interval.} 
\label{t:plasmode_simulation_year_race_S1}
\begingroup\tiny
\begin{tabular}{ccccrrrrr}
  \hline
  \multicolumn{1}{c}{Scenario} & \multicolumn{1}{c}{Year}  & \multicolumn{1}{c}{Race and Ethnicity} & \multicolumn{1}{c}{Truth} & \multicolumn{1}{c}{Method}& \multicolumn{1}{c}{Ave. Bias} & \multicolumn{1}{c}{MSE} & \multicolumn{1}{c}{Coverage} & \multicolumn{1}{c}{Ave. Length} \\
 \hline
S1 & 2018 & AS & 1.777 & BHAV & 0.129 & 0.050 & 0.886 & 0.688 \\ 
  S1 & 2018 & AS & 1.777 & PUGV & 0.097 & 0.034 & 0.912 & 0.576 \\ 
  S1 & 2018 & AS & 1.777 & MRP & \textbf{0.026} & 0.019 & 0.891 & 0.467 \\ 
  S1 & 2018 & AS & 1.777 & Na\"ive & -0.489 & 0.254 & 0.054 & 0.507 \\ 
    \hline S1 & 2018 & BA & 3.055 & BHAV & -0.192 & 0.049 & 0.570 & 0.439 \\ 
  S1 & 2018 & BA & 3.055 & PUGV & -0.104 & 0.021 & 0.844 & 0.422 \\ 
  S1 & 2018 & BA & 3.053 & MRP & \textbf{0.009} & 0.012 & 0.905 & 0.388 \\ 
  S1 & 2018 & BA & 3.053 & Na\"ive & -0.611 & 0.381 & 0.000 & 0.373 \\ 
    \hline S1 & 2018 & HP & 2.950 & BHAV & \textbf{0.011} & 0.010 & 0.953 & 0.404 \\ 
  S1 & 2018 & HP & 2.950 & PUGV & -0.152 & 0.030 & 0.701 & 0.376 \\ 
  S1 & 2018 & HP & 2.950 & MRP & -0.062 & 0.010 & 0.932 & 0.347 \\ 
  S1 & 2018 & HP & 2.950 & Na\"ive & -0.515 & 0.271 & 0.000 & 0.337 \\ 
    \hline S1 & 2018 & WH & 5.001 & BHAV & -0.044 & 0.021 & 0.906 & 0.503 \\ 
  S1 & 2018 & WH & 5.002 & PUGV & \textbf{0.015} & 0.018 & 0.932 & 0.510 \\ 
  S1 & 2018 & WH & 5.001 & MRP & -0.130 & 0.033 & 0.810 & 0.481 \\ 
  S1 & 2018 & WH & 5.001 & Na\"ive & -0.983 & 0.971 & 0.000 & 0.322 \\ 
    \hline S1 & 2022 & AS & 1.728 & BHAV & 0.184 & 0.064 & 0.725 & 0.616 \\ 
  S1 & 2022 & AS & 1.727 & PUGV & 0.098 & 0.031 & 0.884 & 0.543 \\ 
  S1 & 2022 & AS & 1.729 & MRP & \textbf{0.003} & 0.015 & 0.925 & 0.445 \\ 
  S1 & 2022 & AS & 1.729 & Na\"ive & -0.282 & 0.093 & 0.388 & 0.474 \\ 
    \hline S1 & 2022 & BA & 3.010 & BHAV & 0.035 & 0.013 & 0.933 & 0.429 \\ 
  S1 & 2022 & BA & 3.011 & PUGV & -0.051 & 0.011 & 0.952 & 0.408 \\ 
  S1 & 2022 & BA & 3.011 & MRP & \textbf{-0.011} & 0.008 & 0.952 & 0.379 \\ 
  S1 & 2022 & BA & 3.011 & Na\"ive & -0.334 & 0.117 & 0.027 & 0.373 \\ 
    \hline S1 & 2022 & HP & 2.853 & BHAV & -0.033 & 0.007 & 0.966 & 0.349 \\ 
  S1 & 2022 & HP & 2.853 & PUGV & -0.115 & 0.020 & 0.782 & 0.352 \\ 
  S1 & 2022 & HP & 2.853 & MRP & \textbf{-0.017} & 0.006 & 0.959 & 0.328 \\ 
  S1 & 2022 & HP & 2.853 & Na\"ive & -0.501 & 0.254 & 0.000 & 0.293 \\ 
    \hline S1 & 2022 & WH & 4.760 & BHAV & -0.054 & 0.015 & 0.966 & 0.483 \\ 
  S1 & 2022 & WH & 4.761 & PUGV & 0.102 & 0.024 & 0.898 & 0.512 \\ 
  S1 & 2022 & WH & 4.761 & MRP & \textbf{-0.003} & 0.013 & 0.973 & 0.483 \\ 
  S1 & 2022 & WH & 4.761 & Na\"ive & -0.984 & 0.973 & 0.000 & 0.296 \\ \hline
  S2 & 2018 & AS & 2.017 & BHAV & \textbf{-0.113} & 0.046 & 0.843 & 0.685 \\ 
  S2 & 2018 & AS & 2.017 & PUGV & -0.146 & 0.046 & 0.796 & 0.574 \\ 
  S2 & 2018 & AS & 2.017 & MRP & -0.213 & 0.064 & 0.558 & 0.467 \\ 
  S2 & 2018 & AS & 2.017 & Na\"ive & -0.728 & 0.545 & 0.000 & 0.507 \\ 
    \hline S2 & 2018 & BA & 2.939 & BHAV & -0.074 & 0.017 & 0.936 & 0.438 \\ 
  S2 & 2018 & BA & 2.940 & PUGV & \textbf{0.011} & 0.011 & 0.973 & 0.423 \\ 
  S2 & 2018 & BA & 2.938 & MRP & 0.125 & 0.027 & 0.735 & 0.388 \\ 
  S2 & 2018 & BA & 2.938 & Na\"ive & -0.495 & 0.253 & 0.000 & 0.373 \\ 
    \hline S2 & 2018 & HP & 2.932 & BHAV & \textbf{0.029} & 0.011 & 0.936 & 0.403 \\ 
  S2 & 2018 & HP & 2.930 & PUGV & -0.132 & 0.025 & 0.776 & 0.376 \\ 
  S2 & 2018 & HP & 2.932 & MRP & -0.044 & 0.008 & 0.952 & 0.347 \\ 
  S2 & 2018 & HP & 2.932 & Na\"ive & -0.496 & 0.252 & 0.000 & 0.337 \\ 
    \hline S2 & 2018 & WH & 4.906 & BHAV & 0.048 & 0.022 & 0.929 & 0.503 \\ 
  S2 & 2018 & WH & 4.904 & PUGV & 0.113 & 0.031 & 0.905 & 0.507 \\ 
  S2 & 2018 & WH & 4.904 & MRP & \textbf{-0.033} & 0.017 & 0.925 & 0.481 \\ 
  S2 & 2018 & WH & 4.904 & Na\"ive & -0.886 & 0.789 & 0.000 & 0.322 \\ 
    \hline S2 & 2022 & AS & 1.987 & BHAV & \textbf{-0.073} & 0.036 & 0.879 & 0.615 \\ 
  S2 & 2022 & AS & 1.984 & PUGV & -0.162 & 0.048 & 0.748 & 0.539 \\ 
  S2 & 2022 & AS & 1.987 & MRP & -0.254 & 0.080 & 0.469 & 0.445 \\ 
  S2 & 2022 & AS & 1.987 & Na\"ive & -0.539 & 0.305 & 0.007 & 0.474 \\ 
    \hline S2 & 2022 & BA & 2.959 & BHAV & 0.083 & 0.019 & 0.900 & 0.428 \\ 
  S2 & 2022 & BA & 2.961 & PUGV & \textbf{-0.002} & 0.008 & 0.986 & 0.408 \\ 
  S2 & 2022 & BA & 2.961 & MRP & 0.039 & 0.010 & 0.952 & 0.379 \\ 
  S2 & 2022 & BA & 2.961 & Na\"ive & -0.284 & 0.086 & 0.122 & 0.373 \\ 
    \hline S2 & 2022 & HP & 2.907 & BHAV & -0.085 & 0.013 & 0.871 & 0.349 \\ 
  S2 & 2022 & HP & 2.908 & PUGV & -0.169 & 0.036 & 0.544 & 0.353 \\ 
  S2 & 2022 & HP & 2.908 & MRP & \textbf{-0.072} & 0.011 & 0.864 & 0.328 \\ 
  S2 & 2022 & HP & 2.908 & Na\"ive & -0.556 & 0.312 & 0.000 & 0.293 \\ 
    \hline S2 & 2022 & WH & 4.808 & BHAV & -0.098 & 0.022 & 0.893 & 0.482 \\ 
  S2 & 2022 & WH & 4.807 & PUGV & 0.056 & 0.017 & 0.959 & 0.510 \\ 
  S2 & 2022 & WH & 4.808 & MRP & \textbf{-0.050} & 0.016 & 0.932 & 0.483 \\ 
  S2 & 2022 & WH & 4.808 & Na\"ive & -1.031 & 1.068 & 0.000 & 0.296 \\ 
    \hline  \hline
\end{tabular}
\endgroup
\end{table}

\clearpage

\begin{table}[ht]
\centering
\caption{Plasmode simulation results for estimation of T1D prevalence (T1D cases per 1,000 persons) by year, sex, and scenario under the \textbf{BHAV}, \textbf{PUGV}, \textbf{MRP}, and \textbf{Na\"ive} methods. BHAV, PUGV, and MRP refer to Bayesian Hierarchical Analysis of Variance, Partitioning Unobserved Geographic Variations, and Multilevel Regression and Poststratification, respectively. The method with the least bias is in bold. Coverage refers to the 95\% coverage probability. Length refers to the average length of the 95\% uncertainty interval.} 
\label{t:plasmode_simulation_year_sex_S1}
\begingroup\tiny
\begin{tabular}{ccccrrrrr}
  \hline
  \multicolumn{1}{c}{Scenario} & \multicolumn{1}{c}{Year}  & \multicolumn{1}{c}{Sex} & \multicolumn{1}{c}{Truth} & \multicolumn{1}{c}{Method}& \multicolumn{1}{c}{Ave. Bias} & \multicolumn{1}{c}{MSE} & \multicolumn{1}{c}{Coverage} & \multicolumn{1}{c}{Ave. Length} \\
 \hline
S1 & 2018 & F & 3.708 & BHAV & \textbf{-0.017} & 0.010 & 0.946 & 0.380 \\ 
  S1 & 2018 & F & 3.709 & PUGV & -0.057 & 0.010 & 0.898 & 0.337 \\ 
  S1 & 2018 & F & 3.708 & MRP & -0.079 & 0.014 & 0.857 & 0.351 \\ 
  S1 & 2018 & F & 3.708 & Na\"ive & -0.567 & 0.325 & 0.000 & 0.281 \\ 
    \hline S1 & 2018 & M & 3.745 & BHAV & -0.063 & 0.014 & 0.879 & 0.375 \\ 
  S1 & 2018 & M & 3.745 & PUGV & \textbf{-0.041} & 0.009 & 0.898 & 0.337 \\ 
  S1 & 2018 & M & 3.744 & MRP & -0.059 & 0.012 & 0.871 & 0.353 \\ 
  S1 & 2018 & M & 3.744 & Na\"ive & -0.667 & 0.448 & 0.000 & 0.268 \\ 
    \hline S1 & 2022 & F & 3.533 & BHAV & 0.002 & 0.007 & 0.973 & 0.352 \\ 
  S1 & 2022 & F & 3.533 & PUGV & \textbf{-0.001} & 0.006 & 0.986 & 0.318 \\ 
  S1 & 2022 & F & 3.532 & MRP & -0.014 & 0.006 & 0.980 & 0.339 \\ 
  S1 & 2022 & F & 3.532 & Na\"ive & -0.539 & 0.293 & 0.000 & 0.253 \\ 
    \hline S1 & 2022 & M & 3.582 & BHAV & -0.016 & 0.007 & 0.980 & 0.347 \\ 
  S1 & 2022 & M & 3.582 & PUGV & 0.016 & 0.006 & 0.973 & 0.320 \\ 
  S1 & 2022 & M & 3.583 & MRP & \textbf{-0.002} & 0.007 & 0.966 & 0.341 \\ 
  S1 & 2022 & M & 3.583 & Na\"ive & -0.589 & 0.350 & 0.000 & 0.246 \\ \hline
  S2 & 2018 & F & 3.685 & BHAV & \textbf{0.006} & 0.009 & 0.943 & 0.380 \\ 
  S2 & 2018 & F & 3.684 & PUGV & -0.033 & 0.008 & 0.925 & 0.335 \\ 
  S2 & 2018 & F & 3.684 & MRP & -0.054 & 0.011 & 0.932 & 0.351 \\ 
  S2 & 2018 & F & 3.684 & Na\"ive & -0.542 & 0.298 & 0.000 & 0.281 \\ 
    \hline S2 & 2018 & M & 3.678 & BHAV & \textbf{0.002} & 0.010 & 0.936 & 0.375 \\ 
  S2 & 2018 & M & 3.677 & PUGV & 0.027 & 0.008 & 0.939 & 0.337 \\ 
  S2 & 2018 & M & 3.677 & MRP & 0.009 & 0.008 & 0.946 & 0.353 \\ 
  S2 & 2018 & M & 3.677 & Na\"ive & -0.599 & 0.362 & 0.000 & 0.268 \\ 
    \hline S2 & 2022 & F & 3.605 & BHAV & \textbf{-0.070} & 0.012 & 0.907 & 0.352 \\ 
  S2 & 2022 & F & 3.606 & PUGV & -0.074 & 0.012 & 0.850 & 0.318 \\ 
  S2 & 2022 & F & 3.606 & MRP & -0.087 & 0.014 & 0.803 & 0.339 \\ 
  S2 & 2022 & F & 3.606 & Na\"ive & -0.612 & 0.378 & 0.000 & 0.253 \\ 
    \hline S2 & 2022 & M & 3.613 & BHAV & -0.045 & 0.009 & 0.936 & 0.347 \\ 
  S2 & 2022 & M & 3.613 & PUGV & \textbf{-0.015} & 0.006 & 0.952 & 0.320 \\ 
  S2 & 2022 & M & 3.614 & MRP & -0.033 & 0.008 & 0.959 & 0.341 \\ 
  S2 & 2022 & M & 3.614 & Na\"ive & -0.620 & 0.388 & 0.000 & 0.246 \\ 
    \hline  \hline
\end{tabular}
\endgroup
\end{table}

\clearpage

\begin{table}[ht]
\centering
\caption{Plasmode simulation results for estimation of T1D prevalence (T1D cases per 1,000 persons) by year, health system center, and scenario under the \textbf{BHAV}, \textbf{PUGV}, \textbf{MRP}, and \textbf{Na\"ive} methods. BHAV, PUGV, and MRP refer to Bayesian Hierarchical Analysis of Variance, Partitioning Unobserved Geographic Variations, and Multilevel Regression and Poststratification, respectively. LUR refers to Lurie Children's Hospital; PED, PEDSNet; and UFL, OneFlorida$+$. The method with the least bias is in bold. The method with the least bias is in bold. Coverage refers to the 95\% coverage probability. Length refers to the average length of the 95\% uncertainty interval.} 
\label{t:plasmode_simulation_year_site_scenario}
\begingroup\footnotesize
\begin{tabular}{ccccrrrrr}
  \hline
  \multicolumn{1}{c}{Scenario} & \multicolumn{1}{c}{Year} & \multicolumn{1}{c}{Center} & \multicolumn{1}{c}{Truth} & \multicolumn{1}{c}{Method}& \multicolumn{1}{c}{Ave. Bias} & \multicolumn{1}{c}{MSE} & \multicolumn{1}{c}{Coverage} & \multicolumn{1}{c}{Ave. Length} \\
 \hline
S1 & 2018 & LUR & 3.205 & BHAV & 0.016 & 0.008 & 0.946 & 0.324 \\ 
  S1 & 2018 & LUR & 3.204 & PUGV & 0.026 & 0.009 & 0.925 & 0.327 \\ 
  S1 & 2018 & LUR & 3.204 & MRP & \textbf{0.007} & 0.009 & 0.939 & 0.351 \\ 
  S1 & 2018 & LUR & 3.204 & Na\"ive & -0.223 & 0.055 & 0.143 & 0.282 \\ 
    \hline S1 & 2022 & LUR & 3.019 & BHAV & -0.010 & 0.005 & 0.946 & 0.272 \\ 
  S1 & 2022 & LUR & 3.019 & PUGV & \textbf{-0.005} & 0.005 & 0.946 & 0.274 \\ 
  S1 & 2022 & LUR & 3.020 & MRP & -0.027 & 0.006 & 0.959 & 0.300 \\ 
  S1 & 2022 & LUR & 3.020 & Na\"ive & -0.136 & 0.021 & 0.381 & 0.249 \\ 
    \hline S2 & 2018 & LUR & 3.229 & BHAV & -0.008 & 0.008 & 0.936 & 0.324 \\ 
  S2 & 2018 & LUR & 3.228 & PUGV & \textbf{0.003} & 0.008 & 0.925 & 0.329 \\ 
  S2 & 2018 & LUR & 3.228 & MRP & -0.017 & 0.009 & 0.946 & 0.351 \\ 
  S2 & 2018 & LUR & 3.228 & Na\"ive & -0.247 & 0.066 & 0.082 & 0.282 \\ 
    \hline S2 & 2022 & LUR & 3.031 & BHAV & -0.019 & 0.005 & 0.936 & 0.272 \\ 
  S2 & 2022 & LUR & 3.031 & PUGV & \textbf{-0.017} & 0.005 & 0.946 & 0.274 \\ 
  S2 & 2022 & LUR & 3.032 & MRP & -0.039 & 0.006 & 0.946 & 0.300 \\ 
  S2 & 2022 & LUR & 3.032 & Na\"ive & -0.148 & 0.025 & 0.306 & 0.249 \\ 
    \hline S1 & 2018 & PED & 3.433 & BHAV & 0.038 & 0.005 & 0.919 & 0.239 \\ 
  S1 & 2018 & PED & 3.434 & PUGV & 0.059 & 0.007 & 0.837 & 0.243 \\ 
  S1 & 2018 & PED & 3.430 & MRP & \textbf{0.034} & 0.005 & 1.000 & 0.353 \\ 
  S1 & 2018 & PED & 3.430 & Na\"ive & -0.172 & 0.032 & 0.388 & 0.325 \\ 
    \hline S1 & 2022 & PED & 3.102 & BHAV & -0.016 & 0.003 & 0.913 & 0.188 \\ 
  S1 & 2022 & PED & 3.102 & PUGV & \textbf{0.010} & 0.003 & 0.932 & 0.193 \\ 
  S1 & 2022 & PED & 3.103 & MRP & -0.016 & 0.003 & 1.000 & 0.313 \\ 
  S1 & 2022 & PED & 3.103 & Na\"ive & -0.071 & 0.007 & 0.946 & 0.305 \\ 
    \hline S2 & 2018 & PED & 3.457 & BHAV & 0.014 & 0.004 & 0.943 & 0.239 \\ 
  S2 & 2018 & PED & 3.457 & PUGV & 0.037 & 0.005 & 0.925 & 0.243 \\ 
  S2 & 2018 & PED & 3.454 & MRP & \textbf{0.011} & 0.004 & 1.000 & 0.353 \\ 
  S2 & 2018 & PED & 3.454 & Na\"ive & -0.196 & 0.041 & 0.293 & 0.325 \\ 
    \hline S2 & 2022 & PED & 3.077 & BHAV & 0.006 & 0.002 & 0.957 & 0.188 \\ 
  S2 & 2022 & PED & 3.080 & PUGV & 0.032 & 0.003 & 0.891 & 0.193 \\ 
  S2 & 2022 & PED & 3.081 & MRP & \textbf{0.005} & 0.002 & 0.993 & 0.313 \\ 
  S2 & 2022 & PED & 3.081 & Na\"ive & -0.050 & 0.005 & 1.000 & 0.305 \\ 
    \hline S1 & 2018 & UFL & 4.875 & BHAV & \textbf{-0.190} & 0.090 & 0.799 & 0.821 \\ 
  S1 & 2018 & UFL & 4.876 & PUGV & -0.250 & 0.109 & 0.762 & 0.796 \\ 
  S1 & 2018 & UFL & 4.877 & MRP & -0.267 & 0.121 & 0.701 & 0.825 \\ 
  S1 & 2018 & UFL & 4.877 & Na\"ive & -1.755 & 3.092 & 0.000 & 0.465 \\ 
    \hline S1 & 2022 & UFL & 4.739 & BHAV & \textbf{0.003} & 0.036 & 0.980 & 0.788 \\ 
  S1 & 2022 & UFL & 4.740 & PUGV & 0.029 & 0.036 & 0.966 & 0.774 \\ 
  S1 & 2022 & UFL & 4.738 & MRP & 0.031 & 0.041 & 0.959 & 0.821 \\ 
  S1 & 2022 & UFL & 4.738 & Na\"ive & -1.520 & 2.322 & 0.000 & 0.433 \\ 
    \hline S2 & 2018 & UFL & 4.660 & BHAV & \textbf{0.021} & 0.055 & 0.929 & 0.818 \\ 
  S2 & 2018 & UFL & 4.659 & PUGV & -0.033 & 0.049 & 0.946 & 0.792 \\ 
  S2 & 2018 & UFL & 4.660 & MRP & -0.050 & 0.053 & 0.946 & 0.825 \\ 
  S2 & 2018 & UFL & 4.660 & Na\"ive & -1.538 & 2.376 & 0.000 & 0.465 \\ 
    \hline S2 & 2022 & UFL & 4.898 & BHAV & -0.156 & 0.061 & 0.900 & 0.787 \\ 
  S2 & 2022 & UFL & 4.897 & PUGV & -0.129 & 0.053 & 0.898 & 0.771 \\ 
  S2 & 2022 & UFL & 4.896 & MRP & \textbf{-0.127} & 0.058 & 0.918 & 0.821 \\ 
  S2 & 2022 & UFL & 4.896 & Na\"ive & -1.678 & 2.826 & 0.000 & 0.433 \\ 
    \hline  \hline
\end{tabular}
\endgroup
\end{table}

\clearpage

\begin{table}[ht]
\centering
\caption{\tiny Plasmode simulation results for estimation of T1D prevalence (T1D cases per 1,000 persons) by year, health system center, and age for scenario S1 under the \textbf{BHAV}, \textbf{PUGV}, \textbf{MRP}, and \textbf{Na\"ive} methods. BHAV, PUGV, and MRP refer to Bayesian Hierarchical Analysis of Variance, Partitioning Unobserved Geographic Variations, and Multilevel Regression and Poststratification, respectively. LUR refers to Lurie Children's Hospital; PED, PEDSNet; and UFL, OneFlorida$+$. The method with the least bias is in bold. Coverage refers to the 95\% coverage probability. Length refers to the average length of the 95\% uncertainty interval.} 
\label{t:plasmode_simulation_year_site_age_S1}
\begingroup\tiny
\begin{tabular}{cccccrrrrr}
  \hline
  \multicolumn{1}{c}{Scenario} & \multicolumn{1}{c}{Year} & \multicolumn{1}{c}{Center} & \multicolumn{1}{c}{Age} &  \multicolumn{1}{c}{Truth} & \multicolumn{1}{c}{Method}& \multicolumn{1}{c}{Ave. Bias} & \multicolumn{1}{c}{MSE} & \multicolumn{1}{c}{Coverage} & \multicolumn{1}{c}{Ave. Length} \\
 \hline
S1 & 2018 & LUR & [0, 5) & 0.516 & BHAV & -0.076 & 0.007 & 0.483 & 0.142 \\ 
  S1 & 2018 & LUR & [0, 5) & 0.516 & PUGV & -0.041 & 0.003 & 0.810 & 0.146 \\ 
  S1 & 2018 & LUR & [0, 5) & 0.515 & MRP & \textbf{0.037} & 0.003 & 0.707 & 0.115 \\ 
  S1 & 2018 & LUR & [0, 5) & 0.515 & Na\"ive & -0.098 & 0.011 & 0.463 & 0.183 \\ 
    \hline S1 & 2018 & LUR & [5, 10) & 2.183 & BHAV & -0.007 & 0.010 & 0.946 & 0.395 \\ 
  S1 & 2018 & LUR & [5, 10) & 2.182 & PUGV & 0.064 & 0.014 & 0.939 & 0.416 \\ 
  S1 & 2018 & LUR & [5, 10) & 2.182 & MRP & \textbf{0.003} & 0.008 & 0.898 & 0.313 \\ 
  S1 & 2018 & LUR & [5, 10) & 2.182 & Na\"ive & 0.030 & 0.011 & 0.973 & 0.483 \\ 
    \hline S1 & 2018 & LUR & [10, 15) & 4.849 & BHAV & 0.133 & 0.050 & 0.859 & 0.654 \\ 
  S1 & 2018 & LUR & [10, 15) & 4.848 & PUGV & 0.157 & 0.055 & 0.844 & 0.678 \\ 
  S1 & 2018 & LUR & [10, 15) & 4.848 & MRP & \textbf{-0.033} & 0.026 & 0.946 & 0.586 \\ 
  S1 & 2018 & LUR & [10, 15) & 4.848 & Na\"ive & 0.184 & 0.069 & 0.837 & 0.729 \\ 
    \hline S1 & 2018 & LUR & [15, 18) & 6.448 & BHAV & \textbf{0.005} & 0.062 & 0.933 & 0.920 \\ 
  S1 & 2018 & LUR & [15, 18) & 6.446 & PUGV & -0.146 & 0.074 & 0.891 & 0.899 \\ 
  S1 & 2018 & LUR & [15, 18) & 6.447 & MRP & 0.032 & 0.054 & 0.925 & 0.812 \\ 
  S1 & 2018 & LUR & [15, 18) & 6.447 & Na\"ive & -0.271 & 0.140 & 0.782 & 0.998 \\ 
    \hline S1 & 2022 & LUR & [0, 5) & 0.503 & BHAV & -0.008 & 0.001 & 0.899 & 0.137 \\ 
  S1 & 2022 & LUR & [0, 5) & 0.504 & PUGV & -0.036 & 0.002 & 0.810 & 0.130 \\ 
  S1 & 2022 & LUR & [0, 5) & 0.503 & MRP & 0.002 & 0.001 & 0.905 & 0.104 \\ 
  S1 & 2022 & LUR & [0, 5) & 0.503 & Na\"ive & \textbf{-0.000} & 0.002 & 0.973 & 0.185 \\ 
    \hline S1 & 2022 & LUR & [5, 10) & 2.034 & BHAV & 0.073 & 0.014 & 0.846 & 0.352 \\ 
  S1 & 2022 & LUR & [5, 10) & 2.032 & PUGV & 0.042 & 0.011 & 0.946 & 0.365 \\ 
  S1 & 2022 & LUR & [5, 10) & 2.033 & MRP & \textbf{-0.032} & 0.006 & 0.925 & 0.275 \\ 
  S1 & 2022 & LUR & [5, 10) & 2.033 & Na\"ive & 0.088 & 0.018 & 0.912 & 0.426 \\ 
    \hline S1 & 2022 & LUR & [10, 15) & 4.462 & BHAV & 0.069 & 0.024 & 0.919 & 0.539 \\ 
  S1 & 2022 & LUR & [10, 15) & 4.462 & PUGV & 0.100 & 0.031 & 0.898 & 0.577 \\ 
  S1 & 2022 & LUR & [10, 15) & 4.464 & MRP & \textbf{-0.064} & 0.022 & 0.912 & 0.501 \\ 
  S1 & 2022 & LUR & [10, 15) & 4.464 & Na\"ive & 0.189 & 0.054 & 0.864 & 0.621 \\ 
    \hline S1 & 2022 & LUR & [15, 18) & 5.900 & BHAV & -0.270 & 0.112 & 0.691 & 0.728 \\ 
  S1 & 2022 & LUR & [15, 18) & 5.900 & PUGV & -0.203 & 0.074 & 0.823 & 0.753 \\ 
  S1 & 2022 & LUR & [15, 18) & 5.902 & MRP & \textbf{-0.003} & 0.029 & 0.946 & 0.693 \\ 
  S1 & 2022 & LUR & [15, 18) & 5.902 & Na\"ive & -0.194 & 0.073 & 0.898 & 0.842 \\ 
    \hline S1 & 2018 & PED & [0, 5) & 0.651 & BHAV & 0.047 & 0.004 & 0.805 & 0.173 \\ 
  S1 & 2018 & PED & [0, 5) & 0.652 & PUGV & 0.063 & 0.006 & 0.680 & 0.177 \\ 
  S1 & 2018 & PED & [0, 5) & 0.649 & MRP & -0.064 & 0.007 & 0.524 & 0.125 \\ 
  S1 & 2018 & PED & [0, 5) & 0.649 & Na\"ive & \textbf{0.019} & 0.002 & 1.000 & 0.272 \\ 
    \hline S1 & 2018 & PED & [5, 10) & 2.323 & BHAV & \textbf{-0.021} & 0.008 & 0.953 & 0.324 \\ 
  S1 & 2018 & PED & [5, 10) & 2.325 & PUGV & 0.046 & 0.009 & 0.891 & 0.336 \\ 
  S1 & 2018 & PED & [5, 10) & 2.321 & MRP & 0.166 & 0.041 & 0.544 & 0.350 \\ 
  S1 & 2018 & PED & [5, 10) & 2.321 & Na\"ive & -0.131 & 0.024 & 0.925 & 0.525 \\ 
    \hline S1 & 2018 & PED & [10, 15) & 4.928 & BHAV & -0.056 & 0.017 & 0.940 & 0.495 \\ 
  S1 & 2018 & PED & [10, 15) & 4.930 & PUGV & \textbf{-0.027} & 0.015 & 0.952 & 0.498 \\ 
  S1 & 2018 & PED & [10, 15) & 4.926 & MRP & 0.227 & 0.069 & 0.707 & 0.597 \\ 
  S1 & 2018 & PED & [10, 15) & 4.926 & Na\"ive & -0.272 & 0.088 & 0.782 & 0.744 \\ 
    \hline S1 & 2018 & PED & [15, 18) & 6.996 & BHAV & 0.243 & 0.098 & 0.738 & 0.780 \\ 
  S1 & 2018 & PED & [15, 18) & 6.998 & PUGV & 0.196 & 0.078 & 0.803 & 0.768 \\ 
  S1 & 2018 & PED & [15, 18) & 6.991 & MRP & -0.271 & 0.120 & 0.707 & 0.797 \\ 
  S1 & 2018 & PED & [15, 18) & 6.991 & Na\"ive & \textbf{-0.038} & 0.035 & 1.000 & 1.129 \\ 
    \hline S1 & 2022 & PED & [0, 5) & 0.675 & BHAV & 0.079 & 0.008 & 0.510 & 0.173 \\ 
  S1 & 2022 & PED & [0, 5) & 0.675 & PUGV & \textbf{0.048} & 0.004 & 0.803 & 0.158 \\ 
  S1 & 2022 & PED & [0, 5) & 0.675 & MRP & -0.150 & 0.024 & 0.014 & 0.112 \\ 
  S1 & 2022 & PED & [0, 5) & 0.675 & Na\"ive & 0.060 & 0.006 & 0.980 & 0.288 \\ 
    \hline S1 & 2022 & PED & [5, 10) & 2.295 & BHAV & \textbf{0.004} & 0.005 & 0.946 & 0.278 \\ 
  S1 & 2022 & PED & [5, 10) & 2.297 & PUGV & 0.013 & 0.005 & 0.946 & 0.279 \\ 
  S1 & 2022 & PED & [5, 10) & 2.298 & MRP & -0.121 & 0.025 & 0.653 & 0.301 \\ 
  S1 & 2022 & PED & [5, 10) & 2.298 & Na\"ive & -0.018 & 0.006 & 1.000 & 0.501 \\ 
    \hline S1 & 2022 & PED & [10, 15) & 4.519 & BHAV & -0.120 & 0.025 & 0.792 & 0.388 \\ 
  S1 & 2022 & PED & [10, 15) & 4.516 & PUGV & \textbf{-0.076} & 0.017 & 0.871 & 0.398 \\ 
  S1 & 2022 & PED & [10, 15) & 4.518 & MRP & 0.211 & 0.061 & 0.721 & 0.542 \\ 
  S1 & 2022 & PED & [10, 15) & 4.518 & Na\"ive & -0.215 & 0.057 & 0.912 & 0.690 \\ 
    \hline S1 & 2022 & PED & [15, 18) & 6.106 & BHAV & -0.040 & 0.027 & 0.933 & 0.629 \\ 
  S1 & 2022 & PED & [15, 18) & 6.106 & PUGV & 0.076 & 0.033 & 0.918 & 0.651 \\ 
  S1 & 2022 & PED & [15, 18) & 6.107 & MRP & \textbf{0.019} & 0.034 & 0.952 & 0.731 \\ 
  S1 & 2022 & PED & [15, 18) & 6.107 & Na\"ive & -0.266 & 0.097 & 0.925 & 1.018 \\ 
    \hline S1 & 2018 & UFL & [0, 5) & 0.782 & BHAV & -0.211 & 0.051 & 0.208 & 0.272 \\ 
  S1 & 2018 & UFL & [0, 5) & 0.782 & PUGV & -0.101 & 0.016 & 0.748 & 0.303 \\ 
  S1 & 2018 & UFL & [0, 5) & 0.781 & MRP & \textbf{0.001} & 0.006 & 0.810 & 0.196 \\ 
  S1 & 2018 & UFL & [0, 5) & 0.781 & Na\"ive & -0.383 & 0.152 & 0.000 & 0.282 \\ 
    \hline S1 & 2018 & UFL & [5, 10) & 3.325 & BHAV & -0.369 & 0.185 & 0.617 & 0.840 \\ 
  S1 & 2018 & UFL & [5, 10) & 3.326 & PUGV & -0.275 & 0.119 & 0.789 & 0.879 \\ 
  S1 & 2018 & UFL & [5, 10) & 3.325 & MRP & \textbf{-0.181} & 0.067 & 0.748 & 0.629 \\ 
  S1 & 2018 & UFL & [5, 10) & 3.325 & Na\"ive & -1.127 & 1.306 & 0.000 & 0.760 \\ 
    \hline S1 & 2018 & UFL & [10, 15) & 7.254 & BHAV & \textbf{-0.005} & 0.196 & 0.926 & 1.556 \\ 
  S1 & 2018 & UFL & [10, 15) & 7.256 & PUGV & -0.138 & 0.193 & 0.925 & 1.584 \\ 
  S1 & 2018 & UFL & [10, 15) & 7.257 & MRP & -0.427 & 0.308 & 0.680 & 1.276 \\ 
  S1 & 2018 & UFL & [10, 15) & 7.257 & Na\"ive & -1.854 & 3.532 & 0.000 & 1.245 \\ 
    \hline S1 & 2018 & UFL & [15, 18) & 9.650 & BHAV & \textbf{-0.191} & 0.340 & 0.946 & 2.148 \\ 
  S1 & 2018 & UFL & [15, 18) & 9.651 & PUGV & -0.629 & 0.640 & 0.796 & 2.052 \\ 
  S1 & 2018 & UFL & [15, 18) & 9.656 & MRP & -0.549 & 0.524 & 0.701 & 1.718 \\ 
  S1 & 2018 & UFL & [15, 18) & 9.656 & Na\"ive & -2.395 & 5.926 & 0.000 & 1.808 \\ 
    \hline S1 & 2022 & UFL & [0, 5) & 0.787 & BHAV & -0.080 & 0.011 & 0.826 & 0.302 \\ 
  S1 & 2022 & UFL & [0, 5) & 0.788 & PUGV & -0.059 & 0.007 & 0.898 & 0.301 \\ 
  S1 & 2022 & UFL & [0, 5) & 0.788 & MRP & \textbf{0.008} & 0.004 & 0.871 & 0.196 \\ 
  S1 & 2022 & UFL & [0, 5) & 0.788 & Na\"ive & -0.211 & 0.049 & 0.272 & 0.320 \\ 
    \hline S1 & 2022 & UFL & [5, 10) & 3.203 & BHAV & -0.059 & 0.041 & 0.966 & 0.860 \\ 
  S1 & 2022 & UFL & [5, 10) & 3.206 & PUGV & -0.085 & 0.044 & 0.966 & 0.888 \\ 
  S1 & 2022 & UFL & [5, 10) & 3.202 & MRP & \textbf{0.027} & 0.031 & 0.952 & 0.626 \\ 
  S1 & 2022 & UFL & [5, 10) & 3.202 & Na\"ive & -1.030 & 1.089 & 0.000 & 0.703 \\ 
    \hline S1 & 2022 & UFL & [10, 15) & 6.987 & BHAV & 0.251 & 0.214 & 0.906 & 1.489 \\ 
  S1 & 2022 & UFL & [10, 15) & 6.986 & PUGV & 0.280 & 0.220 & 0.884 & 1.564 \\ 
  S1 & 2022 & UFL & [10, 15) & 6.987 & MRP & \textbf{0.002} & 0.112 & 0.932 & 1.265 \\ 
  S1 & 2022 & UFL & [10, 15) & 6.987 & Na\"ive & -1.660 & 2.823 & 0.000 & 1.111 \\ 
    \hline S1 & 2022 & UFL & [15, 18) & 9.147 & BHAV & -0.182 & 0.248 & 0.940 & 1.929 \\ 
  S1 & 2022 & UFL & [15, 18) & 9.150 & PUGV & \textbf{-0.074} & 0.195 & 0.980 & 1.980 \\ 
  S1 & 2022 & UFL & [15, 18) & 9.145 & MRP & 0.115 & 0.187 & 0.966 & 1.680 \\ 
  S1 & 2022 & UFL & [15, 18) & 9.145 & Na\"ive & -2.277 & 5.323 & 0.000 & 1.554 \\ 
    \hline  \hline
\end{tabular}
\endgroup
\end{table}

\clearpage

\begin{table}[ht]
\centering
\caption{\tiny Plasmode simulation results for estimation of T1D prevalence (T1D cases per 1,000 persons) by year, health system center, and age for scenario S2 under the \textbf{BHAV}, \textbf{PUGV}, \textbf{MRP}, and \textbf{Na\"ive} methods. BHAV, PUGV, and MRP refer to Bayesian Hierarchical Analysis of Variance, Partitioning Unobserved Geographic Variations, and Multilevel Regression and Poststratification, respectively. LUR refers to Lurie Children's Hospital; PED, PEDSNet; and UFL, OneFlorida$+$. The method with the least bias is in bold. Coverage refers to the 95\% coverage probability. Length refers to the average length of the 95\% uncertainty interval.} 
\label{t:plasmode_simulation_year_site_age_S2}
\begingroup\tiny
\begin{tabular}{cccccrrrrr}
  \hline
  \multicolumn{1}{c}{Scenario} & \multicolumn{1}{c}{Year} & \multicolumn{1}{c}{Center} & \multicolumn{1}{c}{Age} & \multicolumn{1}{c}{Truth} & \multicolumn{1}{c}{Method}& \multicolumn{1}{c}{Ave. Bias} & \multicolumn{1}{c}{MSE} & \multicolumn{1}{c}{Coverage} & \multicolumn{1}{c}{Ave. Length} \\
 \hline
S2 & 2018 & LUR & [0, 5) & 0.468 & BHAV & -0.028 & 0.002 & 0.864 & 0.143 \\ 
  S2 & 2018 & LUR & [0, 5) & 0.467 & PUGV & \textbf{0.007} & 0.001 & 0.952 & 0.146 \\ 
  S2 & 2018 & LUR & [0, 5) & 0.467 & MRP & 0.085 & 0.009 & 0.184 & 0.115 \\ 
  S2 & 2018 & LUR & [0, 5) & 0.467 & Na\"ive & -0.050 & 0.004 & 0.782 & 0.183 \\ 
    \hline S2 & 2018 & LUR & [5, 10) & 2.231 & BHAV & -0.058 & 0.013 & 0.936 & 0.395 \\ 
  S2 & 2018 & LUR & [5, 10) & 2.231 & PUGV & \textbf{0.015} & 0.011 & 0.966 & 0.417 \\ 
  S2 & 2018 & LUR & [5, 10) & 2.232 & MRP & -0.047 & 0.010 & 0.871 & 0.313 \\ 
  S2 & 2018 & LUR & [5, 10) & 2.232 & Na\"ive & -0.019 & 0.011 & 0.986 & 0.483 \\ 
    \hline S2 & 2018 & LUR & [10, 15) & 5.008 & BHAV & -0.023 & 0.033 & 0.914 & 0.654 \\ 
  S2 & 2018 & LUR & [10, 15) & 5.006 & PUGV & \textbf{0.001} & 0.030 & 0.946 & 0.681 \\ 
  S2 & 2018 & LUR & [10, 15) & 5.005 & MRP & -0.190 & 0.062 & 0.735 & 0.586 \\ 
  S2 & 2018 & LUR & [10, 15) & 5.005 & Na\"ive & 0.027 & 0.036 & 0.946 & 0.729 \\ 
    \hline S2 & 2018 & LUR & [15, 18) & 6.319 & BHAV & 0.131 & 0.080 & 0.893 & 0.919 \\ 
  S2 & 2018 & LUR & [15, 18) & 6.318 & PUGV & \textbf{-0.019} & 0.054 & 0.952 & 0.900 \\ 
  S2 & 2018 & LUR & [15, 18) & 6.320 & MRP & 0.159 & 0.079 & 0.844 & 0.812 \\ 
  S2 & 2018 & LUR & [15, 18) & 6.320 & Na\"ive & -0.144 & 0.086 & 0.891 & 0.998 \\ 
    \hline S2 & 2022 & LUR & [0, 5) & 0.461 & BHAV & 0.037 & 0.003 & 0.836 & 0.137 \\ 
  S2 & 2022 & LUR & [0, 5) & 0.461 & PUGV & \textbf{0.007} & 0.001 & 0.952 & 0.129 \\ 
  S2 & 2022 & LUR & [0, 5) & 0.461 & MRP & 0.045 & 0.003 & 0.571 & 0.104 \\ 
  S2 & 2022 & LUR & [0, 5) & 0.461 & Na\"ive & 0.043 & 0.003 & 0.918 & 0.185 \\ 
    \hline S2 & 2022 & LUR & [5, 10) & 2.070 & BHAV & 0.041 & 0.010 & 0.929 & 0.352 \\ 
  S2 & 2022 & LUR & [5, 10) & 2.071 & PUGV & \textbf{0.004} & 0.009 & 0.959 & 0.367 \\ 
  S2 & 2022 & LUR & [5, 10) & 2.071 & MRP & -0.069 & 0.010 & 0.823 & 0.275 \\ 
  S2 & 2022 & LUR & [5, 10) & 2.071 & Na\"ive & 0.050 & 0.013 & 0.939 & 0.426 \\ 
    \hline S2 & 2022 & LUR & [10, 15) & 4.586 & BHAV & -0.052 & 0.022 & 0.914 & 0.539 \\ 
  S2 & 2022 & LUR & [10, 15) & 4.584 & PUGV & \textbf{-0.022} & 0.022 & 0.959 & 0.578 \\ 
  S2 & 2022 & LUR & [10, 15) & 4.587 & MRP & -0.188 & 0.054 & 0.694 & 0.501 \\ 
  S2 & 2022 & LUR & [10, 15) & 4.587 & Na\"ive & 0.066 & 0.024 & 0.952 & 0.621 \\ 
    \hline S2 & 2022 & LUR & [15, 18) & 5.771 & BHAV & -0.141 & 0.061 & 0.864 & 0.728 \\ 
  S2 & 2022 & LUR & [15, 18) & 5.771 & PUGV & -0.076 & 0.039 & 0.939 & 0.754 \\ 
  S2 & 2022 & LUR & [15, 18) & 5.774 & MRP & 0.125 & 0.045 & 0.912 & 0.693 \\ 
  S2 & 2022 & LUR & [15, 18) & 5.774 & Na\"ive & \textbf{-0.066} & 0.040 & 0.946 & 0.842 \\ 
    \hline S2 & 2018 & PED & [0, 5) & 0.734 & BHAV & -0.038 & 0.003 & 0.864 & 0.173 \\ 
  S2 & 2018 & PED & [0, 5) & 0.736 & PUGV & \textbf{-0.020} & 0.003 & 0.898 & 0.177 \\ 
  S2 & 2018 & PED & [0, 5) & 0.733 & MRP & -0.147 & 0.025 & 0.054 & 0.125 \\ 
  S2 & 2018 & PED & [0, 5) & 0.733 & Na\"ive & -0.064 & 0.006 & 0.918 & 0.272 \\ 
    \hline S2 & 2018 & PED & [5, 10) & 2.334 & BHAV & \textbf{-0.031} & 0.008 & 0.943 & 0.324 \\ 
  S2 & 2018 & PED & [5, 10) & 2.334 & PUGV & 0.037 & 0.009 & 0.925 & 0.336 \\ 
  S2 & 2018 & PED & [5, 10) & 2.331 & MRP & 0.155 & 0.037 & 0.551 & 0.350 \\ 
  S2 & 2018 & PED & [5, 10) & 2.331 & Na\"ive & -0.141 & 0.027 & 0.918 & 0.525 \\ 
    \hline S2 & 2018 & PED & [10, 15) & 4.825 & BHAV & \textbf{0.052} & 0.016 & 0.950 & 0.495 \\ 
  S2 & 2018 & PED & [10, 15) & 4.819 & PUGV & 0.086 & 0.022 & 0.878 & 0.498 \\ 
  S2 & 2018 & PED & [10, 15) & 4.815 & MRP & 0.338 & 0.131 & 0.395 & 0.597 \\ 
  S2 & 2018 & PED & [10, 15) & 4.815 & Na\"ive & -0.161 & 0.039 & 0.946 & 0.744 \\ 
    \hline S2 & 2018 & PED & [15, 18) & 7.137 & BHAV & 0.100 & 0.049 & 0.936 & 0.780 \\ 
  S2 & 2018 & PED & [15, 18) & 7.143 & PUGV & \textbf{0.052} & 0.043 & 0.952 & 0.767 \\ 
  S2 & 2018 & PED & [15, 18) & 7.138 & MRP & -0.418 & 0.224 & 0.490 & 0.797 \\ 
  S2 & 2018 & PED & [15, 18) & 7.138 & Na\"ive & -0.185 & 0.068 & 0.980 & 1.129 \\ 
    \hline S2 & 2022 & PED & [0, 5) & 0.739 & BHAV & 0.015 & 0.002 & 0.907 & 0.173 \\ 
  S2 & 2022 & PED & [0, 5) & 0.738 & PUGV & -0.017 & 0.002 & 0.918 & 0.158 \\ 
  S2 & 2022 & PED & [0, 5) & 0.739 & MRP & -0.214 & 0.048 & 0.000 & 0.112 \\ 
  S2 & 2022 & PED & [0, 5) & 0.739 & Na\"ive & \textbf{-0.004} & 0.002 & 1.000 & 0.288 \\ 
    \hline S2 & 2022 & PED & [5, 10) & 2.272 & BHAV & 0.029 & 0.006 & 0.921 & 0.278 \\ 
  S2 & 2022 & PED & [5, 10) & 2.273 & PUGV & 0.037 & 0.006 & 0.925 & 0.279 \\ 
  S2 & 2022 & PED & [5, 10) & 2.274 & MRP & -0.098 & 0.019 & 0.694 & 0.301 \\ 
  S2 & 2022 & PED & [5, 10) & 2.274 & Na\"ive & \textbf{0.005} & 0.005 & 1.000 & 0.501 \\ 
    \hline S2 & 2022 & PED & [10, 15) & 4.378 & BHAV & \textbf{0.014} & 0.011 & 0.914 & 0.388 \\ 
  S2 & 2022 & PED & [10, 15) & 4.383 & PUGV & 0.058 & 0.014 & 0.884 & 0.398 \\ 
  S2 & 2022 & PED & [10, 15) & 4.384 & MRP & 0.346 & 0.136 & 0.286 & 0.542 \\ 
  S2 & 2022 & PED & [10, 15) & 4.384 & Na\"ive & -0.080 & 0.017 & 0.993 & 0.690 \\ 
    \hline S2 & 2022 & PED & [15, 18) & 6.113 & BHAV & -0.057 & 0.030 & 0.929 & 0.628 \\ 
  S2 & 2022 & PED & [15, 18) & 6.119 & PUGV & 0.063 & 0.032 & 0.932 & 0.651 \\ 
  S2 & 2022 & PED & [15, 18) & 6.121 & MRP & \textbf{0.006} & 0.033 & 0.952 & 0.731 \\ 
  S2 & 2022 & PED & [15, 18) & 6.121 & Na\"ive & -0.280 & 0.105 & 0.905 & 1.018 \\ 
    \hline S2 & 2018 & UFL & [0, 5) & 0.605 & BHAV & \textbf{-0.033} & 0.006 & 0.893 & 0.272 \\ 
  S2 & 2018 & UFL & [0, 5) & 0.604 & PUGV & 0.075 & 0.010 & 0.857 & 0.301 \\ 
  S2 & 2018 & UFL & [0, 5) & 0.603 & MRP & 0.180 & 0.037 & 0.068 & 0.196 \\ 
  S2 & 2018 & UFL & [0, 5) & 0.603 & Na\"ive & -0.205 & 0.047 & 0.218 & 0.282 \\ 
    \hline S2 & 2018 & UFL & [5, 10) & 3.030 & BHAV & -0.081 & 0.055 & 0.914 & 0.837 \\ 
  S2 & 2018 & UFL & [5, 10) & 3.031 & PUGV & \textbf{0.019} & 0.042 & 0.966 & 0.878 \\ 
  S2 & 2018 & UFL & [5, 10) & 3.030 & MRP & 0.115 & 0.047 & 0.857 & 0.629 \\ 
  S2 & 2018 & UFL & [5, 10) & 3.030 & Na\"ive & -0.831 & 0.725 & 0.007 & 0.760 \\ 
    \hline S2 & 2018 & UFL & [10, 15) & 7.220 & BHAV & \textbf{0.016} & 0.194 & 0.929 & 1.551 \\ 
  S2 & 2018 & UFL & [10, 15) & 7.219 & PUGV & -0.103 & 0.184 & 0.952 & 1.580 \\ 
  S2 & 2018 & UFL & [10, 15) & 7.222 & MRP & -0.392 & 0.281 & 0.694 & 1.276 \\ 
  S2 & 2018 & UFL & [10, 15) & 7.222 & Na\"ive & -1.819 & 3.401 & 0.000 & 1.245 \\ 
    \hline S2 & 2018 & UFL & [15, 18) & 9.194 & BHAV & 0.274 & 0.392 & 0.893 & 2.148 \\ 
  S2 & 2018 & UFL & [15, 18) & 9.186 & PUGV & -0.163 & 0.276 & 0.939 & 2.049 \\ 
  S2 & 2018 & UFL & [15, 18) & 9.192 & MRP & \textbf{-0.085} & 0.228 & 0.939 & 1.718 \\ 
  S2 & 2018 & UFL & [15, 18) & 9.192 & Na\"ive & -1.931 & 3.917 & 0.020 & 1.808 \\ 
    \hline S2 & 2022 & UFL & [0, 5) & 0.668 & BHAV & \textbf{0.041} & 0.007 & 0.936 & 0.302 \\ 
  S2 & 2022 & UFL & [0, 5) & 0.669 & PUGV & 0.058 & 0.007 & 0.959 & 0.301 \\ 
  S2 & 2022 & UFL & [0, 5) & 0.669 & MRP & 0.127 & 0.020 & 0.293 & 0.196 \\ 
  S2 & 2022 & UFL & [0, 5) & 0.669 & Na\"ive & -0.092 & 0.012 & 0.796 & 0.320 \\ 
    \hline S2 & 2022 & UFL & [5, 10) & 3.163 & BHAV & \textbf{-0.014} & 0.039 & 0.979 & 0.861 \\ 
  S2 & 2022 & UFL & [5, 10) & 3.163 & PUGV & -0.041 & 0.040 & 0.959 & 0.886 \\ 
  S2 & 2022 & UFL & [5, 10) & 3.160 & MRP & 0.069 & 0.035 & 0.884 & 0.626 \\ 
  S2 & 2022 & UFL & [5, 10) & 3.160 & Na\"ive & -0.988 & 1.004 & 0.000 & 0.703 \\ 
    \hline S2 & 2022 & UFL & [10, 15) & 7.514 & BHAV & -0.286 & 0.228 & 0.886 & 1.486 \\ 
  S2 & 2022 & UFL & [10, 15) & 7.510 & PUGV & \textbf{-0.244} & 0.208 & 0.898 & 1.564 \\ 
  S2 & 2022 & UFL & [10, 15) & 7.511 & MRP & -0.521 & 0.388 & 0.626 & 1.265 \\ 
  S2 & 2022 & UFL & [10, 15) & 7.511 & Na\"ive & -2.184 & 4.839 & 0.000 & 1.111 \\ 
    \hline S2 & 2022 & UFL & [15, 18) & 9.424 & BHAV & -0.447 & 0.421 & 0.857 & 1.929 \\ 
  S2 & 2022 & UFL & [15, 18) & 9.425 & PUGV & -0.346 & 0.319 & 0.912 & 1.973 \\ 
  S2 & 2022 & UFL & [15, 18) & 9.419 & MRP & \textbf{-0.159} & 0.207 & 0.939 & 1.680 \\ 
  S2 & 2022 & UFL & [15, 18) & 9.419 & Na\"ive & -2.552 & 6.652 & 0.000 & 1.554 \\ 
    \hline  \hline
\end{tabular}
\endgroup
\end{table}

\clearpage

\vspace{-0.1cm}

\begin{table}[htbp!]
\centering
\caption{\tiny Plasmode simulation results for estimation of T1D prevalence (T1D cases per 1,000 persons) by year, health system center, and race and ethnicity for scenario S1 under the \textbf{BHAV}, \textbf{PUGV}, \textbf{MRP}, and \textbf{Na\"ive} methods. BHAV, PUGV, and MRP refer to Bayesian Hierarchical Analysis of Variance, Partitioning Unobserved Geographic Variations, and Multilevel Regression and Poststratification, respectively. LUR refers to Lurie Children's Hospital; PED, PEDSNet; and UFL, OneFlorida$+$. The method with the least bias is in bold. Coverage refers to the 95\% coverage probability. Length refers to the average length of the 95\% uncertainty interval.} 
\label{t:plasmode_simulation_year_site_race_S1}
\begingroup\tiny
\begin{tabular}{cccccrrrrr}
  \hline
  \multicolumn{1}{c}{Scenario} & \multicolumn{1}{c}{Year} & \multicolumn{1}{c}{Center} & \multicolumn{1}{c}{Race and Ethnicity} & \multicolumn{1}{c}{Truth} & \multicolumn{1}{c}{Method}& \multicolumn{1}{c}{Ave. Bias} & \multicolumn{1}{c}{MSE} & \multicolumn{1}{c}{Coverage} & \multicolumn{1}{c}{Ave. Length} \\
 \hline
S1 & 2018 & LUR & AS & 1.726 & BHAV & 0.124 & 0.052 & 0.899 & 0.728 \\ 
  S1 & 2018 & LUR & AS & 1.726 & PUGV & 0.148 & 0.057 & 0.850 & 0.675 \\ 
  S1 & 2018 & LUR & AS & 1.727 & MRP & \textbf{0.013} & 0.022 & 0.884 & 0.470 \\ 
  S1 & 2018 & LUR & AS & 1.727 & Na\"ive & -0.400 & 0.193 & 0.374 & 0.718 \\ 
    \hline S1 & 2018 & LUR & BA & 2.959 & BHAV & -0.089 & 0.026 & 0.879 & 0.532 \\ 
  S1 & 2018 & LUR & BA & 2.957 & PUGV & \textbf{-0.001} & 0.016 & 0.952 & 0.541 \\ 
  S1 & 2018 & LUR & BA & 2.956 & MRP & 0.038 & 0.017 & 0.918 & 0.431 \\ 
  S1 & 2018 & LUR & BA & 2.956 & Na\"ive & -0.231 & 0.074 & 0.701 & 0.625 \\ 
    \hline S1 & 2018 & LUR & HP & 2.748 & BHAV & \textbf{0.022} & 0.012 & 0.940 & 0.442 \\ 
  S1 & 2018 & LUR & HP & 2.747 & PUGV & -0.144 & 0.032 & 0.789 & 0.436 \\ 
  S1 & 2018 & LUR & HP & 2.748 & MRP & -0.024 & 0.010 & 0.946 & 0.380 \\ 
  S1 & 2018 & LUR & HP & 2.748 & Na\"ive & -0.188 & 0.046 & 0.626 & 0.460 \\ 
    \hline S1 & 2018 & LUR & WH & 4.209 & BHAV & 0.028 & 0.018 & 0.946 & 0.509 \\ 
  S1 & 2018 & LUR & WH & 4.210 & PUGV & 0.153 & 0.046 & 0.796 & 0.550 \\ 
  S1 & 2018 & LUR & WH & 4.209 & MRP & \textbf{0.017} & 0.017 & 0.966 & 0.505 \\ 
  S1 & 2018 & LUR & WH & 4.209 & Na\"ive & -0.530 & 0.292 & 0.007 & 0.481 \\ 
    \hline S1 & 2022 & LUR & AS & 1.687 & BHAV & 0.159 & 0.059 & 0.805 & 0.647 \\ 
  S1 & 2022 & LUR & AS & 1.685 & PUGV & 0.125 & 0.044 & 0.830 & 0.629 \\ 
  S1 & 2022 & LUR & AS & 1.687 & MRP & \textbf{-0.021} & 0.018 & 0.912 & 0.444 \\ 
  S1 & 2022 & LUR & AS & 1.687 & Na\"ive & -0.169 & 0.061 & 0.776 & 0.687 \\ 
    \hline S1 & 2022 & LUR & BA & 2.806 & BHAV & 0.084 & 0.020 & 0.933 & 0.476 \\ 
  S1 & 2022 & LUR & BA & 2.807 & PUGV & -0.018 & 0.011 & 0.980 & 0.474 \\ 
  S1 & 2022 & LUR & BA & 2.808 & MRP & -0.040 & 0.014 & 0.891 & 0.383 \\ 
  S1 & 2022 & LUR & BA & 2.808 & Na\"ive & \textbf{0.011} & 0.014 & 0.986 & 0.571 \\ 
    \hline S1 & 2022 & LUR & HP & 2.602 & BHAV & -0.069 & 0.012 & 0.899 & 0.360 \\ 
  S1 & 2022 & LUR & HP & 2.601 & PUGV & -0.171 & 0.037 & 0.612 & 0.379 \\ 
  S1 & 2022 & LUR & HP & 2.601 & MRP & \textbf{-0.025} & 0.008 & 0.932 & 0.338 \\ 
  S1 & 2022 & LUR & HP & 2.601 & Na\"ive & -0.252 & 0.070 & 0.252 & 0.386 \\ 
    \hline S1 & 2022 & LUR & WH & 3.978 & BHAV & -0.062 & 0.017 & 0.906 & 0.432 \\ 
  S1 & 2022 & LUR & WH & 3.978 & PUGV & 0.110 & 0.028 & 0.830 & 0.479 \\ 
  S1 & 2022 & LUR & WH & 3.980 & MRP & \textbf{-0.024} & 0.012 & 0.952 & 0.440 \\ 
  S1 & 2022 & LUR & WH & 3.980 & Na\"ive & -0.383 & 0.157 & 0.075 & 0.437 \\ 
    \hline S1 & 2018 & PED & AS & 1.597 & BHAV & \textbf{-0.101} & 0.038 & 0.893 & 0.622 \\ 
  S1 & 2018 & PED & AS & 1.600 & PUGV & -0.127 & 0.040 & 0.871 & 0.581 \\ 
  S1 & 2018 & PED & AS & 1.594 & MRP & 0.192 & 0.063 & 0.592 & 0.481 \\ 
  S1 & 2018 & PED & AS & 1.594 & Na\"ive & -0.437 & 0.216 & 0.361 & 0.717 \\ 
    \hline S1 & 2018 & PED & BA & 2.552 & BHAV & 0.031 & 0.007 & 0.960 & 0.358 \\ 
  S1 & 2018 & PED & BA & 2.554 & PUGV & 0.085 & 0.015 & 0.850 & 0.376 \\ 
  S1 & 2018 & PED & BA & 2.548 & MRP & \textbf{-0.001} & 0.014 & 0.864 & 0.367 \\ 
  S1 & 2018 & PED & BA & 2.548 & Na\"ive & -0.192 & 0.043 & 0.850 & 0.579 \\ 
    \hline S1 & 2018 & PED & HP & 2.441 & BHAV & \textbf{-0.001} & 0.020 & 0.899 & 0.537 \\ 
  S1 & 2018 & PED & HP & 2.440 & PUGV & 0.097 & 0.033 & 0.905 & 0.578 \\ 
  S1 & 2018 & PED & HP & 2.438 & MRP & 0.164 & 0.048 & 0.537 & 0.386 \\ 
  S1 & 2018 & PED & HP & 2.438 & Na\"ive & -0.189 & 0.058 & 0.898 & 0.803 \\ 
    \hline S1 & 2018 & PED & WH & 4.318 & BHAV & 0.077 & 0.013 & 0.839 & 0.336 \\ 
  S1 & 2018 & PED & WH & 4.319 & PUGV & 0.077 & 0.012 & 0.864 & 0.338 \\ 
  S1 & 2018 & PED & WH & 4.316 & MRP & \textbf{-0.014} & 0.008 & 1.000 & 0.456 \\ 
  S1 & 2018 & PED & WH & 4.316 & Na\"ive & -0.249 & 0.067 & 0.361 & 0.475 \\ 
    \hline S1 & 2022 & PED & AS & 1.504 & BHAV & \textbf{-0.093} & 0.026 & 0.872 & 0.492 \\ 
  S1 & 2022 & PED & AS & 1.502 & PUGV & -0.139 & 0.033 & 0.762 & 0.462 \\ 
  S1 & 2022 & PED & AS & 1.506 & MRP & 0.101 & 0.030 & 0.762 & 0.426 \\ 
  S1 & 2022 & PED & AS & 1.506 & Na\"ive & -0.263 & 0.085 & 0.633 & 0.646 \\ 
    \hline S1 & 2022 & PED & BA & 2.546 & BHAV & 0.070 & 0.012 & 0.866 & 0.342 \\ 
  S1 & 2022 & PED & BA & 2.547 & PUGV & \textbf{0.056} & 0.011 & 0.891 & 0.350 \\ 
  S1 & 2022 & PED & BA & 2.548 & MRP & -0.184 & 0.047 & 0.456 & 0.345 \\ 
  S1 & 2022 & PED & BA & 2.548 & Na\"ive & -0.110 & 0.020 & 0.980 & 0.600 \\ 
    \hline S1 & 2022 & PED & HP & 2.426 & BHAV & -0.054 & 0.015 & 0.926 & 0.420 \\ 
  S1 & 2022 & PED & HP & 2.424 & PUGV & 0.083 & 0.022 & 0.891 & 0.479 \\ 
  S1 & 2022 & PED & HP & 2.427 & MRP & \textbf{-0.028} & 0.021 & 0.762 & 0.349 \\ 
  S1 & 2022 & PED & HP & 2.427 & Na\"ive & \textbf{-0.028} & 0.015 & 1.000 & 0.753 \\ 
    \hline S1 & 2022 & PED & WH & 3.827 & BHAV & -0.017 & 0.005 & 0.953 & 0.263 \\ 
  S1 & 2022 & PED & WH & 3.827 & PUGV & \textbf{0.011} & 0.005 & 0.939 & 0.268 \\ 
  S1 & 2022 & PED & WH & 3.826 & MRP & 0.016 & 0.007 & 0.973 & 0.407 \\ 
  S1 & 2022 & PED & WH & 3.826 & Na\"ive & -0.151 & 0.027 & 0.878 & 0.444 \\ 
    \hline S1 & 2018 & UFL & AS & 2.374 & BHAV & 0.529 & 0.708 & 0.872 & 2.320 \\ 
  S1 & 2018 & UFL & AS & 2.373 & PUGV & \textbf{0.132} & 0.176 & 0.973 & 1.736 \\ 
  S1 & 2018 & UFL & AS & 2.376 & MRP & -0.153 & 0.102 & 0.701 & 0.696 \\ 
  S1 & 2018 & UFL & AS & 2.376 & Na\"ive & -0.156 & 0.631 & 0.912 & 3.188 \\ 
    \hline S1 & 2018 & UFL & BA & 3.450 & BHAV & -0.464 & 0.264 & 0.443 & 0.860 \\ 
  S1 & 2018 & UFL & BA & 3.451 & PUGV & -0.359 & 0.169 & 0.646 & 0.875 \\ 
  S1 & 2018 & UFL & BA & 3.450 & MRP & \textbf{-0.033} & 0.048 & 0.891 & 0.678 \\ 
  S1 & 2018 & UFL & BA & 3.450 & Na\"ive & -1.390 & 1.973 & 0.000 & 0.759 \\ 
    \hline S1 & 2018 & UFL & HP & 3.489 & BHAV & \textbf{-0.008} & 0.047 & 0.953 & 0.817 \\ 
  S1 & 2018 & UFL & HP & 3.491 & PUGV & -0.228 & 0.088 & 0.796 & 0.792 \\ 
  S1 & 2018 & UFL & HP & 3.491 & MRP & -0.193 & 0.062 & 0.735 & 0.613 \\ 
  S1 & 2018 & UFL & HP & 3.491 & Na\"ive & -1.202 & 1.472 & 0.000 & 0.625 \\ 
    \hline S1 & 2018 & UFL & WH & 6.707 & BHAV & \textbf{-0.239} & 0.201 & 0.846 & 1.326 \\ 
  S1 & 2018 & UFL & WH & 6.708 & PUGV & -0.243 & 0.191 & 0.905 & 1.360 \\ 
  S1 & 2018 & UFL & WH & 6.710 & MRP & -0.440 & 0.308 & 0.667 & 1.229 \\ 
  S1 & 2018 & UFL & WH & 6.710 & Na\"ive & -1.753 & 3.129 & 0.000 & 1.012 \\ 
    \hline S1 & 2022 & UFL & AS & 2.321 & BHAV & 0.766 & 0.960 & 0.752 & 2.324 \\ 
  S1 & 2022 & UFL & AS & 2.320 & PUGV & 0.312 & 0.261 & 0.905 & 1.748 \\ 
  S1 & 2022 & UFL & AS & 2.322 & MRP & \textbf{-0.008} & 0.072 & 0.816 & 0.718 \\ 
  S1 & 2022 & UFL & AS & 2.322 & Na\"ive & 0.634 & 1.070 & 0.952 & 3.395 \\ 
    \hline S1 & 2022 & UFL & BA & 3.523 & BHAV & \textbf{-0.054} & 0.066 & 0.933 & 0.918 \\ 
  S1 & 2022 & UFL & BA & 3.523 & PUGV & -0.148 & 0.061 & 0.918 & 0.888 \\ 
  S1 & 2022 & UFL & BA & 3.523 & MRP & 0.109 & 0.049 & 0.871 & 0.693 \\ 
  S1 & 2022 & UFL & BA & 3.523 & Na\"ive & -0.740 & 0.589 & 0.054 & 0.849 \\ 
    \hline S1 & 2022 & UFL & HP & 3.424 & BHAV & 0.041 & 0.031 & 0.953 & 0.763 \\ 
  S1 & 2022 & UFL & HP & 3.424 & PUGV & -0.052 & 0.037 & 0.939 & 0.785 \\ 
  S1 & 2022 & UFL & HP & 3.423 & MRP & \textbf{0.002} & 0.023 & 0.946 & 0.611 \\ 
  S1 & 2022 & UFL & HP & 3.423 & Na\"ive & -1.091 & 1.208 & 0.000 & 0.559 \\ 
    \hline S1 & 2022 & UFL & WH & 6.474 & BHAV & -0.065 & 0.090 & 0.973 & 1.285 \\ 
  S1 & 2022 & UFL & WH & 6.476 & PUGV & 0.146 & 0.123 & 0.946 & 1.369 \\ 
  S1 & 2022 & UFL & WH & 6.473 & MRP & \textbf{0.017} & 0.093 & 0.952 & 1.242 \\ 
  S1 & 2022 & UFL & WH & 6.473 & Na\"ive & -1.755 & 3.123 & 0.000 & 0.919 \\ 
    \hline  \hline
\end{tabular}
\endgroup
\end{table}

\clearpage

\begin{table}[htbp!]
\centering
\caption{\tiny Plasmode simulation results for estimation of T1D prevalence (T1D cases per 1,000 persons) by year, health system center, and race and ethnicity for scenario S2 under the \textbf{BHAV}, \textbf{PUGV}, \textbf{MRP}, and \textbf{Na\"ive} methods. BHAV, PUGV, and MRP refer to Bayesian Hierarchical Analysis of Variance, Partitioning Unobserved Geographic Variations, and Multilevel Regression and Poststratification, respectively. LUR refers to Lurie Children's Hospital; PED, PEDSNet; and UFL, OneFlorida$+$. The method with the least bias is in bold. The method with the least bias is in bold. Coverage refers to the 95\% coverage probability. Length refers to the average length of the 95\% uncertainty interval.} 
\label{t:plasmode_simulation_year_site_race_S2}
\begingroup\tiny
\begin{tabular}{cccccrrrrr}
  \hline
  \multicolumn{1}{c}{Scenario} & \multicolumn{1}{c}{Year} & \multicolumn{1}{c}{Center} & \multicolumn{1}{c}{Race and Ethnicity} & \multicolumn{1}{c}{Truth} & \multicolumn{1}{c}{Method}& \multicolumn{1}{c}{Ave. Bias} & \multicolumn{1}{c}{MSE} & \multicolumn{1}{c}{Coverage} & \multicolumn{1}{c}{Ave. Length} \\
 \hline
S2 & 2018 & LUR & AS & 1.918 & BHAV & -0.071 & 0.043 & 0.907 & 0.725 \\ 
  S2 & 2018 & LUR & AS & 1.918 & PUGV & \textbf{-0.047} & 0.038 & 0.905 & 0.675 \\ 
  S2 & 2018 & LUR & AS & 1.918 & MRP & -0.178 & 0.055 & 0.639 & 0.470 \\ 
  S2 & 2018 & LUR & AS & 1.918 & Na\"ive & -0.591 & 0.384 & 0.095 & 0.718 \\ 
    \hline S2 & 2018 & LUR & BA & 2.979 & BHAV & -0.108 & 0.030 & 0.857 & 0.532 \\ 
  S2 & 2018 & LUR & BA & 2.979 & PUGV & -0.022 & 0.017 & 0.959 & 0.543 \\ 
  S2 & 2018 & LUR & BA & 2.978 & MRP & \textbf{0.017} & 0.016 & 0.918 & 0.431 \\ 
  S2 & 2018 & LUR & BA & 2.978 & Na\"ive & -0.252 & 0.085 & 0.646 & 0.625 \\ 
    \hline S2 & 2018 & LUR & HP & 2.751 & BHAV & \textbf{0.019} & 0.012 & 0.936 & 0.442 \\ 
  S2 & 2018 & LUR & HP & 2.749 & PUGV & -0.145 & 0.032 & 0.789 & 0.435 \\ 
  S2 & 2018 & LUR & HP & 2.751 & MRP & -0.027 & 0.011 & 0.939 & 0.380 \\ 
  S2 & 2018 & LUR & HP & 2.751 & Na\"ive & -0.191 & 0.048 & 0.612 & 0.460 \\ 
    \hline S2 & 2018 & LUR & WH & 4.199 & BHAV & 0.037 & 0.018 & 0.943 & 0.509 \\ 
  S2 & 2018 & LUR & WH & 4.199 & PUGV & 0.165 & 0.050 & 0.769 & 0.552 \\ 
  S2 & 2018 & LUR & WH & 4.199 & MRP & \textbf{0.027} & 0.017 & 0.959 & 0.505 \\ 
  S2 & 2018 & LUR & WH & 4.199 & Na\"ive & -0.520 & 0.281 & 0.007 & 0.481 \\ 
    \hline S2 & 2022 & LUR & AS & 1.862 & BHAV & \textbf{-0.014} & 0.034 & 0.929 & 0.647 \\ 
  S2 & 2022 & LUR & AS & 1.857 & PUGV & -0.050 & 0.032 & 0.932 & 0.628 \\ 
  S2 & 2022 & LUR & AS & 1.860 & MRP & -0.194 & 0.056 & 0.599 & 0.444 \\ 
  S2 & 2022 & LUR & AS & 1.860 & Na\"ive & -0.342 & 0.150 & 0.531 & 0.687 \\ 
    \hline S2 & 2022 & LUR & BA & 2.809 & BHAV & 0.077 & 0.019 & 0.936 & 0.475 \\ 
  S2 & 2022 & LUR & BA & 2.814 & PUGV & -0.025 & 0.012 & 0.973 & 0.475 \\ 
  S2 & 2022 & LUR & BA & 2.814 & MRP & -0.047 & 0.014 & 0.891 & 0.383 \\ 
  S2 & 2022 & LUR & BA & 2.814 & Na\"ive & \textbf{0.005} & 0.014 & 0.986 & 0.571 \\ 
    \hline S2 & 2022 & LUR & HP & 2.591 & BHAV & -0.055 & 0.010 & 0.929 & 0.360 \\ 
  S2 & 2022 & LUR & HP & 2.591 & PUGV & -0.161 & 0.033 & 0.653 & 0.379 \\ 
  S2 & 2022 & LUR & HP & 2.592 & MRP & \textbf{-0.016} & 0.008 & 0.952 & 0.338 \\ 
  S2 & 2022 & LUR & HP & 2.592 & Na\"ive & -0.243 & 0.066 & 0.293 & 0.386 \\ 
    \hline S2 & 2022 & LUR & WH & 3.958 & BHAV & -0.036 & 0.014 & 0.943 & 0.432 \\ 
  S2 & 2022 & LUR & WH & 3.957 & PUGV & 0.131 & 0.033 & 0.810 & 0.480 \\ 
  S2 & 2022 & LUR & WH & 3.958 & MRP & \textbf{-0.003} & 0.012 & 0.966 & 0.440 \\ 
  S2 & 2022 & LUR & WH & 3.958 & Na\"ive & -0.361 & 0.141 & 0.102 & 0.437 \\ 
    \hline S2 & 2018 & PED & AS & 1.421 & BHAV & 0.072 & 0.030 & 0.929 & 0.621 \\ 
  S2 & 2018 & PED & AS & 1.427 & PUGV & \textbf{0.047} & 0.024 & 0.939 & 0.579 \\ 
  S2 & 2018 & PED & AS & 1.424 & MRP & 0.362 & 0.156 & 0.163 & 0.481 \\ 
  S2 & 2018 & PED & AS & 1.424 & Na\"ive & -0.267 & 0.095 & 0.741 & 0.717 \\ 
    \hline S2 & 2018 & PED & BA & 2.629 & BHAV & -0.041 & 0.008 & 0.950 & 0.358 \\ 
  S2 & 2018 & PED & BA & 2.629 & PUGV & \textbf{0.010} & 0.008 & 0.973 & 0.377 \\ 
  S2 & 2018 & PED & BA & 2.623 & MRP & -0.075 & 0.020 & 0.796 & 0.367 \\ 
  S2 & 2018 & PED & BA & 2.623 & Na\"ive & -0.267 & 0.077 & 0.633 & 0.579 \\ 
    \hline S2 & 2018 & PED & HP & 2.399 & BHAV & \textbf{0.044} & 0.021 & 0.921 & 0.537 \\ 
  S2 & 2018 & PED & HP & 2.400 & PUGV & 0.141 & 0.043 & 0.857 & 0.577 \\ 
  S2 & 2018 & PED & HP & 2.398 & MRP & 0.204 & 0.063 & 0.463 & 0.386 \\ 
  S2 & 2018 & PED & HP & 2.398 & Na\"ive & -0.149 & 0.043 & 0.932 & 0.803 \\ 
    \hline S2 & 2018 & PED & WH & 4.378 & BHAV & \textbf{0.016} & 0.006 & 0.964 & 0.336 \\ 
  S2 & 2018 & PED & WH & 4.377 & PUGV & 0.020 & 0.007 & 0.952 & 0.338 \\ 
  S2 & 2018 & PED & WH & 4.374 & MRP & -0.072 & 0.013 & 0.973 & 0.456 \\ 
  S2 & 2018 & PED & WH & 4.374 & Na\"ive & -0.307 & 0.100 & 0.197 & 0.475 \\ 
    \hline S2 & 2022 & PED & AS & 1.334 & BHAV & 0.076 & 0.021 & 0.893 & 0.491 \\ 
  S2 & 2022 & PED & AS & 1.338 & PUGV & \textbf{0.028} & 0.012 & 0.966 & 0.461 \\ 
  S2 & 2022 & PED & AS & 1.338 & MRP & 0.269 & 0.089 & 0.245 & 0.426 \\ 
  S2 & 2022 & PED & AS & 1.338 & Na\"ive & -0.094 & 0.022 & 0.959 & 0.646 \\ 
    \hline S2 & 2022 & PED & BA & 2.580 & BHAV & 0.030 & 0.008 & 0.936 & 0.342 \\ 
  S2 & 2022 & PED & BA & 2.584 & PUGV & \textbf{0.017} & 0.008 & 0.952 & 0.349 \\ 
  S2 & 2022 & PED & BA & 2.587 & MRP & -0.223 & 0.064 & 0.333 & 0.345 \\ 
  S2 & 2022 & PED & BA & 2.587 & Na\"ive & -0.149 & 0.029 & 0.952 & 0.600 \\ 
    \hline S2 & 2022 & PED & HP & 2.372 & BHAV & \textbf{-0.001} & 0.012 & 0.957 & 0.420 \\ 
  S2 & 2022 & PED & HP & 2.370 & PUGV & 0.139 & 0.034 & 0.782 & 0.478 \\ 
  S2 & 2022 & PED & HP & 2.371 & MRP & 0.028 & 0.020 & 0.728 & 0.349 \\ 
  S2 & 2022 & PED & HP & 2.371 & Na\"ive & 0.028 & 0.014 & 1.000 & 0.753 \\ 
    \hline S2 & 2022 & PED & WH & 3.823 & BHAV & -0.016 & 0.005 & 0.943 & 0.263 \\ 
  S2 & 2022 & PED & WH & 3.826 & PUGV & \textbf{0.011} & 0.005 & 0.946 & 0.268 \\ 
  S2 & 2022 & PED & WH & 3.827 & MRP & 0.016 & 0.007 & 0.973 & 0.407 \\ 
  S2 & 2022 & PED & WH & 3.827 & Na\"ive & -0.151 & 0.027 & 0.837 & 0.444 \\ 
    \hline S2 & 2018 & UFL & AS & 3.568 & BHAV & \textbf{-0.671} & 0.882 & 0.693 & 2.308 \\ 
  S2 & 2018 & UFL & AS & 3.566 & PUGV & -1.064 & 1.310 & 0.381 & 1.714 \\ 
  S2 & 2018 & UFL & AS & 3.565 & MRP & -1.342 & 1.893 & 0.000 & 0.696 \\ 
  S2 & 2018 & UFL & AS & 3.565 & Na\"ive & -1.346 & 2.425 & 0.626 & 3.188 \\ 
    \hline S2 & 2018 & UFL & BA & 3.023 & BHAV & \textbf{-0.035} & 0.048 & 0.936 & 0.858 \\ 
  S2 & 2018 & UFL & BA & 3.025 & PUGV & 0.065 & 0.045 & 0.952 & 0.873 \\ 
  S2 & 2018 & UFL & BA & 3.024 & MRP & 0.393 & 0.201 & 0.388 & 0.678 \\ 
  S2 & 2018 & UFL & BA & 3.024 & Na\"ive & -0.964 & 0.969 & 0.000 & 0.759 \\ 
    \hline S2 & 2018 & UFL & HP & 3.432 & BHAV & \textbf{0.046} & 0.048 & 0.943 & 0.814 \\ 
  S2 & 2018 & UFL & HP & 3.430 & PUGV & -0.168 & 0.066 & 0.871 & 0.796 \\ 
  S2 & 2018 & UFL & HP & 3.432 & MRP & -0.135 & 0.043 & 0.884 & 0.613 \\ 
  S2 & 2018 & UFL & HP & 3.432 & Na\"ive & -1.144 & 1.336 & 0.000 & 0.625 \\ 
    \hline S2 & 2018 & UFL & WH & 6.373 & BHAV & \textbf{0.088} & 0.155 & 0.921 & 1.323 \\ 
  S2 & 2018 & UFL & WH & 6.370 & PUGV & 0.096 & 0.143 & 0.918 & 1.351 \\ 
  S2 & 2018 & UFL & WH & 6.371 & MRP & -0.101 & 0.126 & 0.925 & 1.229 \\ 
  S2 & 2018 & UFL & WH & 6.371 & Na\"ive & -1.414 & 2.056 & 0.000 & 1.012 \\ 
    \hline S2 & 2022 & UFL & AS & 3.734 & BHAV & \textbf{-0.648} & 0.830 & 0.729 & 2.315 \\ 
  S2 & 2022 & UFL & AS & 3.740 & PUGV & -1.110 & 1.423 & 0.354 & 1.730 \\ 
  S2 & 2022 & UFL & AS & 3.740 & MRP & -1.426 & 2.129 & 0.007 & 0.718 \\ 
  S2 & 2022 & UFL & AS & 3.740 & Na\"ive & -0.784 & 1.307 & 0.782 & 3.395 \\ 
    \hline S2 & 2022 & UFL & BA & 3.352 & BHAV & 0.115 & 0.072 & 0.921 & 0.915 \\ 
  S2 & 2022 & UFL & BA & 3.348 & PUGV & \textbf{0.026} & 0.038 & 0.966 & 0.885 \\ 
  S2 & 2022 & UFL & BA & 3.348 & MRP & 0.285 & 0.117 & 0.592 & 0.693 \\ 
  S2 & 2022 & UFL & BA & 3.348 & Na\"ive & -0.565 & 0.358 & 0.286 & 0.849 \\ 
    \hline S2 & 2022 & UFL & HP & 3.623 & BHAV & \textbf{-0.160} & 0.055 & 0.886 & 0.761 \\ 
  S2 & 2022 & UFL & HP & 3.625 & PUGV & -0.253 & 0.099 & 0.810 & 0.788 \\ 
  S2 & 2022 & UFL & HP & 3.623 & MRP & -0.199 & 0.062 & 0.735 & 0.611 \\ 
  S2 & 2022 & UFL & HP & 3.623 & Na\"ive & -1.292 & 1.687 & 0.000 & 0.559 \\ 
    \hline S2 & 2022 & UFL & WH & 6.653 & BHAV & -0.239 & 0.148 & 0.907 & 1.283 \\ 
  S2 & 2022 & UFL & WH & 6.651 & PUGV & \textbf{-0.028} & 0.108 & 0.980 & 1.364 \\ 
  S2 & 2022 & UFL & WH & 6.649 & MRP & -0.158 & 0.123 & 0.939 & 1.242 \\ 
  S2 & 2022 & UFL & WH & 6.649 & Na\"ive & -1.931 & 3.772 & 0.000 & 0.919 \\ 
    \hline  \hline
\end{tabular}
\endgroup
\end{table}

\clearpage

\begin{table}[ht]
\centering
\caption{Plasmode simulation results for estimation of T1D prevalence (T1D cases per 1,000 persons) by year, health system center, and sex for scenario S1 under the \textbf{BHAV}, \textbf{PUGV}, \textbf{MRP}, and \textbf{Na\"ive} methods. BHAV, PUGV, and MRP refer to Bayesian Hierarchical Analysis of Variance, Partitioning Unobserved Geographic Variations, and Multilevel Regression and Poststratification, respectively. LUR refers to Lurie Children's Hospital; PED, PEDSNet; and UFL, OneFlorida$+$. The method with the least bias is in bold. Coverage refers to the 95\% coverage probability. Length refers to the average length of the 95\% uncertainty interval.} 
\label{t:plasmode_simulation_year_site_sex_S1}
\begingroup\tiny
\begin{tabular}{cccccrrrrr}
  \hline
  \multicolumn{1}{c}{Scenario} & \multicolumn{1}{c}{Year} & \multicolumn{1}{c}{Center} & \multicolumn{1}{c}{Sex} &  \multicolumn{1}{c}{Truth} & \multicolumn{1}{c}{Method}& \multicolumn{1}{c}{Ave. Bias} & \multicolumn{1}{c}{MSE} & \multicolumn{1}{c}{Coverage} & \multicolumn{1}{c}{Ave. Length} \\
 \hline
S1 & 2018 & LUR & F & 3.224 & BHAV & 0.019 & 0.011 & 0.926 & 0.401 \\ 
  S1 & 2018 & LUR & F & 3.223 & PUGV & \textbf{0.002} & 0.010 & 0.925 & 0.366 \\ 
  S1 & 2018 & LUR & F & 3.223 & MRP & -0.031 & 0.011 & 0.925 & 0.376 \\ 
  S1 & 2018 & LUR & F & 3.223 & Na\"ive & -0.123 & 0.024 & 0.803 & 0.414 \\ 
    \hline S1 & 2018 & LUR & M & 3.187 & BHAV & \textbf{0.013} & 0.011 & 0.940 & 0.393 \\ 
  S1 & 2018 & LUR & M & 3.186 & PUGV & 0.050 & 0.011 & 0.925 & 0.367 \\ 
  S1 & 2018 & LUR & M & 3.186 & MRP & 0.042 & 0.012 & 0.946 & 0.378 \\ 
  S1 & 2018 & LUR & M & 3.186 & Na\"ive & -0.317 & 0.107 & 0.102 & 0.384 \\ 
    \hline S1 & 2022 & LUR & F & 3.013 & BHAV & \textbf{-0.017} & 0.009 & 0.953 & 0.339 \\ 
  S1 & 2022 & LUR & F & 3.012 & PUGV & -0.024 & 0.008 & 0.918 & 0.313 \\ 
  S1 & 2022 & LUR & F & 3.012 & MRP & -0.042 & 0.009 & 0.925 & 0.326 \\ 
  S1 & 2022 & LUR & F & 3.012 & Na\"ive & -0.131 & 0.024 & 0.748 & 0.359 \\ 
    \hline S1 & 2022 & LUR & M & 3.025 & BHAV & \textbf{-0.003} & 0.008 & 0.946 & 0.332 \\ 
  S1 & 2022 & LUR & M & 3.026 & PUGV & 0.013 & 0.006 & 0.959 & 0.312 \\ 
  S1 & 2022 & LUR & M & 3.027 & MRP & -0.013 & 0.007 & 0.959 & 0.326 \\ 
  S1 & 2022 & LUR & M & 3.027 & Na\"ive & -0.141 & 0.026 & 0.673 & 0.346 \\ 
    \hline S1 & 2018 & PED & F & 3.300 & BHAV & \textbf{0.028} & 0.007 & 0.946 & 0.318 \\ 
  S1 & 2018 & PED & F & 3.302 & PUGV & 0.063 & 0.010 & 0.844 & 0.306 \\ 
  S1 & 2018 & PED & F & 3.296 & MRP & 0.113 & 0.022 & 0.776 & 0.380 \\ 
  S1 & 2018 & PED & F & 3.296 & Na\"ive & -0.182 & 0.039 & 0.728 & 0.458 \\ 
    \hline S1 & 2018 & PED & M & 3.561 & BHAV & 0.048 & 0.009 & 0.940 & 0.316 \\ 
  S1 & 2018 & PED & M & 3.562 & PUGV & 0.055 & 0.009 & 0.898 & 0.305 \\ 
  S1 & 2018 & PED & M & 3.559 & MRP & \textbf{-0.041} & 0.008 & 0.966 & 0.386 \\ 
  S1 & 2018 & PED & M & 3.559 & Na\"ive & -0.168 & 0.034 & 0.755 & 0.461 \\ 
    \hline S1 & 2022 & PED & F & 3.054 & BHAV & -0.030 & 0.005 & 0.906 & 0.254 \\ 
  S1 & 2022 & PED & F & 3.053 & PUGV & \textbf{0.008} & 0.004 & 0.946 & 0.250 \\ 
  S1 & 2022 & PED & F & 3.055 & MRP & -0.019 & 0.006 & 0.973 & 0.335 \\ 
  S1 & 2022 & PED & F & 3.055 & Na\"ive & -0.057 & 0.008 & 0.986 & 0.433 \\ 
    \hline S1 & 2022 & PED & M & 3.149 & BHAV & \textbf{-0.001} & 0.004 & 0.946 & 0.252 \\ 
  S1 & 2022 & PED & M & 3.150 & PUGV & 0.012 & 0.004 & 0.952 & 0.247 \\ 
  S1 & 2022 & PED & M & 3.150 & MRP & -0.014 & 0.006 & 0.973 & 0.344 \\ 
  S1 & 2022 & PED & M & 3.150 & Na\"ive & -0.086 & 0.011 & 0.993 & 0.429 \\ 
    \hline S1 & 2018 & UFL & F & 4.847 & BHAV & \textbf{-0.109} & 0.076 & 0.926 & 0.954 \\ 
  S1 & 2018 & UFL & F & 4.848 & PUGV & -0.234 & 0.100 & 0.789 & 0.841 \\ 
  S1 & 2018 & UFL & F & 4.850 & MRP & -0.273 & 0.123 & 0.701 & 0.844 \\ 
  S1 & 2018 & UFL & F & 4.850 & Na\"ive & -1.544 & 2.408 & 0.000 & 0.688 \\ 
    \hline S1 & 2018 & UFL & M & 4.903 & BHAV & -0.268 & 0.139 & 0.792 & 0.942 \\ 
  S1 & 2018 & UFL & M & 4.903 & PUGV & -0.265 & 0.126 & 0.755 & 0.846 \\ 
  S1 & 2018 & UFL & M & 4.903 & MRP & \textbf{-0.262} & 0.128 & 0.728 & 0.854 \\ 
  S1 & 2018 & UFL & M & 4.903 & Na\"ive & -1.953 & 3.837 & 0.000 & 0.628 \\ 
    \hline S1 & 2022 & UFL & F & 4.696 & BHAV & 0.053 & 0.050 & 0.960 & 0.915 \\ 
  S1 & 2022 & UFL & F & 4.698 & PUGV & \textbf{0.037} & 0.039 & 0.952 & 0.818 \\ 
  S1 & 2022 & UFL & F & 4.695 & MRP & 0.039 & 0.044 & 0.952 & 0.842 \\ 
  S1 & 2022 & UFL & F & 4.695 & Na\"ive & -1.409 & 2.005 & 0.000 & 0.629 \\ 
    \hline S1 & 2022 & UFL & M & 4.780 & BHAV & -0.045 & 0.047 & 0.980 & 0.906 \\ 
  S1 & 2022 & UFL & M & 4.780 & PUGV & \textbf{0.021} & 0.040 & 0.973 & 0.824 \\ 
  S1 & 2022 & UFL & M & 4.780 & MRP & 0.023 & 0.047 & 0.959 & 0.852 \\ 
  S1 & 2022 & UFL & M & 4.780 & Na\"ive & -1.626 & 2.664 & 0.000 & 0.597 \\ 
    \hline  \hline
\end{tabular}
\endgroup
\end{table}

\clearpage

\begin{table}[ht]
\centering
\caption{Plasmode simulation results for estimation of T1D prevalence (T1D cases per 1,000 persons) by year, health system center, and sex for scenario S2 under the \textbf{BHAV}, \textbf{PUGV}, \textbf{MRP}, and \textbf{Na\"ive} methods. BHAV, PUGV, and MRP refer to Bayesian Hierarchical Analysis of Variance, Partitioning Unobserved Geographic Variations, and Multilevel Regression and Poststratification, respectively. LUR refers to Lurie Children's Hospital; PED, PEDSNet; and UFL, OneFlorida$+$. The method with the least bias is in bold. Coverage refers to the 95\% coverage probability. Length refers to the average length of the 95\% uncertainty interval.} 
\label{t:plasmode_simulation_year_site_sex_S2}
\begingroup\tiny
\begin{tabular}{cccccrrrrr}
  \hline
  \multicolumn{1}{c}{Scenario} & \multicolumn{1}{c}{Year} & \multicolumn{1}{c}{Center} & \multicolumn{1}{c}{Sex} &  \multicolumn{1}{c}{Truth} & \multicolumn{1}{c}{Method}& \multicolumn{1}{c}{Ave. Bias} & \multicolumn{1}{c}{MSE} & \multicolumn{1}{c}{Coverage} & \multicolumn{1}{c}{Ave. Length} \\
 \hline
S2 & 2018 & LUR & F & 3.251 & BHAV & \textbf{-0.007} & 0.011 & 0.929 & 0.401 \\ 
  S2 & 2018 & LUR & F & 3.250 & PUGV & -0.025 & 0.011 & 0.918 & 0.367 \\ 
  S2 & 2018 & LUR & F & 3.250 & MRP & -0.058 & 0.014 & 0.905 & 0.376 \\ 
  S2 & 2018 & LUR & F & 3.250 & Na\"ive & -0.150 & 0.031 & 0.701 & 0.414 \\ 
    \hline S2 & 2018 & LUR & M & 3.207 & BHAV & \textbf{-0.009} & 0.010 & 0.950 & 0.392 \\ 
  S2 & 2018 & LUR & M & 3.206 & PUGV & 0.030 & 0.010 & 0.946 & 0.369 \\ 
  S2 & 2018 & LUR & M & 3.207 & MRP & 0.022 & 0.010 & 0.952 & 0.378 \\ 
  S2 & 2018 & LUR & M & 3.207 & Na\"ive & -0.338 & 0.121 & 0.075 & 0.384 \\ 
    \hline S2 & 2022 & LUR & F & 3.026 & BHAV & \textbf{-0.028} & 0.009 & 0.943 & 0.339 \\ 
  S2 & 2022 & LUR & F & 3.026 & PUGV & -0.039 & 0.009 & 0.918 & 0.313 \\ 
  S2 & 2022 & LUR & F & 3.027 & MRP & -0.057 & 0.010 & 0.884 & 0.326 \\ 
  S2 & 2022 & LUR & F & 3.027 & Na\"ive & -0.145 & 0.028 & 0.680 & 0.359 \\ 
    \hline S2 & 2022 & LUR & M & 3.036 & BHAV & -0.010 & 0.008 & 0.957 & 0.332 \\ 
  S2 & 2022 & LUR & M & 3.035 & PUGV & \textbf{0.004} & 0.006 & 0.973 & 0.312 \\ 
  S2 & 2022 & LUR & M & 3.037 & MRP & -0.023 & 0.007 & 0.952 & 0.326 \\ 
  S2 & 2022 & LUR & M & 3.037 & Na\"ive & -0.151 & 0.029 & 0.626 & 0.346 \\ 
    \hline S2 & 2018 & PED & F & 3.309 & BHAV & \textbf{0.021} & 0.007 & 0.929 & 0.317 \\ 
  S2 & 2018 & PED & F & 3.310 & PUGV & 0.055 & 0.009 & 0.871 & 0.306 \\ 
  S2 & 2018 & PED & F & 3.305 & MRP & 0.104 & 0.020 & 0.816 & 0.380 \\ 
  S2 & 2018 & PED & F & 3.305 & Na\"ive & -0.191 & 0.042 & 0.707 & 0.458 \\ 
    \hline S2 & 2018 & PED & M & 3.600 & BHAV & \textbf{0.008} & 0.007 & 0.964 & 0.316 \\ 
  S2 & 2018 & PED & M & 3.599 & PUGV & 0.019 & 0.006 & 0.966 & 0.305 \\ 
  S2 & 2018 & PED & M & 3.597 & MRP & -0.079 & 0.013 & 0.925 & 0.386 \\ 
  S2 & 2018 & PED & M & 3.597 & Na\"ive & -0.206 & 0.049 & 0.619 & 0.461 \\ 
    \hline S2 & 2022 & PED & F & 3.020 & BHAV & \textbf{0.001} & 0.004 & 0.943 & 0.254 \\ 
  S2 & 2022 & PED & F & 3.020 & PUGV & 0.041 & 0.006 & 0.912 & 0.250 \\ 
  S2 & 2022 & PED & F & 3.022 & MRP & 0.014 & 0.005 & 0.986 & 0.335 \\ 
  S2 & 2022 & PED & F & 3.022 & Na\"ive & -0.025 & 0.005 & 0.993 & 0.433 \\ 
    \hline S2 & 2022 & PED & M & 3.134 & BHAV & 0.011 & 0.004 & 0.950 & 0.252 \\ 
  S2 & 2022 & PED & M & 3.140 & PUGV & 0.023 & 0.004 & 0.932 & 0.247 \\ 
  S2 & 2022 & PED & M & 3.139 & MRP & \textbf{-0.003} & 0.006 & 0.980 & 0.344 \\ 
  S2 & 2022 & PED & M & 3.139 & Na\"ive & -0.075 & 0.009 & 0.993 & 0.429 \\ 
    \hline S2 & 2018 & UFL & F & 4.709 & BHAV & \textbf{0.023} & 0.065 & 0.943 & 0.951 \\ 
  S2 & 2018 & UFL & F & 4.709 & PUGV & -0.095 & 0.056 & 0.925 & 0.836 \\ 
  S2 & 2018 & UFL & F & 4.710 & MRP & -0.133 & 0.068 & 0.925 & 0.844 \\ 
  S2 & 2018 & UFL & F & 4.710 & Na\"ive & -1.404 & 1.998 & 0.000 & 0.688 \\ 
    \hline S2 & 2018 & UFL & M & 4.614 & BHAV & \textbf{0.019} & 0.069 & 0.921 & 0.940 \\ 
  S2 & 2018 & UFL & M & 4.611 & PUGV & 0.027 & 0.057 & 0.912 & 0.842 \\ 
  S2 & 2018 & UFL & M & 4.612 & MRP & 0.030 & 0.060 & 0.905 & 0.854 \\ 
  S2 & 2018 & UFL & M & 4.612 & Na\"ive & -1.661 & 2.782 & 0.000 & 0.628 \\ 
    \hline S2 & 2022 & UFL & F & 4.926 & BHAV & \textbf{-0.180} & 0.080 & 0.893 & 0.912 \\ 
  S2 & 2022 & UFL & F & 4.928 & PUGV & -0.193 & 0.078 & 0.850 & 0.816 \\ 
  S2 & 2022 & UFL & F & 4.925 & MRP & -0.191 & 0.081 & 0.850 & 0.842 \\ 
  S2 & 2022 & UFL & F & 4.925 & Na\"ive & -1.638 & 2.706 & 0.000 & 0.629 \\ 
    \hline S2 & 2022 & UFL & M & 4.872 & BHAV & -0.133 & 0.063 & 0.936 & 0.906 \\ 
  S2 & 2022 & UFL & M & 4.868 & PUGV & -0.067 & 0.046 & 0.946 & 0.821 \\ 
  S2 & 2022 & UFL & M & 4.868 & MRP & \textbf{-0.065} & 0.052 & 0.932 & 0.852 \\ 
  S2 & 2022 & UFL & M & 4.868 & Na\"ive & -1.715 & 2.961 & 0.000 & 0.597 \\ 
    \hline  \hline
\end{tabular}
\endgroup
\end{table}

\clearpage

\begin{table}[ht]
\centering
\caption{Summary of estimation performance for T1D prevalence for all estimands (except county-level) across scenarios in the plasmode simulation study. BHAV, PUGV, and MRP refer to Bayesian Hierarchical Analysis of Variance, Partitioning Unobserved Geographic Variations, and Multilevel Regression and Poststratification, respectively. LAB indicates least average bias. Coverage refers to attainment of the 95\% coverage probability. To assess attainment of the 95\% coverage probability, a threshold of greater than 0.91 was used while excluding coverage probabilities of 1. Boldfacing indicates the best performing method.} 
\label{t:summary_plasmode_simulation_scenario_method_overall}
\begingroup\normalsize
\begin{tabular}{llccc}
  \hline
  \multicolumn{1}{c}{Scenario} & \multicolumn{1}{c}{Method} & \multicolumn{1}{c}{No.} & \multicolumn{1}{c}{No. LAB}& \multicolumn{1}{c}{No. LAB \& Coverage}   \\
 \hline
S1 and S2 & BHAV & 176 &  61 & \textbf{46} \\ 
  S1 and S2 & PUGV & 176 &  51 & 43 \\ 
  S1 and S2 & MRP & 176 &  55 & 37 \\ 
  S1 and S2 & Na\"ive & 176 &   9 & 4 \\ 
    \hline S1 & BHAV &  88 &  26 & 20 \\ 
  S1 & PUGV &  88 &  19 & 15 \\ 
  S1 & MRP &  88 &  38 & \textbf{23} \\ 
  S1 & Na\"ive &  88 &   5 & 2 \\ 
    \hline S2 & BHAV &  88 &  35 & 26 \\ 
  S2 & PUGV &  88 &  32 & \textbf{28} \\ 
  S2 & MRP &  88 &  17 & 14 \\ 
  S2 & Na\"ive &  88 &   4 & 2 \\ 
    \hline  \hline
\end{tabular}
\endgroup
\end{table}

\begin{table}[ht]
\centering
\caption{Summary of estimation performance for T1D prevalence by year and by demographic group and year across scenarios in the plasmode simulation study. BHAV, PUGV, and MRP refer to Bayesian Hierarchical Analysis of Variance, Partitioning Unobserved Geographic Variations, and Multilevel Regression and Poststratification, respectively. LAB indicates least average bias. Coverage refers to attainment of the 95\% coverage probability. To assess attainment of the 95\% coverage probability, a threshold of greater than 0.91 was used while excluding coverage probabilities of 1. Boldfacing indicates the best performing method.} 
\label{t:summary_plasmode_simulation_scenario_method_year}
\begingroup\normalsize
\begin{tabular}{llccc}
  \hline
  \multicolumn{1}{c}{Scenario} & \multicolumn{1}{c}{Method} & \multicolumn{1}{c}{No.} & \multicolumn{1}{c}{No. LAB}& \multicolumn{1}{c}{No. LAB \& Coverage}   \\
 \hline
S1 and S2 & BHAV &  44 &  13 & 8 \\ 
  S1 and S2 & PUGV &  44 &  15 & \textbf{13} \\ 
  S1 and S2 & MRP &  44 &  16 & 12 \\ 
  S1 and S2 & Na\"ive &  44 &   0 & 0 \\ 
    \hline S1 & BHAV &  22 &   6 & 5 \\ 
  S1 & PUGV &  22 &   5 & 4 \\ 
  S1 & MRP &  22 &  11 & \textbf{8} \\ 
  S1 & Na\"ive &  22 &   0 & 0 \\ 
    \hline S2 & BHAV &  22 &   7 & 3 \\ 
  S2 & PUGV &  22 &  10 & \textbf{9} \\ 
  S2 & MRP &  22 &   5 & 4 \\ 
  S2 & Na\"ive &  22 &   0 & 0 \\ 
    \hline  \hline
\end{tabular}
\endgroup
\end{table}

\clearpage

\begin{landscape}
    
\begin{table}[ht]
\centering
\caption{Summary of estimation performance for T1D prevalence by year and health system center, and by year, health system center, and demographic group across scenarios in the plasmode simulation study. BHAV, PUGV, and MRP refer to Bayesian Hierarchical Analysis of Variance, Partitioning Unobserved Geographic Variations, and Multilevel Regression and Poststratification, respectively. LUR refers to Lurie Children's Hospital; PED, PEDSNet; and UFL, OneFlorida$+$. LAB indicates least average bias. Coverage refers to attainment of the 95\% coverage probability. To assess attainment of the 95\% coverage probability, a threshold of greater than 0.91 was used while excluding coverage probabilities of 1. Boldfacing indicates the best performing method.} 
\label{t:summary_plasmode_simulation_scenario_method_year_center}
\begingroup\tiny
\begin{tabular}{llcccccccccccc}
  \hline
    & & \multicolumn{3}{c}{Overall} & \multicolumn{3}{c}{LUR} & \multicolumn{3}{c}{PED} & \multicolumn{3}{c}{UFL} \\
  \multicolumn{1}{c}{Scenario} & \multicolumn{1}{c}{Method} & \multicolumn{1}{c}{No.} & \multicolumn{1}{c}{No. LAB}& \multicolumn{1}{c}{No. LAB \& Coverage} &  \multicolumn{1}{c}{No.} & \multicolumn{1}{c}{No. LAB}& \multicolumn{1}{c}{No. LAB \& Coverage} &  \multicolumn{1}{c}{No.} & \multicolumn{1}{c}{No. LAB}& \multicolumn{1}{c}{No. LAB \& Coverage} &  \multicolumn{1}{c}{No.} & \multicolumn{1}{c}{No. LAB}& \multicolumn{1}{c}{No. LAB \& Coverage}  \\
 \hline
S1 and S2 & BHAV & 132 &  48 & \textbf{38} &  44 &  10 & 10 &  44 &  16 & \textbf{13} &  44 &  22 & \textbf{15} \\ 
  S1 and S2 & PUGV & 132 &  36 & 30 &  44 &  14 & \textbf{13} &  44 &  14 & 10 &  44 &   8 & 7 \\ 
  S1 and S2 & MRP & 132 &  39 & 25 &  44 &  16 & \textbf{13} &  44 &   9 & 4 &  44 &  14 & 8 \\ 
  S1 and S2 & Na\"ive & 132 &   9 & 4 &  44 &   4 & 4 &  44 &   5 & 0 &  44 &   0 & 0 \\ 
    \hline S1 & BHAV &  66 &  20 & \textbf{15} &  22 &   5 & 5 &  22 &   7 & \textbf{4} &  22 &   8 & \textbf{6} \\ 
  S1 & PUGV &  66 &  14 & 11 &  22 &   3 & 3 &  22 &   7 & \textbf{4} &  22 &   4 & 4 \\ 
  S1 & MRP &  66 &  27 & \textbf{15} &  22 &  12 & \textbf{9} &  22 &   5 & 2 &  22 &  10 & 4 \\ 
  S1 & Na\"ive &  66 &   5 & 2 &  22 &   2 & 2 &  22 &   3 & 0 &  22 &   0 & 0 \\ 
    \hline S2 & BHAV &  66 &  28 & \textbf{23} &  22 &   5 & 5 &  22 &   9 & \textbf{9} &  22 &  14 & \textbf{9} \\ 
  S2 & PUGV &  66 &  22 & 19 &  22 &  11 & \textbf{10} &  22 &   7 & 6 &  22 &   4 & 3 \\ 
  S2 & MRP &  66 &  12 & 10 &  22 &   4 & 4 &  22 &   4 & 2 &  22 &   4 & 4 \\ 
  S2 & Na\"ive &  66 &   4 & 2 &  22 &   2 & 2 &  22 &   2 & 0 &  22 &   0 & 0 \\ 
    \hline  \hline
\end{tabular}
\endgroup
\end{table}

\end{landscape}

\clearpage

\begin{table}[ht]
\centering
\caption{Summary of estimation performance for T1D prevalence by scenario for non-Hispanic Asian, non-Hispanic Black, and Hispanic by year and health system center in PED and UFL in the plasmode simulation study. BHAV, PUGV, and MRP refer to Bayesian Hierarchical Analysis of Variance, Partitioning Unobserved Geographic Variations, and Multilevel Regression and Poststratification, respectively. PED refers to PEDSNet; and UFL, OneFlorida$+$. LAB indicates least average bias. Coverage refers to attainment of the 95\% coverage probability. To assess attainment of the 95\% coverage probability, a threshold of greater than 0.91 was used while excluding coverage probabilities of 1. Boldfacing indicates the best performing method.} 
\label{t:summary_plasmode_simulation_scenario_method_sparse_race}
\begingroup\tiny
\begin{tabular}{llccccccccc}
  \hline
  & & \multicolumn{3}{c}{Overall} & \multicolumn{3}{c}{PED} & \multicolumn{3}{c}{UFL} \\
  \multicolumn{1}{c}{Scenario} & \multicolumn{1}{c}{Method} & \multicolumn{1}{c}{No.} & \multicolumn{1}{c}{No. LAB}& \multicolumn{1}{c}{No. LAB \& Coverage}  & \multicolumn{1}{c}{No.} & \multicolumn{1}{c}{No. LAB}& \multicolumn{1}{c}{No. LAB \& Coverage} &  \multicolumn{1}{c}{No.} & \multicolumn{1}{c}{No. LAB}& \multicolumn{1}{c}{No. LAB \& Coverage} \\
 \hline
S1 and S2 & BHAV &  24 &  12 & \textbf{6} &  12 &   5 & 2 &  12 &   7 & \textbf{4} \\ 
  S1 and S2 & PUGV &  24 &   7 & \textbf{6} &  12 &   5 & \textbf{4} &  12 &   2 & 2 \\ 
  S1 and S2 & MRP &  24 &   4 & 1 &  12 &   1 & 0 &  12 &   3 & 1 \\ 
  S1 and S2 & Na\"ive &  24 &   1 & 0 &  12 &   1 & 0 &  12 &   0 & 0 \\ 
    \hline S1 & BHAV &  12 &   5 & \textbf{2} &   6 &   3 & \textbf{0} &   6 &   2 & \textbf{2} \\ 
  S1 & PUGV &  12 &   2 & 1 &   6 &   1 & \textbf{0} &   6 &   1 & 1 \\ 
  S1 & MRP &  12 &   4 & 1 &   6 &   1 & \textbf{0} &   6 &   3 & 1 \\ 
  S1 & Na\"ive &  12 &   1 & 0 &   6 &   1 & \textbf{0} &   6 &   0 & 0 \\ 
    \hline S2 & BHAV &  12 &   7 & 4 &   6 &   2 & 2 &   6 &   5 & \textbf{2} \\ 
  S2 & PUGV &  12 &   5 & \textbf{5} &   6 &   4 & \textbf{4} &   6 &   1 & 1 \\ 
  S2 & MRP &  12 &   0 & 0 &   6 &   0 & 0 &   6 &   0 & 0 \\ 
  S2 & Na\"ive &  12 &   0 & 0 &   6 &   0 & 0 &   6 &   0 & 0 \\ 
    \hline  \hline
\end{tabular}
\endgroup
\end{table}

\clearpage

\begin{landscape}

\begin{table}[ht]
\centering
\caption{Summary of estimation performance for county-level T1D prevalence by scenario and health system center in the plasmode simulation study. BHAV, PUGV, and MRP refer to Bayesian Hierarchical Analysis of Variance, Partitioning Unobserved Geographic Variations, and Multilevel Regression and Poststratification, respectively. LUR refers to Lurie Children's Hospital; PED, PEDSNet; and UFL, OneFlorida$+$. LAB indicates least average bias. Coverage refers to attainment of the 95\% coverage probability. To assess attainment of the 95\% coverage probability, a threshold of greater than 0.91 was used while excluding coverage probabilities of 1. Boldfacing indicates the best performing method.} 
\label{t:summary_plasmode_simulation_scenario_method_county_center}
\begingroup\tiny
\begin{tabular}{llcccccccccccc}
  \hline
     & & \multicolumn{3}{c}{Overall} & \multicolumn{3}{c}{LUR} & \multicolumn{3}{c}{PED} & \multicolumn{3}{c}{UFL} \\
  \multicolumn{1}{c}{Scenario} & \multicolumn{1}{c}{Method} & \multicolumn{1}{c}{No.} & \multicolumn{1}{c}{No. LAB}& \multicolumn{1}{c}{No. LAB \& Coverage} & \multicolumn{1}{c}{No.} & \multicolumn{1}{c}{No. LAB}& \multicolumn{1}{c}{No. LAB \& Coverage} & \multicolumn{1}{c}{No.} & \multicolumn{1}{c}{No. LAB}& \multicolumn{1}{c}{No. LAB \& Coverage} & \multicolumn{1}{c}{No.} & \multicolumn{1}{c}{No. LAB}& \multicolumn{1}{c}{No. LAB \& Coverage} \\
 \hline
S1 and S2 & BHAV & 540 & 223 & \textbf{176} & 144 &  55 & \textbf{50} & 128 &  53 & \textbf{39} & 268 & 115 & \textbf{87} \\ 
  S1 and S2 & PUGV & 540 &  89 & 70 & 144 &  25 & 20 & 128 &  31 & 18 & 268 &  33 & 32 \\ 
  S1 and S2 & MRP & 540 &  96 & 66 & 144 &  28 & 27 & 128 &  13 & 9 & 268 &  55 & 30 \\ 
  S1 and S2 & Naive & 540 & 132 & 57 & 144 &  36 & 24 & 128 &  31 & 11 & 268 &  65 & 22 \\ 
    \hline S1 & BHAV & 270 &  96 & \textbf{73} &  72 &  27 & \textbf{24} &  64 &  25 & \textbf{18} & 134 &  44 & \textbf{31} \\ 
  S1 & PUGV & 270 &  45 & 36 &  72 &  11 & 9 &  64 &  17 & 10 & 134 &  17 & 17 \\ 
  S1 & MRP & 270 &  48 & 31 &  72 &  15 & 14 &  64 &   6 & 4 & 134 &  27 & 13 \\ 
  S1 & Naive & 270 &  81 & 32 &  72 &  19 & 10 &  64 &  16 & 6 & 134 &  46 & 16 \\ 
    \hline S2 & BHAV & 270 & 127 & \textbf{103} &  72 &  28 & \textbf{26} &  64 &  28 & \textbf{21} & 134 &  71 & \textbf{56} \\ 
  S2 & PUGV & 270 &  44 & 34 &  72 &  14 & 11 &  64 &  14 & 8 & 134 &  16 & 15 \\ 
  S2 & MRP & 270 &  48 & 35 &  72 &  13 & 13 &  64 &   7 & 5 & 134 &  28 & 17 \\ 
  S2 & Naive & 270 &  51 & 25 &  72 &  17 & 14 &  64 &  15 & 5 & 134 &  19 & 6 \\ 
    \hline  \hline
\end{tabular}
\endgroup
\end{table}

\end{landscape}


\clearpage

\begin{figure}[htbp!]
\centering
\includegraphics[scale = 0.5]{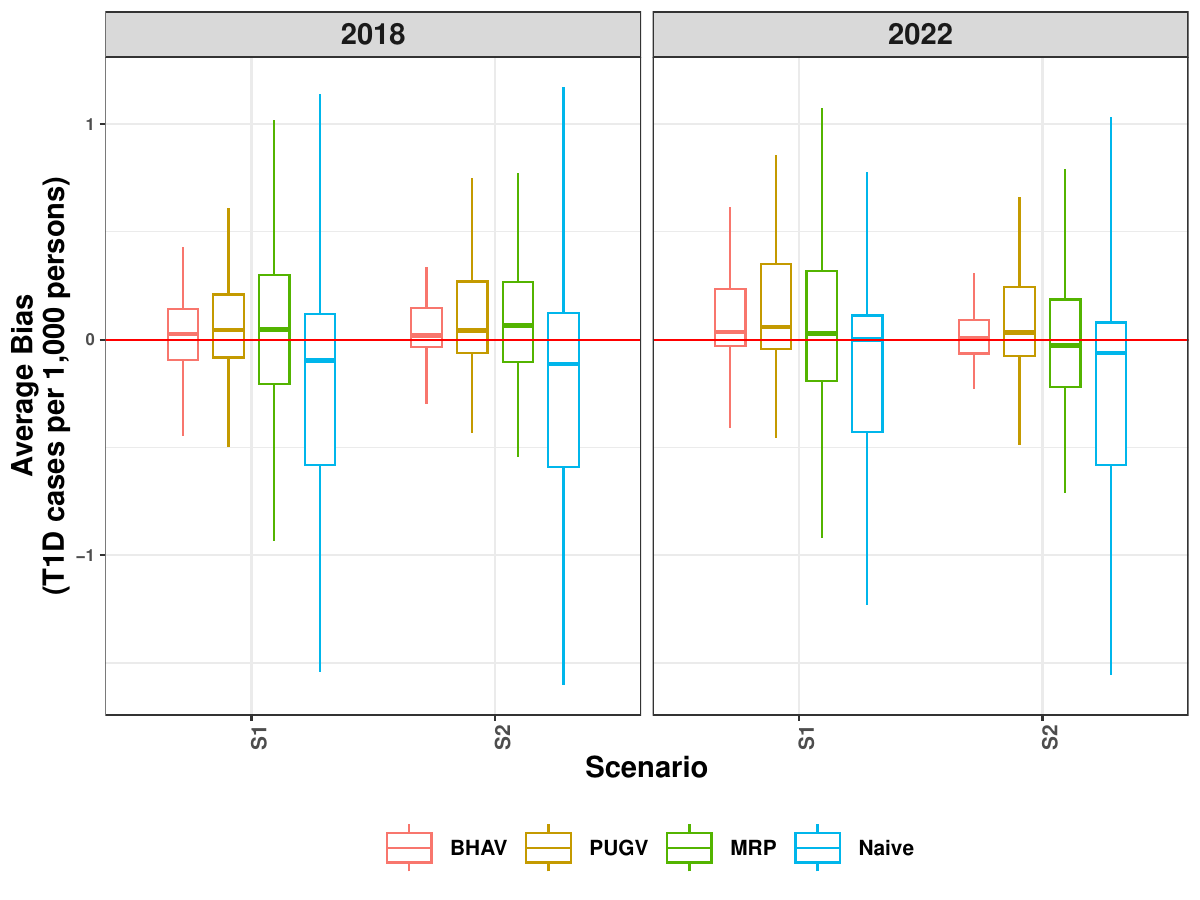}
\caption{Box plots of the average bias in T1D prevalence by county and year according to the superpopulation models and Na\"ive estimation method by scenario S1 and S2. BHAV, PUGV, and MRP refer to Bayesian Hierarchical Analysis of Variance, Partitioning Unobserved Geographic Variations, and Multilevel Regression and Poststratification, respectively.}
\label{f:Fig_plasmode_sim_ave_bias_fips_county_code_index_year_a.pdf}
\end{figure}

\begin{figure}[htbp!]
\centering
\includegraphics[scale = 0.5]{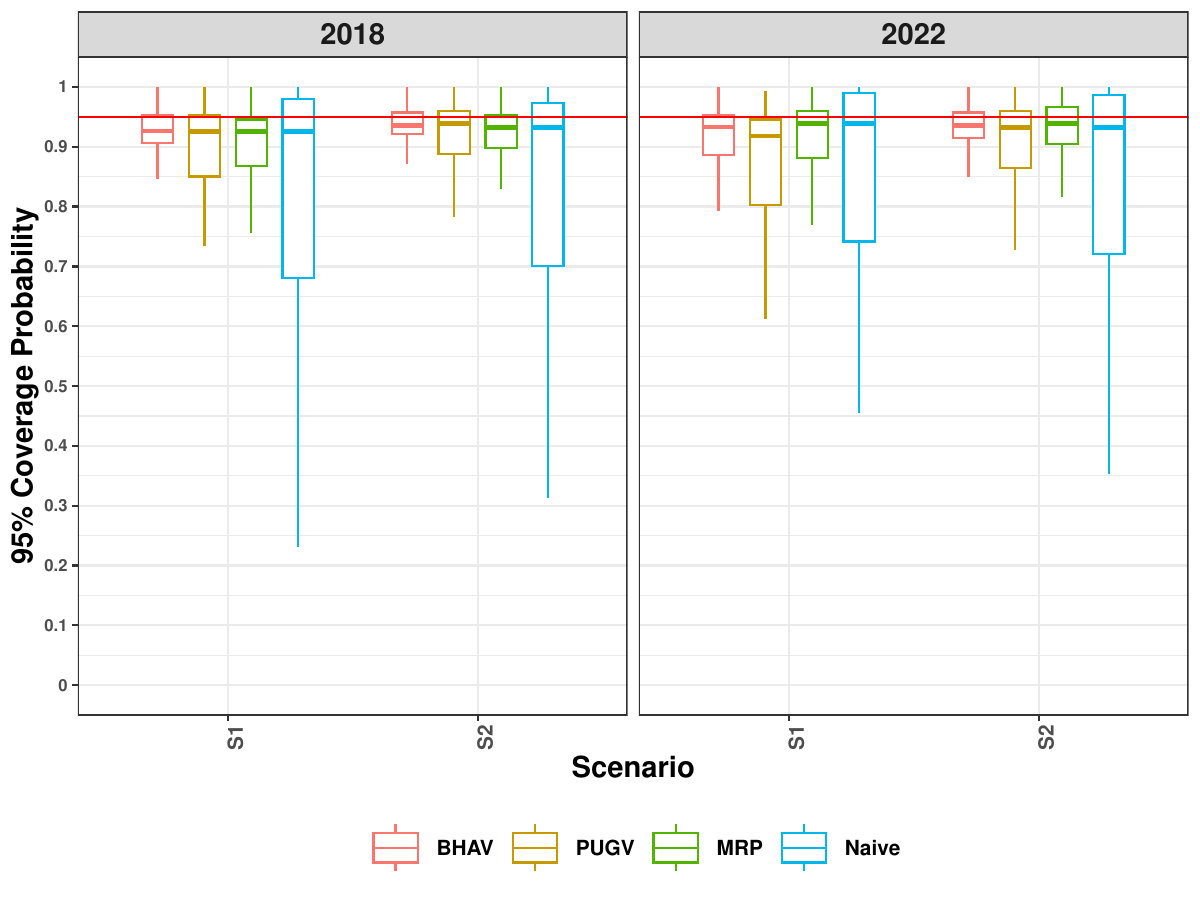}
\caption{Box plots of the 95\% coverage probabilities for T1D prevalence by county and year according to the superpopulation models and Na\"ive estimation method by scenario S1 and S2. BHAV, PUGV, and MRP refer to Bayesian Hierarchical Analysis of Variance, Partitioning Unobserved Geographic Variations, and Multilevel Regression and Poststratification, respectively.}
\label{f:Fig_plasmode_sim_coverage_fips_county_code_index_year_a.pdf}
\end{figure}

\clearpage

\subsection*{DiCAYA Study Group}

Collaborators of the DiCAYA Study Group are listed below.

\begin{description}
\item[Children’s Hospital of Philadelphia/PEDSnet:] L. Charles Bailey, MD, PhD (MPI); Christopher B. Forrest, MD, PhD (MPI); Levon H. Utidjian, MD, MBI; Mitchell Maltenfort, PhD; Amy S. Shah, MD, MS (for CCHMC/ PEDSnet); Eneida A. Mendonca, MD, PhD (for CCHMC/PEDSnet); G. Todd Alonso, MD (for Colorado/PEDSnet); Sara Deakyne Davies, MPH (for Colorado/PEDSnet); H. Timothy Bunnell, PhD (for Nemours/PEDSnet); Anne Kazak, PhD (for Nemours/PEDSnet); Melody Kitzmiller, BS (for NCH/PEDSnet); Manmohan Kamboj, MD (for NCH/PEDSnet); Dimitri Christakis, MD, MPH (for SCH/PEDSnet); Daksha Ranade, MPH, MBA (for SCH/PEDSnet). 

\item[Geisinger:] Annemarie G. Hirsch, PhD, MPH (PI); Joseph J. DeWalle, BS; H. Lester Kirchner, PhD; Meredith Lewis, MS; Dione G. Mercer, BS, BA; Cara M. Nordberg, MPH; Amy Poissant, BS; Brian S. Schwartz MD, MS.

\item[Indiana University/Regenstrief:] Brian E. Dixon, PhD, MPA (PI); Katie S. Allen, PhD (for Regenstrief Institute); Muchiri Wandai, PhD, MPH (for Regenstrief Institute); Nimish Valvi, DrPH, MPH, MBBS (for Regenstrief Institute); Jeff Warvel (for Regenstrief Institute); Acatia R. Greenwell, MPH (for Regenstrief Institute); John Price (for Regenstrief Institute); Ashley Wiensch, MPH (for Regenstrief Institute); Jiali Han, PhD (for IU Fairbanks School of Public Health); Tamara S Hannon, MD (for IU School of Medicine); Shaun J. Grannis, MD, MS (for IU School of Medicine).

\item[Kaiser Permanente of Southern California:] Kristi Reynolds, PhD, MPH (PI); Jenny Chang, MPH; John Chang, MPH; Teresa N. Harrison, SM; Eva Lustigova, MPH; Don McCarthy, MA; Matthew T. Mefford, PhD; Rong Wei, MA; Hui Zhou, PhD.

\item[Lurie Children's Hospital:] Marc Rosenman, MD (PI); Lu Zhang, PhD; George Lales, MS; Anthony Wong, PhD; Yuan Luo, PhD (for Northwestern University and the University of Washington); Allison Zelinski, MS (for Children's Hospital of Philadelphia); Mark Weiner, MD (for Weill Cornell Medicine); Pedro Rivera, MS (for OCHIN); Thuy Le, MPH (for OCHIN); Thomas Carton, PhD, MS (for Louisiana Public Health Institute); Elizabeth Nauman, MPH, PhD (for Louisiana Public Health Institute); Harold P. Lehmann, MD, PhD (for Johns Hopkins University).

\item[NYU Langone Health:] Jasmin Divers, MS, PhD (MPI); Lorna E. Thorpe, MPH, PhD (MPI); Rebecca Anthopolos, DrPH; Sarah Conderino, MPH; Andrew Fair, ScM, MS; Jessica Guillaume, MPH; Alan Jacobson, MD; Soo Young Kim, MA; David C. Lee, MD, MS; Xiaochun Li, PhD; Anand Rajan, MPH.

\item[University of Colorado Denver:] Tessa Crume, PhD, MPH (MPI); Dana Dabelea, MD, PhD (MPI); Toan Ong, PhD; Rebecca Conway, PhD, MPH; Anna Bellatorre, PhD, MA; Shawna Burgett, PhD; Jack Pattee, PhD; Bethelhem (Betty) Shiferaw, MPH; Theresa Anderson, MS.

\item[University of Florida:] Jiang Bian, PhD (MPI); Yi Guo, PhD (MPI); Hui Shao, MD, PhD (MPI); Elizabeth A. Shenkman, PhD (MPI); Sarah J. Bost, MSLS; William T. Donahoo, MD; William R. Hogan, MD, PhD; Piaopiao Li; Tianchen Lyu, MS; Mattia Prosperi, PhD; Yonghui Wu, PhD. 

\item[University of South Carolina:] Angela D. Liese, PhD, MPH (PI); Bo Cai, PhD, MSc; Caroline Rudisill, MSc, PhD; Rabins Wosti, PhD; Lisa Knight, MD, MBA. For Medical University of South Carolina: Deborah Bowlby, MD; Jihad S. Obeid, MD; Terry Headley. For Prisma Health: Elaine Apperson MD; Alex Ewing, PhD, MPH.  
\item[Centers for Disease Control and Prevention, Division of Diabetes Translation:] Ibrahim Zaganjor, PhD; Meda Pavkov, MD, PhD. 
\end{description}

\end{document}